\documentclass[universe,review,accept,moreauthors,pdftex,10pt,a4paper]{mdpi}

\usepackage{amsfonts}
\usepackage{cancel}

\DeclareMathOperator{\extdm}{d}
\newcommand{\extd}{\extdm \!}

\providecommand{\Lt}{{\tt L}}
\renewcommand{\Lt}{{\tt L}}
\providecommand{\Mt}{{\tt Mt}}
\renewcommand{\Mt}{{\tt M}}
\providecommand{\Ut}{{\tt U}}
\renewcommand{\Ut}{{\tt U}}
\providecommand{\Vt}{{\tt V}}
\renewcommand{\Vt}{{\tt V}}
\providecommand{\Gt}{{\tt G}}
\renewcommand{\Gt}{{\tt G}}

\providecommand{\Jt}{{\tt J}}
\renewcommand{\Jt}{{\tt J}}
\providecommand{\Pt}{{\tt P}}
\renewcommand{\Pt}{{\tt P}}
\providecommand{\Tt}{{\tt T}}
\renewcommand{\Tt}{{\tt T}}
\providecommand{\St}{{\tt S}}
\renewcommand{\St}{{\tt S}}
\providecommand{\Ht}{{\tt H}}
\renewcommand{\Ht}{{\tt H}}
\providecommand{\Kt}{{\tt K}}
\renewcommand{\Kt}{{\tt K}}
\providecommand{\IW}{{\.In\"on\"u--Wigner }}
\renewcommand{\IW}{{\.In\"on\"u--Wigner }}

\usepackage{todonotes} 
\usepackage{tikz}
\usepackage{tikz-cd}
\usepackage{tikz-3dplot}
\newcommand\parmp{\mathbin{\vcenter{\hbox{%
  \oalign{$\scriptstyle({-})$\cr
          \noalign{\kern-0.8ex}
          \hfil$\scriptscriptstyle+$\hfil\cr}%
}}}}
\def\comma{,}
\usepackage{tensor}

\firstpage{1}
\articlenumber{20}
\doinum{10.3390/universe4010020}
\pubvolume{4}
\pubyear{2018}
\copyrightyear{2018}
\history{Received: 31 October 2017 ; Accepted: 9 January 2018; Published: 19 January 2018}

\Title{Higher Spins Without (Anti-)de Sitter}

\Author{Stefan Prohazka\orcidB{} and Max Riegler\orcidC{}}

\AuthorNames{Stefan Prohazka, Max Riegler}

\address[1]{%
Universit\'e Libre de Bruxelles and International Solvay Institutes\\
Physique Mathématique des Interactions Fondamentales\\
Campus Plaine - CP~231,
B-1050 Bruxelles,
Belgium, Europe\\
stefan.prohzaka@ulb.ac.be, max.riegler@ulb.ac.be}

\abstract{Can the holographic principle be extended beyond the well known AdS/CFT correspondence? During the last couple of years there has been a substantial amount of research trying to find answers for this question. In this work we provide a review of recent developments of three-dimensional theories of gravity with higher spin symmetries. We focus in particular on a proposed holographic duality involving asymptotically flat spacetimes and higher spin extended $\mathfrak{bms}_3$ symmetries. 
In addition we also discuss developments
concerning relativistic and
nonrelativistic higher spin algebras.
As a special case Carroll gravity will be
discussed in detail.
}


\begin{document}

\newpage
\tableofcontents
\newpage

\section{Introduction}
\label{sec:introduction}

Higher spin theories on Anti-de Sitter (AdS) backgrounds provide many useful insights into various aspects of the holographic principle. Many of these works were inspired by the seminal work of Klebanov and Polyakov \cite{Klebanov:2002ja,Mikhailov:2002bp,Sezgin:2002rt} who conjectured a holographic correspondence between the $O(N)$ vector model in three dimensions and Fradkin--Vasiliev higher spin gravity on AdS$_4$ \cite{Fradkin:1987ks,Fradkin:1986qy,Vasiliev:1990en} (see \cite{Sagnotti:2010at,Vasiliev:2012vf,Didenko:2014dwa} for reviews and \cite{Giombi:2009wh,Giombi:2010vg,Koch:2010cy,Giombi:2011ya,Douglas:2010rc,Giombi:2012ms,Sleight:2016dba} for some key developments). There are many features of higher spin theories that make them interesting to study. In the context of holography one of these features is that it is a weak/weak correspondence \cite{Maldacena:2011jn,Maldacena:2012sf}. In contrast, the usual AdS/CFT correspondence \cite{Maldacena:1997re,Gubser:1998bc,Witten:1998qj} is a weak/strong correspondence that makes 
it useful for applications but harder to check in detail since calculations are often feasible only on one side of the correspondence.\\
In particular three-dimensional higher spin theories are useful in this context, since -- in contrast to the higher-dimensional examples -- one can truncate the otherwise infinite tower of higher spin excitations \cite{Aragone:1983sz}. Furthermore, the equations that describe the propagation of a massless field of spin $s$ in three dimensions imply that there are no local degrees of freedom when $s \geq 2$. Thus one can also formulate three-dimensional higher spin theories as Chern-Simons theories \cite{Blencowe:1988gj} with specific boundary conditions \cite{Henneaux:2010xg,Campoleoni:2010zq,Gaberdiel:2011wb,Campoleoni:2011hg,Henneaux:2013dra}. This is a considerable simplification in comparison with the more complicated higher-dimensional case. Developments in three-dimensional higher spin theories in AdS include\footnote{Further examples can be found in \cite{Castro:2010ce,Ammon:2011nk,Castro:2011fm,Ammon:2011ua,Henneaux:2012ny,Campoleoni:2012hp,deBoer:2013gz,Compere:2013nba,deBoer:2014fra,Campoleoni:2013iha,Campoleoni:2014tfa,Castro:2011iw,Castro:2014mza,deBoer:2014sna,Banados:2015tft}} e.g. the discovery of minimal model holography \cite{Gaberdiel:2010pz, Gaberdiel:2012uj,Candu:2012ne}, higher spin black holes \cite{Gutperle:2011kf, Castro:2011fm, Ammon:2012wc,Bunster:2014mua} and higher spin holographic entanglement entropy \cite{Ammon:2013hba, deBoer:2013vca}.\\
Since higher spin holography in AdS backgrounds has lead to many interesting insights a natural question to ask is how to generalize this duality such that it involves other spacetimes or quantum field theories. And, indeed, there are many applications where one has spacetimes that do not asymptote to AdS or do so in a weaker way compared to the Brown-Henneaux boundary conditions \cite{Brown:1986nw}. Some examples include Lobachevsky spacetimes \cite{Riegler:2012fa,Afshar:2012nk,Afshar:2012hc}, null warped AdS and their generalizations Schr\"odinger \cite{Son:2008ye,Balasubramanian:2008dm,Adams:2008wt,Breunhoelder:2015waa,Lei:2015ika}, Lifshitz spacetimes \cite{Kachru:2008yh,Gutperle:2013oxa,Gary:2014mca}, flat space \cite{Afshar:2013vka, Gonzalez:2013oaa,Grumiller:2014lna,Gary:2014ppa} and de Sitter holography
\cite{Krishnan:2013zya,Krishnan:2013cra,Basu:2015exa}. Some of these spacetimes play an important role as gravity duals for nonrelativistic CFTs, which are a common occurrence in e.g. condensed matter physics and thus may be able to provide new insight in these strongly interacting systems. Schr\"odinger spacetimes for example can be used as a holographic dual to describe cold atoms \cite{Son:2008ye,Balasubramanian:2008dm}.\\
Even though non-AdS higher spin holography is a rather new field of research there has been quite a lot of research in this direction during the last couple of years. Our aim with this review is to give an overview of the results and ideas that have been accumulated over the years. A special focus of this review will be on higher spin theories in three-dimensional flat space as well as the construction of new higher spin theories using kinematical algebras as bulk isometries.\\
This review is organized as follows. In Section~\ref{sec:non-ads-through} we present an overview of non-AdS holography that makes use of non-AdS boundary conditions of certain higher spin gravity theories. In Section~\ref{sec:flat-space-higher} we focus on a specific example of non-AdS higher spin holography namely flat space higher spin gravity. Section~\ref{sec:non-ads-through-1} can be read independently of Section~\ref{sec:non-ads-through} and \ref{sec:flat-space-higher} and explains a different approach of studying non-AdS higher spin theories not via boundary conditions but rather by using different choices of gauge algebras realizing certain higher spin theories in the bulk.

\section{Non-AdS Through Boundary Conditions}
\label{sec:non-ads-through}

A lot of the progress in non-AdS higher spin holography has been achieved by imposing suitable boundary conditions that in turn allow to compare physical boundary observables with their bulk counterpart. In three dimensions this can be done rather nicely using a first order formulation of gravity \cite{Achucarro:1987vz,Witten:1988hc}. In order to set the stage for non-AdS higher spin holography we give now a brief review\footnote{Parts of this review are based on \cite{Riegler:2016hah,Riegler:2017fqv,Prohazka:2017pkc}. There is also a slight overlap with \cite{Afshar:2014rwa}.} of this formulation for the case of Einstein gravity as well as AdS higher spin gravity.

\subsection{The (Higher Spin) Chern-Simons Formulation of Gravity}
\label{sec:higher spin-chern}

In many situations it is advantageous to not describe gravity in terms of a metric formulation but rather in terms of local orthonormal Lorentz frames. That is, one exchanges the metric $g_{\mu\nu}$ with a vielbein $e$ and a spin connection $\omega$. In three dimensions the dreibein $e$ and dualized spin connection $\omega$ can have the same index structure in their Lorentz indices. Thus, one can combine these two quantities into a single gauge field
	\begin{equation}\label{eq:RelationAVielbeinSpinConnection}
		\mathcal{A}\equiv e^a\Pt_a+\omega^a\Jt_a,
	\end{equation}
where the generators $\Pt_a$ and $\Jt_a$ generate the following Lie algebra\footnote{We raise and lower indices with $\eta=\mathrm{diag}(-,+,+)$ and $\epsilon_{012}=1$.}
	\begin{equation}\label{eq:AllThreeAlgebras}
		[\Pt_a,\Pt_b]=\mp\frac{1}{\ell^2}\epsilon_{abc}\Jt^c,\quad[\Jt_a,\Jt_b]=\epsilon_{abc}\Jt^c,\quad[\Jt_a,\Pt_b]=\epsilon_{abc}P^c.
	\end{equation}
	\begin{itemize}
		\item For $-\frac{1}{\ell^2}$, i.e.\ de Sitter spacetimes this gauge algebra is $\mathfrak{so}(3,1)$.
		\item For $\ell\rightarrow\infty$, i.e.\ flat spacetimes this gauge algebra is $\mathfrak{isl}(2,\mathbb{R})$.
		\item For $+\frac{1}{\ell^2}$, i.e.\ Anti-de Sitter spacetimes this gauge algebra is\\
		${\mathfrak{so}(2,2)\sim\mathfrak{sl}(2,\mathbb{R})\oplus\mathfrak{sl}(2,\mathbb{R})}$.
	\end{itemize}
It has been shown \cite{Achucarro:1987vz,Witten:1988hc} that the Chern-Simons action
	\begin{equation}\label{eq:ChernSimonsAction}
		S_{\textnormal{CS}}[\mathcal{A}]=\frac{k}{4\pi}\int_{\mathcal{M}}\left<\mathcal{A}\wedge\extd\mathcal{A}+\frac{2}{3}\mathcal{A}\wedge\mathcal{A}\wedge\mathcal{A}\right>,
	\end{equation}
defined on a three-dimensional manifold $\mathcal{M}=\Sigma\times\mathbb{R}$, with the invariant nondegenerate symmetric bilinear form
	\begin{equation}\label{eq:Isl2RBilinFormJanP}
		\left<\Jt_a,\Pt_b\right>=\eta_{ab},\quad\left<\Jt_a,\Jt_b\right>=\left<\Pt_a,\Pt_b\right>=0,
	\end{equation}
        is equivalent (up to boundary terms) to the Einstein-Hilbert-Palatini action\footnote{For a nice and explicit calculation see Appendix A in \cite{deBoer:2013vca}.} with vanishing cosmological constant,
        provided one identifies the Chern-Simons level $k$ with Newton's constant $G$ in three dimensions as
	\begin{equation}
          k=\frac{1}{4G}.
        \end{equation}
The just mentioned bilinear form is also called an invariant metric.
Its properties are important for each component of the Chern-Simons gauge field to have a kinematical term (nondegeneracy) and for the action to be invariant under gauge transformations (invariance). This is the general setup for Einstein gravity in three dimensions using the Chern-Simons formalism.\\
AdS spacetimes in particular have some very nice features in this formalism that allow for a very efficient treatment of many physical questions. Maybe the most important feature from a Chern-Simons perspective is that the isometry algebra of AdS $\mathfrak{so}(2,2)$ is a direct sum of two copies of $\mathfrak{sl}(2,\mathbb{R})$. This also means that one can split the gauge field $\mathcal{A}$ into two parts $A$ and $\bar{A}$. On the level of the generators this split can be made explicit by introducing the generators
	\begin{equation}
		\Tt_a=\frac{1}{2}\left(\Jt_a+\ell \Pt_a\right),\qquad\bar{\Tt}_a=\frac{1}{2}\left(\Jt_a-\ell \Pt_a\right).
	\end{equation}
These new generators satisfy
	\begin{equation}\label{eq:SplitAlgebra}
		\left[\Tt_a,\bar{\Tt}_b\right]=0,\qquad\left[\Tt_a,\Tt_b\right]=\epsilon_{abc}\Tt^{c},\qquad\left[\bar{\Tt}_a,\bar{\Tt}_b\right]=\epsilon_{abc}\bar{\Tt}^{c}.
	\end{equation}
Both $\Tt_a$ and $\bar{\Tt}_a$ satisfy an $\mathfrak{sl}(2,\mathbb{R})$ algebra. From \eqref{eq:Isl2RBilinFormJanP} one can immediately see that the invariant bilinear forms are given by
	\begin{equation}\label{eq:TTbarInvBilform}
		\left<\Tt_a,\Tt_b\right>=\frac{\ell}{2}\eta_{ab},\qquad\left<\bar{\Tt}_a,\bar{\Tt}_b\right>=-\frac{\ell}{2}\eta_{ab}.
	\end{equation}
The gauge field $\mathcal{A}$ in terms of this split can now be written as
	\begin{equation}\label{eq:AdS3GaugeFieldSplit}
		\mathcal{A}=
			A^a\Tt_a+\bar{A}^a\bar{\Tt}_a.
	\end{equation}
Thus, after implementing this explicit split of $\mathfrak{so}(2,2)$ into $\mathfrak{sl}(2,\mathbb{R})\oplus\mathfrak{sl}(2,\mathbb{R})$, the Chern-Simons action \eqref{eq:ChernSimonsAction} also splits into two contributions
	\begin{equation}
		S_{\textnormal{EH}}^{\textnormal{AdS}}[A,\bar{A}]=S_{\textnormal{CS}}[A]+S_{\textnormal{CS}}[\bar{A}],
	\end{equation}
where the invariant bilinear forms appearing in the Chern-Simons action are given by \eqref{eq:TTbarInvBilform}. Since both $\Tt_a$ and $\bar{\Tt}_a$ satisfy an $\mathfrak{sl}(2,\mathbb{R})$ algebra it is usually practical to not distinguish between the two generators, i.e.\ setting $\Tt_a=\bar{\Tt}_a$. This in turn also means that the invariant bilinear form in both sectors will be the same. From \eqref{eq:TTbarInvBilform}, however, one knows that the invariant bilinear form in both sectors should have opposite sign. This is not a real problem since this relative minus sign can be easily introduced by hand by not taking the sum, but rather the difference of the two Chern-Simons actions
	\begin{equation}\label{eq:ChernSimonsActionAAbar}
		S_{\textnormal{EH}}^{\textnormal{AdS}}=S_{\textnormal{CS}}[A]-S_{\textnormal{CS}}[\bar{A}]=\frac{1}{16\pi G}\left[\int_\mathcal{M}\extd ^3x\sqrt{|g|}\left(\mathcal{R}+\frac{2}{\ell^2}\right)-\int_{\partial\mathcal{M}}\omega^a\wedge e_a\right].
	\end{equation}
As the factor of $\ell$ in \eqref{eq:TTbarInvBilform} only yields an overall factor of $\ell$ to the action \eqref{eq:ChernSimonsActionAAbar} one can also absorb this factor simply in the Chern-Simons level as
	\begin{equation}\label{eq:ChernSimonsLevelNewtonConstant}
		k=\frac{\ell}{4G}.
	\end{equation}
The form of the Chern-Simons connection \eqref{eq:ChernSimonsActionAAbar} is usually the one discussed in the literature on AdS holography in three dimensions. The big advantage of this split into an unbarred and a barred part in the case of AdS holography is that usually one only has to explicitly treat one of the two sectors, as the other sector works in complete analogy, up to possible overall minus signs.\\
Aside from this technical simplification there is another reason why the Chern-Simons formulation is very often used in AdS and non-AdS holography alike. 
While a generalization to higher-dimensional gravity
is easier in the metric formulation
higher spin extensions are more straightforward in this setup. Since a Chern-Simons gauge theory with gauge algebra $\mathfrak{sl}(2,\mathbb{R})\oplus\mathfrak{sl}(2,\mathbb{R})$ corresponds to spin-2 gravity with AdS isometries, it is natural to promote the gauge algebra to $\mathfrak{sl}(N,\mathbb{R})\oplus\mathfrak{sl}(N,\mathbb{R})$\footnote{To be more precise: The spectrum of the higher spin gravity theory depends on the specific embedding of $\mathfrak{sl}(2,\mathbb{R})\hookrightarrow\mathfrak{sl}(N,\mathbb{R})$. A very popular choice in the literature on AdS higher spin holography is the principal embedding of $\mathfrak{sl}(2,\mathbb{R})\hookrightarrow\mathfrak{sl}(N,\mathbb{R})$. This is due to the fact that all generators in that particular embedding have a conformal weight greater or equal to two and thus can be interpreted as describing fields with spin $s\geq2$.} in order to describe gravity theories with additional higher spin symmetries. And indeed in \cite{Blencowe:1988gj} it was shown that for $N\geq3$ such a Chern-Simons theory describes the nonlinear interactions of gravity coupled to a finite tower of massless integer spin-$s\leq N$ fields.\\
From a holographic perspective one point of interest is given by the asymptotic symmetries of these higher spin gravity theories for given sets of boundary conditions. The first set of consistent boundary conditions that lead to interesting higher spin extensions of the Virasoro algebra has been worked out in \cite{Henneaux:2010xg,Campoleoni:2010zq}.\\ 
Aside from extending the gauge algebra one also has to take care of the normalization of the Chern-Simons level $k$. This has to be done in such a way that the spin-2 part of the resulting higher spin theory coincides with Einstein gravity. 
In order to give the Chern-Simons description an interpretation in terms of a metric one needs to re-extract the geometric information hidden in the gauge field $\mathcal{A}$. For AdS as well as flat space higher spin theories (in the principal embedding) this can be done via
    \begin{equation}
        g_{\mu\nu}=\#\langle e^z_\mu,e^z_\nu\rangle,
    \end{equation}
where $\#$ is some normalization constant and $e^z_\mu$ is the so called \emph{zuvielbein} that can be seen as a higher spin extension of the dreibein $e_\mu$ encountered previously. The expression \emph{zuvielbein} is a German expression meaning ``too many legs'' to emphasize that the object $e_\mu^z$ now contains more geometric information than the usual dreibein found in spin-2 gravity. In the well known AdS case the previous equation can be equivalently written as \cite{Castro:2011iw,Ammon:2012wc}
	\begin{equation}\label{eq:CStoMetricTrace}
		g_{\mu\nu}=
		\#\left<A_\mu-\bar{A}_\mu,A_\nu-\bar{A}_\nu\right>.
	\end{equation}

\subsection{Boundary Terms and Higher Spins}

After this brief reminder about the Chern-Simons formulation of (AdS) higher spin theories in three dimensions we now want to set the stage for the transition to non-AdS spacetimes. All of the interesting physics aside global properties -- in three-dimensional gravity are governed by degrees of freedom at the boundary. Thus it is of utmost importance to make sure that one can impose consistently fall off conditions on the gauge field\footnote{Or the metric in a second order formulation.} at the asymptotic boundary. Consistent in this context means that one still has a well defined variational principle after imposing said boundary conditions. This is crucial since a consistent variational principle is the core principle underlying the definition of equations of motion of a physical system described by some action. Thus the necessity of having such a well defined variational principle in turn also influences the possible set of boundary conditions that can be consistently imposed.\\
In order to see this take a closer look at the variation of the Chern-Simons action \eqref{eq:ChernSimonsAction}
	\begin{equation}
		\delta S_{\textrm{CS}}[\mathcal{A}]=\frac{k}{2\pi}\int_{\mathcal{M}}\left\langle\delta \mathcal{A}\wedge F\right\rangle+\frac{k}{4\pi}\int_{\partial\mathcal{M}}\left\langle\delta \mathcal{A}\wedge \mathcal{A}\right\rangle.
	\end{equation}
This expression only vanishes on-shell i.e.\\ when $F=0$ if the second term on the right hand side vanishes as well. Assuming that the boundary $\partial\mathcal{M}$ is parametrized by a timelike coordinate $t$ and an angular coordinate $\varphi$ this amounts to
	\begin{equation}
		\frac{k}{4\pi}\int_{\partial\mathcal{M}}\left\langle\delta \mathcal{A}_t\mathcal{A}_\varphi-\delta\mathcal{A}_\varphi\mathcal{A}_t\right\rangle.
	\end{equation}
This term vanishes, for instance, if either $\mathcal{A}_\varphi$ or $\mathcal{A}_t$ are equal to zero at the boundary. This is quite a stringent condition on possible boundary conditions. Thus it would be nice to have a way of enlarging the possible set of consistent boundary conditions. This can be most easily done by adding a boundary term $B[\mathcal{A}]$ to the Chern-Simons action \eqref{eq:ChernSimonsAction}.\\
One could consider for example the following boundary term
	\begin{equation}
		B[\mathcal{A}]=\frac{k}{4\pi}\int_{\partial\mathcal{M}}\left\langle\mathcal{A}_\varphi \mathcal{A}_t\right\rangle.
	\end{equation}
Including this boundary term the total variation of the resulting action is on-shell
	\begin{equation}
		\delta S_{\textrm{CS}}[\mathcal{A}]^{\textrm{Tot}}=\frac{k}{2\pi}\int_{\partial\mathcal{M}}\left\langle\delta \mathcal{A}_t\mathcal{A}_\varphi\right\rangle.
	\end{equation}
Vanishing of the total variation then can be achieved for example via
	\begin{equation}
		\mathcal{A}_\varphi\Bigr|_{\partial\mathcal{M}}=0\quad\textnormal{or}\quad\delta \mathcal{A}_t\Bigr|_{\partial\mathcal{M}}=0.
	\end{equation}
Choosing $\delta \mathcal{A}_t\Bigr|_{\partial\mathcal{M}}=0$, one is thus able to enlarge the possible set of boundary conditions by making sure that the \emph{variation} of a part of the Chern-Simons connection vanishes.
\subsection{Examples of Non-AdS Spacetimes Realized With Higher Spin Symmetries}
\label{sec:examples-non-ads}
Adding a suitable boundary term to the Chern-Simons connection in order to allow for a bigger set of possible boundary conditions is one of the necessary prerequisites for doing non-AdS holography. The second one is due to an observation first made explicit in \cite{Gary:2012ms}. That is, higher spin isometries i.e.\ isometries based on $\mathfrak{sl}(N,\mathbb{R})$, can be used to realize certain non-AdS spacetimes asymptotically.\\
Take for example a direct product of maximally symmetric spacetimes such as AdS$_2\times\mathbb{R}$ or $\mathbb{H}_2\times\mathbb{R}$, where $\mathbb{H}_2$ is the two dimensional Lobachevsky plane. Then assume that the gauge algebra of the Chern-Simons connection is given by a direct sum of an embedding of $\mathfrak{sl}(2,\mathbb{R})\hookrightarrow\mathfrak{sl}(N,\mathbb{R})$ that contains at least one singlet $\St$ with $\textrm{tr}(\St^2)\neq0$ and whose $\mathfrak{sl}(2,\mathbb{R})$ generators are labelled as $\Lt_n$. Furthermore, assume that the manifold $\mathcal{M}$ where the Chern-Simons theory is defined on has the topology of a cylinder with radial coordinate $\rho$ and boundary coordinates $x^1$ and $x^2$. Then, using \eqref{eq:CStoMetricTrace} and the connection
    \begin{equation}
    A = \Lt_0\extd\rho+a_1 e^\rho \Lt_+\extd x^1, \qquad 
    \bar{A}= -\Lt_0\extd\rho+e^\rho\Lt_-\extd x^1+\St\extd x^2,
    \end{equation}
where $a_1$ is some non-zero constant one obtains the following non-vanishing metric components
    \begin{equation}
        g_{\rho\rho}=2\textrm{tr}(\Lt^2_0),\qquad
        g_{11}=-a_1 \textrm{tr}(\Lt_+\Lt_-)e^{2\rho},\qquad
        g_{22}=\frac{1}{2}\textrm{tr}(\St^2).
    \end{equation}
Depending on the sign of $a_1$ this metric is locally and asymptotically either AdS$_2\times\mathbb{R}$ or $\mathbb{H}_2\times\mathbb{R}$.\\
This was a first indication that one can model Lobachevsky spacetimes using higher spin gauge invariant Chern-Simons theories. Following up on this a natural question to ask is whether or not one can introduce boundary conditions in this setup that lead to interesting boundary dynamics. In \cite{Riegler:2012fa,Afshar:2012hc} it was shown that this is, indeed, possible using a very general algorithm\footnote{See also e.g. \cite{Brown:1986nw,Henneaux:1992ig,Blagojevic:2002du}.}. This algorithm can roughly be summarized by the following steps:

\paragraph{Identify Bulk Theory and Variational Principle:}

The first step in this algorithm consists of identifying the bulk theory\footnote{This usually boils down to choosing an appropriate embedding of $\mathfrak{sl}(2,\mathbb{R})\hookrightarrow\mathfrak{sl}(N,\mathbb{R})$ and then fix the Chern-Simons connections $A$ and $\bar{A}$ in such a way that they correctly reproduce the desired gravitational background.} one wants to describe. After that one has to propose a suitable generalized variational principle i.e.\ add appropriate boundary terms that are consistent with the theory under consideration. 

\paragraph{Impose Suitable Boundary Conditions:}

After having chosen the bulk theory the next step in this algorithm is choosing appropriate boundary conditions for the Chern-Simons connection $\mathcal{A}$. This is the most crucial step in the whole analysis as the boundary conditions essentially determine the physical content of the putative dual field theory at the boundary. Since one is dealing with a Chern-Simons gauge theory one also has some gauge freedom left that can be used to simplify computations. Choosing a gauge
	\begin{equation}\label{eq:AdSxRGenrealFormGaugePrescription}
		\mathcal{A}_\mu=b^{-1}\left(\mathfrak{a}_\mu+a^{(0)}_\mu+a^{(1)}_\mu\right)b,\enspace
		b=b(\rho),
	\end{equation}
one can then identify the following three contributions to the Chern-Simons connections:
	\begin{itemize}
		\item $\mathfrak{a}_\mu$ denotes the (fixed) background that was chosen in the previous step.
		\item $a^{(0)}_\mu$ corresponds to state dependent leading contributions in addition to the background that contain all the physical information about the field degrees of freedom at the boundary.
		\item $a^{(1)}_\mu$ are subleading contributions.
	\end{itemize}
Choosing suitable boundary conditions in this context thus means choosing $a^{(0)}_\mu$ and 	$a^{(1)}_\mu$ in such a way that there exist gauge transformations which preserve these boundary conditions i.e.\
	\begin{equation}
		\delta_\varepsilon \mathcal{A}_\mu=\mathcal{O}\left(b^{-1}a^{(0)}_\mu b\right)+\mathcal{O}\left(b^{-1}a^{(1)}_\mu b\right),
	\end{equation}
for some gauge parameter $\varepsilon$ which can also be written as
	\begin{equation}
		\varepsilon=b^{-1}\left(\epsilon^{(0)}+\epsilon^{(1)}\right)b.
	\end{equation}
The transformations $\epsilon^{(0)}$ usually generate the asymptotic symmetry algebra while $\epsilon^{(1)}$ are trivial gauge transformations.
		
\paragraph{Perform Canonical Analysis and Check Consistency of Boundary Conditions:}

Once the boundary conditions and the gauge transformations that preserve these boundary conditions have been fixed one has to determine the canonical boundary charges. This is a standard procedure which is described in great detail for example in \cite{Henneaux:1992ig,Blagojevic:2002du} and is based on the results of \cite{Regge:1974zd}. This procedure eventually leads to the variation of the canonical boundary charge
	\begin{equation}
		\delta\mathcal{Q}[\varepsilon]=\frac{k}{2\pi}\int_{\partial\Sigma}\extd \varphi\left<\epsilon^{(0)}\delta a_\varphi^{(0)}\right>,
	\end{equation}
where $\varphi$ parametrizes the cycle of the boundary cylinder. Of course one also has to check whether or not the boundary conditions chosen at the beginning of the algorithm are actually physically admissible. For the three-dimensionsal higher spin gravity gravity examples that are treated in this review that means that the variation of the canonical boundary charge is finite, conserved in time and integrable in field space. However, we want to stress that there are also other examples such as e.g. \cite{Barnich:2011mi} where one can also have physically interesting boundary conditions where the canonical boundary charges do not necessarily meet all of the previously stated conditions.

\paragraph{Determine Semiclassical Asymptotic Symmetry Algebra:}

This step consists in working out the Dirac brackets between the canonical generators $\mathcal{G}$ that directly yield the semiclassical asymptotic symmetry algebra. There is a well known trick which can be used to simplify calculations at this point. Assume that one has two charges with Dirac bracket $\left\{\mathcal{G}[\varepsilon_1],\mathcal{G}[\varepsilon_2]\right\}$. Then one can exploit the fact that these  brackets generate a gauge transformation as $\left\{\mathcal{G}[\varepsilon_1],\mathcal{G}[\varepsilon_2]\right\}=\delta_{\epsilon_2}\mathcal{G}$, and read of the Dirac brackets by evaluating $\delta_{\epsilon_2}\mathcal{G}$. This relation for the canonical gauge generators is on-shell equivalent to a corresponding relation only involving the canonical boundary charges
	\begin{equation}
		\left\{\mathcal{Q}[\varepsilon_1],\mathcal{Q}[\varepsilon_2]\right\}=\delta_{\epsilon_2}\mathcal{Q},
	\end{equation}
which in most cases is straightforward to calculate. This directly leads to the semiclassical asymptotic symmetry algebra including all possible semiclassical central extensions.
	 
\paragraph{Determine the Quantum Asymptotic Symmetry Algebra:}

This part of the algorithm first appeared in \cite{Henneaux:2010xg}. One insight of this paper was that the asymptotic symmetry algebra derived in the previous steps is only valid for large values of the central charges. For non-linear algebras such as $\mathcal{W}$-algebras that are frequently encountered in higher spin holography that means in particular that one has to think about how normal ordering affects the algebra when passing from a semi-classical to a quantum description of the asymptotic symmetries. One particularly simple way of doing this is to take the semi-classical symmetry algebra, normal order non-linear terms and add all possible deformations to the commutation relations. Requiring that the resulting algebra satisfies the Jacobi identities (see e.g. \cite{Gaberdiel:2012ku}) is usually enough to fix all the structure constants yielding the quantum asymptotic symmetry algebras. 



\paragraph{Identify the Dual Field Theory:}

With the results from all the previous steps one can then proceed in trying to identify or put possible restrictions on a quantum field theory that explicitly realizes these quantum asymptotic symmetries. Once this dual field theory is identified one can perform further nontrivial checks of the holographic conjecture.
\subsubsection{Lobachevsky Spacetimes}
As an explicit example of this algorithm let us consider the Lobachevsky case worked out in \cite{Riegler:2012fa,Afshar:2012hc}. In this work the non-principal embedding of $\mathfrak{sl}(2,\mathbb{R})\hookrightarrow\mathfrak{sl}(3,\mathbb{R})$ was used to describe fluctuations around the background
    \begin{equation}
        \extd s^2=\extd t^2+\extd\rho^2+\sinh^2\rho\extd\varphi^2.
    \end{equation}
In a Chern-Simons formulation this means that one can consider a connection of the form \begin{subequations}\label{AdSxR:BCs1}
		\begin{align}
A_t &=0, \qquad\qquad \bar{A}_t=\sqrt{3}\,\St, \qquad\qquad A_\rho = \Lt_0, \qquad\qquad \bar{A}_\rho=-\Lt_0, \\ 
A_\varphi &= -\frac{1}{4}\,\Lt_1\,e^\rho + \frac{2\pi}{k_{\textrm{\tiny CS}}}\,\big(\mathcal{J}(\varphi)\St+\mathcal{G}^{\pm}(\varphi)\psi^{\pm}_{-\frac{1}{2}}\,e^{-\rho/2}+\mathcal{L}(\varphi)\Lt_{-1}\,e^{-\rho}\big),\\
\bar{A}_\varphi&=-\Lt_{-1}\,e^\rho + \frac{2\pi}{k_{\textrm{\tiny CS}}}\,\bar{\mathcal{J}}(\varphi)\St, 
		\end{align}
\end{subequations}
where the non-principal embedding of $\mathfrak{sl}(2,\mathbb{R})\hookrightarrow\mathfrak{sl}(3,\mathbb{R})$ is characterized by three $\mathfrak{sl}(2)$ generators $\Lt_n$ $(n=-1,0,1)$, two sets of generators $\psi_n^{\pm}$ $(n=-\frac{1}{2},\frac{1}{2})$ and one singlet $\St$.\\
Performing the algorithm described previously one finds that the (quantum) asymptotic symmetry algebra is given by a direct sum of the  Polyakov--Bershadsky algebra \cite{Polyakov:1989dm,Bershadsky:1990bg} and a $\hat u(1)$ current algebra.\\
Defining $\hat{k} = -k-3/2$ and denoting normal ordering with respect to a highest-weight representation by $::$, the asymptotic symmetry algebra is given by 
	\begin{subequations}\label{AdSxR:W32AlgebraQuantumBershadsky}
		\begin{align}
			&[\Jt_n,\Jt_m]=\kappa\,n\,\delta_{n+m,0} = [\bar \Jt_n,\bar \Jt_m],\\
			&[\Jt_n,\Lt_m]=n\Jt_{n+m},\\
			&[\Jt_n,\Gt_m^{\pm}]=\pm \Gt_{m+n}^{\pm},\\
			&[\Lt_n,\Lt_m]=(n-m)\Lt_{m+n}+\frac{c}{12}\,n(n^2-1)\,\delta_{n+m,0},\\
			&[\Lt_n,\Gt_m^{\pm}]=\big(\frac{n}{2}-m\big)\,\Gt_{n+m}^{\pm},\\
			&[\Gt_n^{+},\Gt_m^{-}]= \frac{\lambda}{2}\,\big(n^2-\frac{1}{4}\big)\,\delta_{n+m,0}\nonumber \\
&\qquad\qquad\quad -(\hat{k}+3)\Lt_{m+n}+\frac{3}{2}(\hat{k}+1)(n-m)\Jt_{m+n}
			+3\sum_{p\in\mathbb{Z}}:\Jt_{m+n-p}\Jt_p:,
		\end{align}
	\end{subequations}
with the $\hat{\mathfrak{u}(1)}$ level 
    \begin{equation}
        \kappa = \frac{2\hat{k}+3}{3},
    \end{equation}
the Virasoro central charge
    \begin{equation}
        c=25-\frac{24}{\hat{k}+3}-6(\hat{k}+3), \label{eq:c}
    \end{equation}
and the central term in the $G^\pm$ commutator
    \begin{equation}
        \lambda = (\hat{k} + 1)(2\hat{k} + 3)\,.
    \end{equation}
Looking at unitary representations of this algebra one finds that there is only one value where there are negative norm states are absent and that is for $\hat{k}=-1$ and thus also $c=1$. Hence, a natural guess for a dual quantum field theory is a free boson.\\
Applying the same logic to other non-principally embedded $\mathfrak{sl}(N,\mathbb{R})$ Chern-Simons theories it became quickly clear that the requirement of having no negative norm states is a very simple tool in restricting possible values of the Chern-Simons level. Furthermore, one could also see that with increasing $N$ also the allowed values for the central charges started to grow. In \cite{Afshar:2012nk,Afshar:2014cma} it was shown that a Chern-Simons theory with next-to-principally embedded $\mathfrak{sl}(2,\mathbb{R})\hookrightarrow\mathfrak{sl}(N,\mathbb{R})$ allows for boundary conditions that yield a $\mathcal{W}^{(2)}_N$ Feigin-Semikhatov \cite{Feigin:2004wb} algebra as an asymptotic symmetry algebra. Looking at negative norm states for these algebras one finds again restrictions on the allowed values of the central charge $c$ that depend on $N$ in such a way that the central charge can take arbitrarily large (but finite\footnote{In \cite{Castro:2012bc} it has been shown that any embedding of $\mathfrak{sl}(2,\mathbb{R})\hookrightarrow\mathfrak{sl}(N,\mathbb{R})$ that contains a singlet contains negative norm states for $c\rightarrow\infty$.}) values. This is quite an interesting result since this provides an example of a unitary theory of gravity whose boundary dynamics are covered by a dual quantum field theory that allows both for a semiclassical (large values of the central charge), as well as an ultra quantum (central charge of $\mathcal{O}(1)$) regime. Thus this family of $\mathcal{W}^{(2)}_N$ models provide a novel class of models that may be good candidates for toy models of quantum gravity in three dimensions.
\subsubsection{Lifschitz Spacetimes}
Even though the Lobachevsky case was the first example where higher spin symmetries proved useful for describing asymptotics beyond AdS it by far not the only case considered in the literature so far. Another example that gained quite a bit of attention is the case of asymptotic Lifshitz spacetimes \cite{Kachru:2008yh}
    \begin{equation}\label{eq:LifschitzMetric}
        \extd s^2=\ell^2\left(\frac{\extd r^2+
        \extd x^2}{r^2}-\frac{\extd t^2}{r^{2z}}\right),
    \end{equation}
where $z\in\mathbb{R}$ is a scaling exponent.\\
The authors of \cite{Gutperle:2013oxa} used the Chern-Simons higher spin formulation successfully to describe non-rotating black holes in three-dimensional Lifshitz spacetimes with $z=2$. In addition this allowed them also to study the thermodynamic properties of these black holes in detail.\\
Another very interesting aspect of describing Lifschitz spacetimes using higher spin symmetries has been explored in \cite{Gary:2014mca}. The starting point of the analysis was again an $\mathfrak{sl}(3,\mathbb{R})\oplus\mathfrak{sl}(3,\mathbb{R})$ higher spin Chern-Simons theory with boundary conditions such that the corresonding metric asymptotes to the Lifschitz spacetime \eqref{eq:LifschitzMetric}. Looking at the resulting form of the asymptotic symmetry algebra the authors found two copies of a $\mathcal{W}_3$ algebra with a central charge $c=\frac{3\ell}{2G}$. This is quite an interesting result, since this is exactly what one would get starting with a spin-3 extension of AdS$_3$ \cite{Campoleoni:2010zq,Campoleoni:2011hg}. It was then later argued in \cite{Lei:2015ika} that this may be due to non-invertibility of the zuvielbein in the higher spin Lifschitz case and thus the metric interpretation of \cite{Gary:2014mca} might not be entirely correct.\\
These are not the only interesting features that have been explored in the context of Lifschitz holography using higher spin symmetries. Also very interesting relations to integrable systems have been discovered in \cite{Gutperle:2014aja,Beccaria:2015iwa,Gutperle:2017ewo}.
\subsubsection{Null Warped, Schr\"odinger Spacetimes}
Null warped AdS 
    \begin{equation}
        \extd s^2 = \ell^2\left(\frac{\extd r^2}{4r^2}+2r\extd t\extd\varphi+f(r,z)\extd\varphi^2\right),
    \end{equation}
with $f(r,z)=r^z+\beta r+\alpha^2$ and where $z$ is a real parameter and $\alpha$ and $\beta$ constants of motion, is another case of spacetimes that have been linked to higher spin theories. In \cite{Breunhoelder:2015waa} the authors proposed boundary conditions that asymptote to null warped AdS and found a single copy of the $\mathcal{W}_3^{(2)}$ Polyakov-Berschadsky algebra \eqref{AdSxR:W32AlgebraQuantumBershadsky} as asymptotic symmetries.\\
Last but not least we also want to mention that Schr\"odinger spacetimes
    \begin{equation}
        \extd s^2 = -r^{2z}\extd t^2-2r^2\extd t\extd x^-+\frac{\extd r^2}{r^2}+r^2\extd x^i,
    \end{equation}
can be treated in a higher spin context, both in three dimensions as well as in higher dimensions \cite{Lei:2015gza,Lei:2016pfu}.\\
Even though this review is focused on higher spins without anti-de Sitter we also want to point out some work on higher spins in de Sitter \cite{Krishnan:2013zya,Krishnan:2013cra,Basu:2015exa} as well an example of chiral higher spin theories in AdS\footnote{The boundary conditions in this work can be seen as the spin-3 extension of the boundary conditions found in \cite{Grumiller:2016pqb}.} \cite{Krishnan:2017xct}.
\section{Flat Space Higher Spin Theories as Specific Examples}
\label{sec:flat-space-higher}
Besides the examples of non-AdS higher spin theories that have already been mentioned in the previous section there is another quite prominent example of a holographic correspondence involving higher spins, that is, flat space. Before we go into more details regarding higher spins in flat space we want to give a brief overview of important developments regarding flat space holography in general.\\
The first indications that there might be a holographic correspondence in asymptotically flat spacetimes were worked out in \cite{Polchinski:1999ry,Susskind:1998vk,Giddings:1999jq}. In the last decade there has been a lot of progress in that direction especially in three spacetime dimensions. In 2006 Barnich and Comp{\`e}re \cite{Barnich:2006av} presented a consistent set of boundary conditions for asymptotically flat spacetimes at null infinity\footnote{These are boundary conditions for either future or past null infinity. Thus to be more precise one obtains one copy of $\mathfrak{bms}_3$ on future and another copy on past null infinity. For successful efforts of connecting these two algebras see \cite{Strominger:2013jfa,Kapec:2015vwa,Prohazka:2017equ}.} that extended previous considerations of \cite{Ashtekar:1996cd}. Using these boundary conditions Barnich and Comp{\`e}re were able to show that the corresponding asymptotic symmetry algebra is given by the three-dimensional Bondi-Metzner-Sachs algebra ($\mathfrak{bms}_3$) \cite{Bondi:1962px,Sachs:1962zza}. Since the discovery of the Barnich-Comp{\`e}re boundary conditions many other boundary conditions in asymptotically flat spacetimes have been found leading to either extensions of the $\mathfrak{bms}_3$ algebra as asymptotic symmetry algebra such as \cite{Barnich:2014cwa,Barnich:2015jua,Detournay:2016sfv,Setare:2017mry,Basu:2017aqn,Fuentealba:2017fck}, or to other algebras such as a warped conformal algebra \cite{Afshar:2015wjm}, Heisenberg algebras \cite{Afshar:2016kjj} or an $\mathfrak{isl}(2)_k$ algebra \cite{Grumiller:2017sjh}.\\
In particular the Barnich-Comp{\`e}re boundary conditions and the associated $\mathfrak{bms}_3$ asymptotic symmetries were used quite extensively for various non-trivial checks of a putative holographic correspondence \cite{Bagchi:2012xr,Barnich:2012xq,Barnich:2012aw,Bagchi:2012cy,Bagchi:2013lma,Fareghbal:2013ifa,Krishnan:2013wta,Bagchi:2013qva,Detournay:2014fva,Barnich:2014kra,Bagchi:2014iea,Fareghbal:2014qga,Fareghbal:2014kfa,Barnich:2015mui,Barnich:2015uva,Bagchi:2015wna,Asadi:2016plj,Barnich:2017jgw,Fareghbal:2017ujy}.\\
The previously mentioned developments were mainly focused on either pure Einstein gravity, or supersymmetric extensions thereof. Now, what about (massless) higher spin theories in flat space? In four or higher-dimensional flat space there are in fact quite a number of no-go theorems that forbid non-trivial higher spin interactions such as the Coleman-Mandula theorem \cite{Coleman:1967ad}, its generalization by Pelc and Horwitz \cite{Pelc:1996vg}, the Aragone-Deser no-go result \cite{Aragone:1979hx}, the Weinberg-Witten theorem \cite{Weinberg:1980kq} and others. For a very nice overview of all these various no-go theorems please refer to \cite{Bekaert:2010hw}. This seems like bad news for non-trivial interacting (massless) higher spin theories. However, every no-go theorem is only as good as its premises and as such there are various ways of circumventing these theorems such as e.g. having a non-zero cosmological constant \cite{Vasiliev:1990en}. Interestingly enough, the no-go theorems mentioned previously do not apply in three dimensions and thus it seems possible to have non-trivial interacting (massless) higher spin theories in three dimensions\footnote{Even though (A)dS backgrounds favor interactions of massless higher spin fields, higher spin interactions are not completely ruled out even in higher-dimensional flat space. For recent developments regarding higher spins in four or higher-dimensional flat space see e.g. \cite{Sleight:2016xqq,Ponomarev:2016jqk,Ponomarev:2016lrm,Campoleoni:2017mbt}.}.\\
And indeed, in \cite{Afshar:2013vka,Gonzalez:2013oaa} the first consistent boundary conditions for a higher spin extension of the Poincar\'e algebra were found.

\subsection{Flat Space Spin-3 Gravity}\label{sec:FSSpin-3Gravity}

Higher spin theories in three-dimensional flat space can be described in a very similar fashion as in the AdS$_3$ case, that is, by a suitable Chern-Simons formulation. In the AdS$_3$ case the basic gauge symmetries of the Chern-Simons gauge field are given by a direct sum of two copies of $\mathfrak{sl}(N,\mathbb{R})$ (or more general $\mathfrak{hs}[\lambda]$). In the flat space case the corresponding connections take values in $\mathfrak{isl}(N,\mathbb{R})$. The structure of $\mathfrak{isl}(N,\mathbb{R})$ that of a semidirect sum of $\mathfrak{sl}(N,\mathbb{R})$ with an abelian ideal that is isomorphic to $\mathfrak{sl}(N,\mathbb{R})$ as a vector space. One nice thing about this structure is that it can be straightforwardly obtained by suitable \.In\"on\"u--Wigner contractions \cite{Inonu:1953sp} and thus one has a direct way of obtaining these algebras from the well known AdS$_3$ higher spin gauge symmetries\footnote{Please refer to \cite{Bagchi:2009my,Bagchi:2009ca,Bagchi:2009pe,Bagchi:2010eg,Bagchi:2012cy,Fareghbal:2013ifa,Fareghbal:2014kfa,Fareghbal:2017ujy} for early as well as recent work in flat space holography in three dimensions that rely on contractions.}. These kind of contractions have been used quite successfully to obtain new higher spin algebras in flat space (both isometries and asymptotic symmetries) \cite{Grumiller:2014lna,Campoleoni:2016vsh,Riegler:2016hah} as well as flat space analogues of important formulas like the (spin-3) Cardy formula \cite{Riegler:2014bia,Fareghbal:2014qga}.\\
Thus the starting point for a spin-3 theory in flat space is a Chern-Simons action with gauge algebra\footnote{To be more precise it is the principal embedding of $\mathfrak{isl(2,\mathbb{R})}\hookrightarrow\mathfrak{isl}(3,\mathbb{R})$.} $\mathfrak{isl}(3,\mathbb{R})$ equipped with an appropriate bilinear form. Then one can choose boundary conditions as \cite{Afshar:2013vka,Gonzalez:2013oaa}
	\begin{equation}\label{eq:FlatSpin3Connection}
		\mathcal{A}=b^{-1}\extd b+b^{-1}a(u,\varphi)b,\qquad b=e^{\frac{r}{2} \Mt_{-1}},
	\end{equation}
with
	\begin{equation}\label{eq:FlatSpin3BoundaryConditions}
		a(u,\varphi)=a_\varphi(u,\varphi)\extd\varphi+a_u(u,\varphi)\extd u,
	\end{equation}
where
	\begin{subequations}
	\begin{align}
		a_\varphi(u,\varphi)=&\Lt_1-\frac{\mathcal{M}}{4}\Lt_{-1}-\frac{\mathcal{N}}{2}\Mt_{-1}+\frac{\mathcal{V}}{2}\Ut_{-2}+\mathcal{Z}\Vt_{-2},\\
		a_u(u,\varphi)=&\Mt_1-\frac{\mathcal{M}}{4}\Mt_{-1}+\frac{\mathcal{V}}{2}\Vt_{-2}.
	\end{align}
	\end{subequations}
The operators $\Lt_n,\Mt_n$ with $n=\pm1,0$ and $\Ut_m,\Vt_m$ with $m=\pm2,\pm1,0$ span the $\mathfrak{isl}(3,\mathbb{R})$ algebra\footnote{The commutation relations are identical with the ones in \eqref{eq:FlatSpaceSpin3SemiClassicalASA} after restricting the mode numbers as already mentioned and in addition dropping all non-linear terms.} with invariant bilinear form
    \begin{equation}
        \langle\Lt_n\Mt_m\rangle=-2\eta_{nm},\qquad\langle\Ut_n\Vt_m\rangle=\frac{2}{3}\mathcal{K}_{nm},
    \end{equation}
where $\eta_{nm}=\textrm{antidiag}(1,-\frac{1}{2},1)$ and $\mathcal{K}_{nm}=\textrm{antidiag}(12,-3,2,-3,12)$. The zuvielbein can be extracted from the gauge connection by using that $\mathcal{A}=e^{(2)}_n\Mt_n+e^{(3)}_n\Vt_n+\omega^{(2)}_n\Lt_n+\omega^{(3)}_n\Ut_n$. Using these ingredients one can determine the metric\footnote{The spin-3 field can be determined in analogy by using the cubic Casimir of the $\mathfrak{sl}(3,\mathbb{R})$ subalgebra.} 
    \begin{equation}
        g_{\mu\nu}=\eta_{ab}e^{(2)a}_\mu e^{(2)b}_\nu+\mathcal{K}_{ab}e^{(3)a}_\mu e^{(3)b}_\nu.
    \end{equation}
Thus consequently these boundary conditions describe the following metric and spin-3 field
    \begin{equation}
        \extd s^2=\mathcal{M}\extd u^2-2\extd r\extd u+2\mathcal{N}\extd u\extd\varphi+r^2\extd\varphi^2,\qquad \Phi_{\mu\nu\lambda}\extd x^\mu\extd x^\nu\extd x^\lambda=2\mathcal{V}\extd u ^3+4\mathcal{Z}\extd u^2\extd\varphi.
    \end{equation}
From a geometric point of view the metric is nothing else than flat space in Eddington-Finkelstein coordinates. Working out the asmyptotic symmetries one finds that these boundary conditions lead to the following non-linear, centrally extended asymptotic symmetry algebra:
	\begin{subequations}\label{eq:FlatSpaceSpin3SemiClassicalASA}
		\begin{align}
		[\Lt_n,\Lt_m]=&(n-m)\Lt_{n-m}+\frac{c_L}{12}n(n^2-1)\delta_{n+m,0},\\
		[\Lt_n,\Mt_m]=&(n-m)\Mt_{n-m}+\frac{c_M}{12}n(n^2-1)\delta_{n+m,0},\\
		[\Lt_n,\Ut_m]=&(2n-m)\Ut_{n+m},\\
		[\Lt_n,\Vt_m]=&(2n-m)\Vt_{n+m},\\
		[\Mt_n,\Ut_m]=&(2n-m)\Vt_{n+m},\\
		[\Ut_n,\Ut_m]=&(n-m)(2n^2+2m^2-nm-8)\Lt_{n+m}+\frac{192}{c_M}(n-m)\Lambda_{n+m}\nonumber\\
				&-\frac{96c_L}{c_M^2}(n-m)\Theta_{n+m}+\frac{c_L}{12}n(n^2-1)(n^2-4)\delta_{n+m,0},\\
		[\Ut_n,\Vt_m]=&(n-m)(2n^2+2m^2-nm-8)\Mt_{n+m}+\frac{96}{c_M}(n-m)\Theta_{n+m}\nonumber\\
				&+\frac{c_M}{12}n(n^2-4)(n^2-1)\delta_{n+m,0},
		\end{align}
	\end{subequations}
with
	\begin{equation}
		\Lambda_{n}=\sum_{p\in\mathbb{Z}}\Lt_p\Mt_{n-p},\qquad\Theta_n=\sum_{p\in\mathbb{Z}}\Mt_p\Mt_{n-p},
	\end{equation}
and
	\begin{equation}
		c_L=0,\qquad c_M=\frac{3}{G}.
	\end{equation}
This algebra is usually denoted by $\mathcal{FW}_3$ to denote its role similar to the $\mathcal{W}_3$ algebra in AdS$_3$ higher spin holography. Furthermore, this algebra can be obtained as a specific \.In\"on\"u--Wigner contraction that can be interpreted as a limit of vanishing cosmological constant of the the AdS$_3$ spin-3 asymptotic $\mathcal{W}_3$ symmetries.\\
Assuming that one starts with two copies of a quantum\footnote{That means that all non-linear terms are normal ordered with respect to some highest-weight representation and the central terms have $\mathcal{O}(1)$ corrections that are necessary to satisfy the Jacobi identities when the non-linear terms are normal ordered.} $\mathcal{W}_3$ algebra \cite{Zamolodchikov:1985wn} whose generators are labelled as $\mathcal{L}_n$, $\bar{\mathcal{L}}_n$ and $\mathcal{W}_n$, $\bar{\mathcal{W}}_n$. Then one can define the following linear combinations:
	\begin{subequations}
	\label{eq:fshsg12}
		\begin{align}
			\Lt_n:=&\,\mathcal{L}_n-\bar{\mathcal{L}}_{-n},\qquad & \Mt_n &:=\frac{1}{\ell}\left(\mathcal{L}_n+\bar{\mathcal{L}}_{-n}\right),\\
			\Ut_n:=&\,\mathcal{W}_n-\bar{\mathcal{W}}_{-n},\qquad & \Vt_n &:=\frac{1}{\ell}\left(\mathcal{W}_n+\bar{\mathcal{W}}_{-n}\right),
		\end{align}
	\end{subequations}    
and in the limit $\ell\rightarrow\infty$ obtain exactly \eqref{eq:FlatSpaceSpin3SemiClassicalASA}. It should also be noted that besides the contraction \eqref{eq:fshsg12} one can also perform a so-called nonrelativistic contraction using the following alternative linear combination:
	\begin{subequations}
	\label{eq:fshsg14}
		\begin{align}
			\Lt_n:=&\,\mathcal{L}_n+\bar{\mathcal{L}}_{n},\qquad & \Mt_n &:=-\epsilon\left(\mathcal{L}_n-\bar{\mathcal{L}}_{n}\right),\\
			\Ut_n:=&\,\mathcal{W}_n+\bar{\mathcal{W}}_{n},\qquad & \Vt_n &:=-\epsilon\left(\mathcal{W}_n-\bar{\mathcal{W}}_{n}\right),
		\end{align}
	\end{subequations}
that in the limit $\epsilon\rightarrow0$ yields another kind of non-linear, centrally extended algebra \cite{Campoleoni:2016vsh} that can be seen as natural (quantum) higher spin extension of the Galilean conformal algebra $\mathfrak{gca}_2$. In the spin-2 case these two limits yield two isomorphic algebras, namely the $\mathfrak{bms}_3$ and $\mathfrak{gca}_2$ algebra respectively. However, as soon as one adds higher spins these two limits do not yield isomorphic algebras anymore. The reason for this is basically that each limit favors different representations. The ultrarelativistic limit that leads to the $\mathfrak{bms}_3$ algebra favors so-called (unitary) induced representations, whereas the nonrelativistic contraction favors (generically non-unitary \cite{Grumiller:2014lna}) highest-weight representations \cite{Campoleoni:2016vsh}. Since normal ordering requires a notion of vacuum what is meant by normal ordering differs as soon as there are non-linear terms present in the algebra and as such also influences the structure constants.

\subsection{Flat Space Cosmologies With Spin-3 Hair}

Cosmological solutions in flat space \cite{Cornalba:2002fi,Cornalba:2003kd} are well known and thoroughly studied objects. As such another very interesting thing to study in the context of higher spin theories are cosmological solutions in flat space that also carry higher spin hair. This has been done successfully first in \cite{Gary:2014ppa} and subsequently also in \cite{Matulich:2014hea}. The basic idea of describing such cosmological solutions is by taking the $\varphi$-part of the connection \eqref{eq:FlatSpin3BoundaryConditions} and extending the $u$-part by arbitrary but fixed chemical potentials in such a way that the equations of motion $F=0$ are satisfied. By imposing suitable holonomy conditions\footnote{Alternatively one can also use a closed Wilson loop wrapped around the horizon \cite{Basu:2015evh} in order to determine the thermal entropy.} one can then determine, the inverse temperature, angular potential and higher spin chemical potentials and subsequently also the thermal entropy of cosmological solutions with additional spin-3 hair. If one denotes the spin-2 charges by $\mathcal{N}$, $\mathcal{M}$, the spin-3 charges by $\mathcal{Z}$, $\mathcal{V}$ and introducing the dimensionless ratios
    \begin{equation}
        \frac{{\cal R}-1}{4{\cal R}^{3/2}}=\frac{|{\cal V}|}{{\cal M}^{3/2}}, \qquad {\cal R} > 3,\qquad\textrm{and}\qquad\mathcal{P} = \frac{{\cal Z}}{\sqrt{\cal M}{\cal N}},
    \end{equation}
one obtains the following formula for the thermal entropy of cosmological solutions with spin-3 hair:
    \begin{equation}\label{eq:FlatSpaceSpin3ThermalEntropyFormula}
        S_{\textrm{Th}} = \frac{\pi}{2G} \frac{|{\cal N}|}{\sqrt{\cal M}}\frac{2{\cal R}-3 - 12{\cal P}\sqrt{\cal R}}{({\cal R}-3)\sqrt{4-3/{\cal R}}}.
    \end{equation}
This result can also be understood in terms of a limiting procedure of the AdS spin-3 results for the thermal entropy of BTZ solutions with spin-3 hair \cite{Riegler:2014bia}. In complete analogy to the limiting procedure of the BTZ black hole entropy one has to consider the following expression that can be seen as a inner horizon entropy formula of spin-3 charged BTZ black holes
	\begin{equation}\label{eq:SinnerBTZSpin3}
		S_{\textrm{\tiny inner}} =2\pi\left|\sqrt{\frac{c\,\mathcal{L}}{6}}\sqrt{1-\frac{3}{4C}}-\sqrt{\frac{\bar{c}\,\bar{\mathcal{L}}}{6}}\sqrt{1-\frac{3}{4\bar{C}}}\right|,
	\end{equation}
where $c=\bar{c}=\frac{3\ell}{2G}$ and the dimensionless ratios $C$ and $\bar{C}$ are given by
    \begin{equation}
        \sqrt{\frac{c}{6\mathcal{L}^3}}\frac{\mathcal{W}}{4}=\frac{C-1}{C^{\frac{3}{2}}},\qquad\sqrt{\frac{\bar{c}}{6\bar{\mathcal{L}}^3}}\frac{\bar{\mathcal{W}}}{4}=\frac{\bar{C}-1}{\bar{C}^{\frac{3}{2}}}.
    \end{equation}
In order to take the limit one also has to introduce suitable relation of the AdS spin-2 and spin-3 charges $\mathcal{L}$, $\bar{\mathcal{L}}$, $\mathcal{W}$, $\bar{\mathcal{W}}$ with their flat space counterparts $\mathcal{N}$, $\mathcal{M}$ and $\mathcal{Z}$, $\mathcal{V}$. If one chooses the relations as
    \begin{subequations}
    \begin{align}
    	\mathcal{M} & =12\left(\frac{\mathcal{L}}{c}+\frac{\bar{\mathcal{L}}}{\bar{c}}\right),& \mathcal{N} &=6\ell\left(\frac{\mathcal{L}}{c}-\frac{\bar{\mathcal{L}}}{\bar{c}}\right),\\
    	\mathcal{V} &=12\left(\frac{\mathcal{W}}{c}+\frac{\bar{\mathcal{W}}}{\bar{c}}\right),& \mathcal{Z} &=6\ell\left(\frac{\mathcal{W}}{c}-\frac{\bar{\mathcal{W}}}{\bar{c}}\right),
    \end{align}
    \end{subequations}
and in addition defines
	\begin{equation}\label{eq:CBarCFlatRelations}
		C=\mathcal{R}+\frac{2}{\ell} D(\mathcal{R},\mathcal{P},\mathcal{M},\mathcal{N}),\qquad \bar{C}=\mathcal{R}-\frac{2}{\ell}D(\mathcal{R},\mathcal{P},\mathcal{M},\mathcal{N}).
	\end{equation}
with
	\begin{equation}
		D(\mathcal{R},\mathcal{P},\mathcal{M},\mathcal{N})=\frac{\mathcal{N}}{\mathcal{M}}\frac{\mathcal{R} \left(\mathcal{R}^{\frac{3}{2}} \mathcal{P}+3 \mathcal{R}-3\right)}{ \left(\mathcal{R}-3\right)},
	\end{equation}
then it is straightforward to show that, indeed, one reproduces the entropy formula \eqref{eq:FlatSpaceSpin3ThermalEntropyFormula} in the limit $\ell\rightarrow\infty$.\\	
Having an entropy formula like \eqref{eq:SinnerBTZSpin3} at hand also allows one to study possible phase transitions of these higher spin cosmological solutions in flat space by looking at the free energy. And, indeed, one finds the usual phase transition to hot flat space first described in \cite{Bagchi:2013lma} plus additional phase transitions because of the additional spin-3 charges. Interestingly and in contrast to the possible phase transitions in AdS \cite{David:2012iu} there can also be first order phase transitions between various thermodynamical phases in the flat space case.

\subsection{Higher Spin Soft Hair in Flat Space}

Soft hair excitations of black holes as possible solutions\footnote{For a contrasting view on the role of soft hair in solving the black hole paradox see e.g. \cite{Bousso:2017dny,Bousso:2017rsx} and references therein.} to the black hole information paradox have attracted quite some research interest lately see e.g. \cite{Hawking:2016msc,Hawking:2016sgy}. Especially three-dimensional gravity proved to be quite an active playground to study soft hair on (higher spin) black holes in AdS \cite{Afshar:2016wfy,Afshar:2016uax,Grumiller:2016kcp,Grumiller:2017jft}, higher-derivative theories of gravity \cite{Setare:2016vhy} as well as flat space \cite{Afshar:2016kjj,Ammon:2017vwt}. What is most intriguing about all these near horizon boundary conditions is that they all lead to a (number of) $\hat{\mathfrak{u}}(1)$ current algebra(s) but the entropy is always given in a very simple form
    \begin{equation}\label{eq:SoftEntropy}
        S_{\textrm{Th}}=2\pi\left(\Jt_0^++\Jt^-_0\right),
    \end{equation}
where $\Jt^\pm_0$ are the spin-2 zero modes of the near horizon symmetry algebras. In the following we give a brief overview on how to obtain this result for the entropy for spin-3 gravity in flat space.\\
The starting point is again a Chern-Simons formulation of gravity with a gauge algebra $\mathfrak{isl}(3,\mathbb{R})$ as in Section~\ref{sec:FSSpin-3Gravity}\footnote{Please note that instead of the retarded time coordinate $u$ it is more natural to use the advanced time coordinate $v$.}. However, one is now interested in describing near horizon boundary conditions of flat space cosmologies with additional spin-3 hair in contrast to the examples previously that focused on the asymptotic symmetries of such configurations. These near horizon boundary conditions can be described by:
    \begin{equation}
\label{eq:gaugeans}
        \mathcal{A}=b^{-1}(a+\extd \,) \, b\,,
    \end{equation}
where the radial dependence is encoded in the group element $b$ as~\cite{Afshar:2016kjj}
    \begin{equation}
\label{eq:gaugeb}
        b=\exp\left( \frac{1}{\mu_\mathcal{P}} \, \Mt_1\right) \, \exp\left( \frac{r}{2} \, \Mt_{-1}\right) \,,
    \end{equation}
and the connection $a$ reads
\begin{equation}
a = a_v  \extd v + a_\varphi \extd \varphi \, ,
\end{equation}
with
    \begin{subequations}\label{eq:FSCSpin3BCs}
    \begin{align}
        a_\varphi&=\mathcal{J} \, \Lt_0+\mathcal{P} \, \Mt_0+\mathcal{J}^{(3)} \, \Ut_0+\mathcal{P}^{(3)} \, \Vt_0 \, ,  \\
        a_v&=\mu_\mathcal{P} \, \Lt_0+\mu_\mathcal{J} \, \Mt_0+\mu_\mathcal{P}^{(3)} \, \Ut_0+\mu_\mathcal{J}^{(3)} \, \Vt_0\,.
    \end{align}
    \end{subequations}
All the functions appearing in \eqref{eq:FSCSpin3BCs} are in principle arbitrary functions of the advanced time $v$ and the angular coordinate $\varphi$. Based on these boundary conditions it is straightforward to determine the near horizon symmetry algebra as
    \begin{equation}\label{eq:FourierASA}
        [\Jt_n,\Pt_m]=k \, n \, \delta_{n+m,0}\,,\qquad\quad [\Jt_n^{(3)},\Pt_m^{(3)}]=\frac{4k}{3} \, n \, \delta_{n+m,0}\,,
    \end{equation}
where $k=\frac{1}{4G}$ that can also be brought into a different form by
   \begin{equation}
        \Jt_{\pm n}^\pm=\frac{1}{2}(\Pt_n\pm \Jt_n)\,,\qquad \Jt_{\pm n}^{(3)\pm}=\frac{1}{2}(\Pt_n^{(3)}\pm \Jt_n^{(3)})\,.
    \end{equation}
The generators $\Jt_n^\pm$ and $\Jt_n^{{(3)\pm}}$ then satisfy
    \begin{subequations}
    \begin{align}
        [\Jt^\pm_n,\Jt^\pm_m]&=\frac{k}{2}n\delta_{n+m,0}\,,\qquad &[\Jt^+_n,\Jt^-_m]=0 \, ,\\
        [\Jt^{(3)\pm}_n,\Jt^{(3)\pm}_m]&=\frac{2k}{3}n\delta_{n+m,0} \, , &[\Jt^{(3)+}_n,\Jt^{(3)-}_m]=0 \,. 
    \end{align}
   \end{subequations}
Thus the near horizon symmetries are given by two pairs of $\hat{\mathfrak{u}}(1)$ current algebras. Calculating both the Hamiltonian (in order to check that these excitations are, indeed, soft) as well as the thermal entropy, which is given by \eqref{eq:SoftEntropy}, is a straightforward exercise and we refer the interested reader to \cite{Ammon:2017vwt} for more details.\\
With a simple result like \eqref{eq:SoftEntropy} for the thermal entropy of a spin-3 charged flat space cosmology and a rather complicated one like \eqref{eq:FlatSpaceSpin3ThermalEntropyFormula} a natural question to ask is: How exactly are these two related? Is there a way to construct the asymptotic state-dependent functions $\mathcal{M}$, $\mathcal{N}$, $\mathcal{V}$ and $\mathcal{Z}$ in terms of the near-horizon state-dependent functions $\mathcal{J}$, $\mathcal{P}$, $\mathcal{J}^{(3)}$ and $\mathcal{P}^{(3)}$?\\
In order to answer these questions one has to find a gauge transformation that maps these two connections into each other without changing the canonical boundary charges. Such a gauge transformation can, indeed, be found and gives the relations
    \begin{subequations}\label{eq:FSMiuraTrafos}
    \begin{align}
        \mathcal{M} &= \mathcal{J}^2+\frac{4}{3}\left(\mathcal{J}^{(3)}\right)^2+2\mathcal{J}',\\
        \mathcal{N} &= \mathcal{J}\mathcal{P}+\frac{4}{3}\mathcal{J}^{(3)}\mathcal{P}^{(3)}+\mathcal{P}',\displaybreak[1]\\
        \mathcal{V} &= \frac{1}{54}\left(18\mathcal{J}^2\mathcal{J}^{(3)}-8\left(\mathcal{J}^{(3)}\right)^3+9\mathcal{J}'\mathcal{J}^{(3)}+27\mathcal{J}\mathcal{J}^{(3)'}+9\mathcal{J}^{(3)''}\right),\\
        \mathcal{Z} &= \frac{1}{36}\left(6\mathcal{J}^2\mathcal{P}^{(3)}-8\mathcal{P}^{(3)}\left(\mathcal{J}^{(3)}\right)^2+3\mathcal{P}^{(3)}\mathcal{J}'+3\mathcal{J}^{(3)}\mathcal{P}'\right.\nonumber\\
        &\qquad\quad\left.+9\mathcal{J}\mathcal{P}^{(3)'}+9\mathcal{P}\mathcal{J}^{(3)'}+12\mathcal{P}\mathcal{J}\mathcal{J}^{(3)}+3\mathcal{P}^{(3)''}\right)\,,
    \end{align}
    \end{subequations}
that are basically (twisted) Sugawara constructions for the spin-2 and spin-3 fields. One can use these relations and an appropriate Fourier expansion in order to solve for the zero modes $\Pt_0=\Jt^+_0+\Jt^-_0$ which gives
    \begin{equation}
        P_0=\Jt^+_0+\Jt^-_0=\frac{1}{4G}\frac{\mathcal{N}\left(4\mathcal{R}-6+3\mathcal{P}\sqrt{\mathcal{R}}\right)}{4\sqrt{\mathcal{M}}(\mathcal{R}-3)\sqrt{1-\frac{3}{4\mathcal{R}}}}\,,
    \end{equation}
and correctly reproduces \eqref{eq:FlatSpaceSpin3ThermalEntropyFormula}. Thus one sees that also for flat space cosmologies there seems to be a much easier way to count the microscopical states contributing to the thermal entropy. That is, in terms of near horizon variables instead of asymptotic ones.

\subsection{One Loop Higher Spin Partition Functions in Flat Space}

One loop partition functions often provide very useful insights on the consistency of the spectrum for a possible interacting quantum field theory. On (A)dS backgrounds this feature has been exploited quite successfully. In three bulk dimensions for example the comparison between bulk and boundary partition functions \cite{Gaberdiel:2010ar,Gaberdiel:2011zw,Creutzig:2011fe} has been an important ingredient in defining the holographic correspondence between higher spin gauge theories and minimal model CFTs \cite{Gaberdiel:2012uj}. In spacetime dimensions higher than three the analysis of one-loop partition functions of infinite sets of higher spin fields provided the first quantum checks \cite{Giombi:2013fka,Giombi:2014iua,Giombi:2014yra,Beccaria:2014xda,Beccaria:2015vaa} of analogous AdS/CFT dualities \cite{Klebanov:2002ja}.\\
Since the study of one-loop higher spin partition functions has proven to be quite a fruitful line of research a natural question to ask is whether or not one can extend such considerations also to higher spin theories in $d$-dimensional flat space. This venue has been successfully pursued in \cite{Campoleoni:2015qrh}, where one-loop partition functions of (supersymmetric) higher spin fields in $d$-dimensional thermal flat space with angular potentials $\vec\theta$ and inverse temperature $\beta$ have been computed for the first time using both a heat kernel as well as a group theoretic approach. Also in this case the three-dimensional case is of special interest since for $d=3$ one can explicitly show that suitable products of massless one-loop partition functions 
    \begin{equation}
    Z[\beta,\vec\theta]
=
e^{\delta_{s,2} \frac{\beta c_M}{24}}
\prod_{n\,=\,s}^{\infty}\frac{1}{|1-e^{in(\theta+i\epsilon)}|^2}\,,
\quad
c_M=3/G,
    \end{equation}
coincide with vacuum
characters of $\mathcal{FW}_N$ algebras
    \begin{equation}
        \chi_{\mathcal{FW}_N}
=
e^{\beta c_M/24}
\prod_{s=2}^N
\left(\prod_{n=s}^{\infty}\frac{1}{|1-e^{in(\theta+i\epsilon)|^2}}\right).    
    \end{equation}
  
\subsection{Further Aspects of Higher Spins in 3D Flat Space}

As some final remarks regarding higher spins in flat space we want to describe a little bit more explicitly the content of the two works \cite{Krishnan:2013tza,Gonzalez:2014tba}.\\
The first work \cite{Krishnan:2013tza} shows how higher spin symmetries could be used to get rid of the causal singularity in the Milne metric in three dimensions \cite{Barnich:2012aw,Barnich:2012xq}. The basic idea here being that one can reformulate the Milne metric equivalently in terms of a Chern-Simons connection and then enlarging the gauge algebra of the Chern-Simons connection from $\mathfrak{isl}(2,\mathbb{R})$ to $\mathfrak{isl}(3,\mathbb{R})$. Requiring that the holonomies of the higher spin connection match those of the original spin-2 connection does place some restrictions on the possible spin-3 extensions of the Chern-Simons gauge field, however, it still leaves enough freedom that can be used to get rid of the causal singularity that is present in the spin-2 case at the level of the Ricci scalar\footnote{It should be noted that there is still the possibility that a possible spin-3 generalization of the Ricci scalar is singular. However, there is at the moment no full geometric interpretation of higher spin symmetries that would be necessary in order to check this.} and in addition have a non-singular spin-3 field supporting the geometry.\\
The second work in this context that we would like to mention explicitly is \cite{Gonzalez:2014tba}. One very important ingredient in establishing a (higher spin) holographic principle in asymptotically flat spacetimes is to find concrete theories that are invariant under the corresponding asymptotic symmetries. For Einstein gravity without cosmological constant and Barnich-Comp{\`e}re boundary conditions this would be the $\mathfrak{bms}_3$ algebra and, indeed, for this case it has been suggested in \cite{Barnich:2012rz} that a flat limit of Liouville theory would be a suitable candidate\footnote{See also \cite{Barnich:2017jgw} for a more group theoretic approach to the problem.}. The work \cite{Gonzalez:2014tba} extended the previous considerations accordingly to a two-dimensional action invariant under a spin-3 extension of the $\mathfrak{bms}_3$ algebra. The corresponding action can also be obtained as a suitable limit of $\mathfrak{sl}(3,\mathbb{R})$ Toda theory as expected.

\section{Non-AdS Through Choice of Gauge Group}
\label{sec:non-ads-through-1}

After the considerations of the preceding sections it is natural to ask if there
are higher spin theories based on Lie algebras beyond (A)dS and Poincar\'e.
This is of interest because it became clear
that
for nonrelativistic holography also nonrelativistic geometries play a fundmanetal role,
for a review see e.g.\ \cite{Taylor:2015glc}.
In Section~\ref{sec:examples-non-ads} the focus was on obtaining geometries beyond (A)dS using
different backgrounds and boundary conditions than (A)dS,
while still working
in a theory given by the gauge group of AdS and its higher spin generalizations.
In this section -- like in Section \ref{sec:flat-space-higher} -- we are going to make non-(A)dS geometries manifest by changing the gauge algebra.\\
Since the tools that were used in the derivation of
kinematical algebras and their
Chern-Simons theories are the same for the spin-2 case and
their spin-3 generalization, we will first focus on the former
and comment afterwards on the latter.

\subsection{Kinematical Algebras}
\label{sec:kinem-lie-algebr}

\begin{figure}[ht]
  \centering
\tdplotsetmaincoords{80}{120}
\begin{tikzpicture}[
tdplot_main_coords,
dot/.style={circle,fill},
linf/.style={ultra thick,->,blue},
cinf/.style={ultra thick,->,red},
tinf/.style={ultra thick,->},
stinf/.style={ultra thick,->,gray},
scale=0.6
]

\node (ads) at (0,10,10) [dot, label=above:(Anti-)de Sitter] {};
\node (p) at (10,10,10) [dot, label=above:Poincar\'e] {};
\node (nh) at (0,10,0) [dot, label=below:Newton--Hooke] {};
\node (pp) at (0,0,10) [dot, label=above:Para-Poincar\'e] {};
\node (g) at (10,10,0) [dot, label=below:Galilei] {};
\node (pg) at (0,0,0) [dot, label=below:Para-Galilei] {};
\node (car) at (10,0,10) [dot, label=above:Carroll] {};
\node (st) at (10,0,0) [dot, label=below:Static] {};

\draw[linf] (ads) -- node [sloped,below
] {\small  Space-time;
  $\infty \leftarrow \ell$
}  (p);

\draw[linf] (nh) -- (g);
\draw[linf] (pp) -- (car);
\draw[linf,dashed] (pg) -- (st);

\draw[cinf,double] (ads) -- node [sloped,above] {\small  Speed-space;
  $c \to \infty$
} (nh);
\draw[cinf,double,dashed] (pp) -- (pg);
\draw[cinf] (p) -- (g);
\draw[cinf] (car) -- (st);

\draw[tinf,double] (ads) -- node [sloped,above] {\small  Speed-time;
  $0 \leftarrow c$
} (pp);
\draw[tinf] (p) -- (car);
\draw[tinf] (g) -- (st);
\draw[tinf,double,dashed] (nh) -- (pg);

\end{tikzpicture}  
\caption{
  This cube summarizes the kinematical Lie algebras~\cite{Bacry:1968zf}.
  Each dot represents a kinematical Lie algebra,
  given explicitly in Appendix~\ref{sec:expl-kinem-algebra},
  and each arrow represents an \IW contraction.
  The double arrows mean that there are two contractions
  since we start with two Lie algebras (AdS and dS).
  The algebras on the back surface and therefore of finite (A)dS radius $\ell$
  can be considered as cosmological algebras.
  The top and bottom surfaces can
  be understood as relative and absolute time Lie algebras,
  connected by the nonrelativistic limit $c \to \infty$.
  Sending the speed of light $c$ to zero, the ultrarelativistic limit,
  leads to absolute space Lie algebras.
  The parameters of the limits should not be taken too serious since,
  one of course can not take the limit of $c \to 0$ and $c \to \infty$
  simultaneously (there are actually three parameters involved).
  They should merely make intuitively clear that the light cone either closes
  in the ultrarelativistic limit
  or opens up as for the nonrelativistic one.
}
\label{fig:cube}
\end{figure}
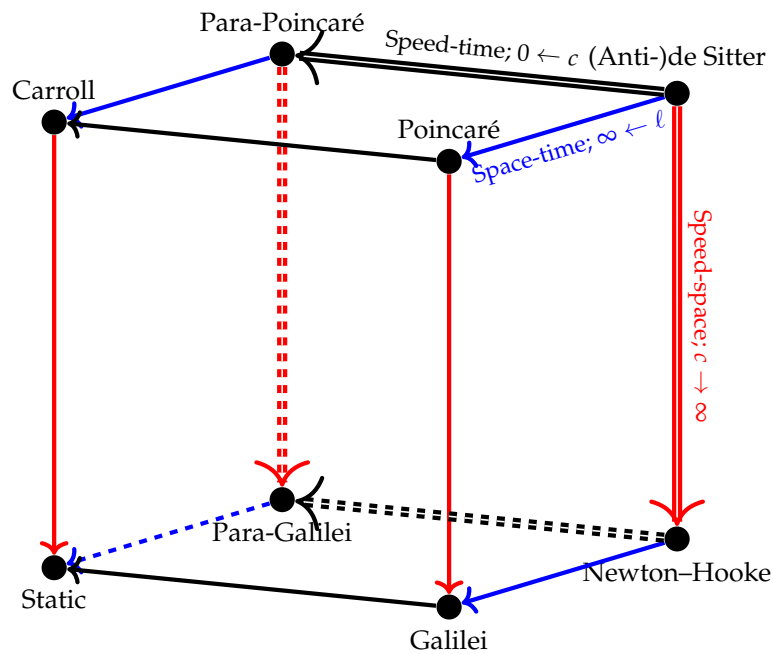

A classification of interesting 
kinematical algebras,
consisting of generators
of time and spatial translations $\Ht$ and $\Pt_{a}$\footnote{
  The indices take now the values $a,b,m=(1,2)$.
}
, rotations $\Jt$ and,
inertial transformations $\Gt_{a}$
has been given 
by Bacry and Levy-Leblond~\cite{Bacry:1968zf}.
The classification was provided under the assumptions that
\begin{enumerate}
\item Space is isotropic.
\item Parity and time-reversal are automorphisms of the kinematical
  groups.
\item Inertial transformations in any given direction form a
  noncompact subgroup.
\end{enumerate}
This analysis led to other Lie algebras besides the
already mentioned (A)dS and Poincar\'e algebras.
Other prominent examples are the Galilei algebra and
Carroll algebra
and their cousins that appear in the context of spacetimes with non-vanishing cosmological constant.
All of them can be conveniently summarized as a cube of \IW contractions\footnote{
  We will use the term \IW contractions here to denote contractions
  of the form originally defined in~\cite{Inonu:1953sp},
  sometimes called simple \IW contractions.
  In contrast to some generalizations like generalized \IW contraction
  they are linear in the contraction parameter.
}
starting from the (A)dS algebras,
see Figure~\ref{fig:cube}.
Since contractions are physically seen as approximations they often
automatically provide insights from the original  
to the contracted theory.
To define an \IW contraction one starts with a Lie algebra $\mathfrak{g}$ which is a vector space
direct sum of $\mathfrak{h}$ and $\mathfrak{i}$, i.e., 
$\mathfrak{g} =\mathfrak{h} \oplus \mathfrak{i}$.
One then rescales  $\mathfrak{i} \mapsto
\frac{1}{\epsilon}\mathfrak{i}$.
The commutation relations before and after the
contraction are then explicitly given by 
\begin{align}
  [\mathfrak{h},\mathfrak{h}] & = \phantom{\epsilon} \mathfrak{h} + \cancel{\frac{1}{\epsilon}\mathfrak{i}},                     &  &                                        & [\mathfrak{h},\mathfrak{h}] & =  \mathfrak{h}, 
 \\                                                                                                             
  [\mathfrak{h},\mathfrak{i}] & = \epsilon \mathfrak{h} +
                                \phantom{\frac{1}{\epsilon}}
                                \mathfrak{i},             &  &
                                                              \overset{\epsilon
                                                              \to 0}{\longrightarrow} & [\mathfrak{h},\mathfrak{i}] & =  \mathfrak{i},                        
 \\                                                                                                             
  [\mathfrak{i},\mathfrak{i}] & =\epsilon \mathfrak{h} + \epsilon^{2} \mathfrak{i}, &  &                                        & [\mathfrak{i},\mathfrak{i}] & =0 \,.
\end{align}
The term on the right hand side of the
$[\mathfrak{h},\mathfrak{h}]$ commutator that has been crossed out
basically shows that the contraction is convergent in the $\epsilon
\to 0$ limes,
and therefore well defined, if and only if
$\mathfrak{h}$ is a Lie subalgebra of $\mathfrak{g}$~\cite{Inonu:1953sp}.
Specifying $\mathfrak{h}$ completely determines the \IW contraction 
(up to an isomorphism)~\cite{Saletan:61} and one can therefore
enumerate possible contractions by specifying a subalgebra.\\
With this knowledge we start with the three-dimensional
(Anti)-de Sitter algebra
(the upper sign is the AdS algebra, the lower for dS)
\begin{align}
  \left[ \Jt , \Gt_{a} \right]    &= \epsilon_{am}  \Gt_{m}, \label{eq:AdsSt}&
  \left[ \Jt , \Pt_{a} \right]    &= \epsilon_{am}  \Pt_{m},\\
  \left[ \Gt_{a} , \Gt_{b} \right] &= - \epsilon_{ab}  \Jt,   &
  \left[ \Gt_{a} , \Ht \right]     &= -\epsilon_{am}  \Pt_{m},  \\
  \left[ \Gt_{a} , \Pt_{b} \right] &= -\epsilon_{ab} \Ht,       &
  \left[ \Ht , \Pt_{a} \right]     &= \pm \epsilon_{am} \Gt_{m},\\
  \left[ \Pt_{a} , \Pt_{b} \right] &= \mp  \epsilon_{ab}\Jt,    \label{eq:AdsEn}
\end{align}
and specify the contractions according to
Table \ref{tab:spin2contr}.
Consecutive \IW contractions then lead to the cube of
Figure~\ref{fig:cube}.
For completeness, all the Lie algebras
are explicitly given in Table \ref{tab:KinSpin2AdS} and \ref{tab:KinSpin2Car} in Appendix~\ref{sec:expl-kinem-algebra}.
For nontrivial contractions, i.e., $\mathfrak{g}\neq \mathfrak{h}$
the resulting algebra is always non-semisimple due to the abelian
ideal spanned by $\mathfrak{i}$.

\begin{table}[H]
  \centering
$
  \begin{array}{l l l l l }
\toprule
    \text{Contraction } & \phantom{aa} & \multicolumn{1}{c}{\mathfrak{h}} & \phantom{aa} & \multicolumn{1}{c}{\mathfrak{i}} \\ \midrule
    \text{Space-time}   &              & \{\Jt,\Gt_{a} \}                 &              & \{\Ht,\Pt_{a} \}                 \\
    \text{Speed-space}  &              & \{\Jt,\Ht \}                     &              & \{\Gt_{a},\Pt_{a} \}             \\
    \text{Speed-time}   &              & \{\Jt, \Pt_{a} \}                &              & \{\Gt_{a}, \Ht \}                \\
     \text{General}     &              & \{\Jt \}                         &              & \{\Ht, \Pt_{a}, \Gt_{a} \}       \\ \bottomrule
  \end{array}
$
\caption{The four different \IW contractions classified in~\cite{Bacry:1968zf}.}
  \label{tab:spin2contr}
\end{table}


\subsection{Carroll Gravity}
\label{sec:carr-grav}

As already discussed in Section \ref{sec:higher spin-chern},
if one wants to write a Chern-Simons theory for the Lie algebras at hand,
it is important for the Lie algebra to admit an invariant metric.
While the three-dimensional Carroll algebra automatically admits an invariant metric,
others like the Galilei algebra do not.
We will discuss in the next section why this is no surprise,
but first we want to show how to construct a Chern-Simons theory with the Carroll
algebra and impose boundary conditions
(we will follow closely \cite{Bergshoeff:2016soe},
to which we also refer to for more details).
Carroll geometries were recently studied
because of their relation to asymptotically flat spacetimes
 \cite{Duval:2014uva,Hartong:2015xda}.\\
The Carroll algebra
\begin{align}\label{eq:CarrollAlgebraBulk}
\left[\,\Jt  \comma \Gt_{a} \,\right]     &= \epsilon_{am}  \Gt_{m}, &
  \left[\, \Jt \comma \Pt_{a} \,\right]     &= \epsilon_{am}  \Pt_{m}, &
  \left[\, \Gt_{a} \comma \Pt_{b} \,\right] &= -\epsilon_{ab} \Ht,    
\end{align}
has the invariant metric
\begin{align}\label{eq:CarrollInvMetric}
\langle \Ht , \Jt \rangle  &=-1, & \langle \Pt_{a}, \Gt_{b}\rangle&=
                                                                  \delta_{ab} \,.
\end{align}
The connection
\begin{align}
  A =  \tau\, \Ht + e^a\, \Pt_a  + \omega\, \Jt + B^a\, \Gt_a,  
\end{align}
is the one-form that takes values in the Carroll algebra
and the action is the usual
Chern-Simons action \eqref{eq:ChernSimonsAction}.
The next step is to construct Brown-Henneaux-like boundary conditions
around the Carroll vacuum configuration
\begin{align}
  e^1_\varphi = \rho,\qquad e^2_\rho = 1,\qquad e^1_\rho=e^2_\varphi = 0,
\end{align}
which one can also write as
\begin{align}
  d s^2_{(2)} = e^a e^b \delta_{ab} = \rho^2 d\varphi^2 + d\rho^2\,.
\end{align}
We assume that $\rho$ is a radial coordinate and $\varphi$
is an angular coordinate that is periodically identified by 
$\varphi\sim\varphi+2\pi$.
Moreover, on the background the time-component should be fixed as
\begin{align}
\tau = d t\,.
\end{align}
This can be accomplished by the gauge transformation
\begin{align}
  A = b^{-1}(\rho)\,\big(d+a(t,\,\varphi)\big)\,b(\rho),
  \qquad b(\rho) = e^{\rho \Pt_2},
\end{align}
and using~\cite{Bergshoeff:2016soe}
\begin{align}
a_\varphi &= -\Jt + h(t,\,\varphi)\, \Ht + p_a(t,\,\varphi)\, \Pt_a + g_a(t,\,\varphi)\, \Gt_a\,,  \label{eq:car9a} \\
a_t &= \mu(t,\,\varphi)\, \Ht\,. \label{eq:car9b}
\end{align}
These off-shell boundary conditions lead to
\begin{align}
d s^2_{(2)}& = \big[\big(\rho + p_1(t,\,\varphi)\big)^2 +
  p_2(t,\,\varphi)^2\big]\,d\varphi^2 + 2p_2(t,\,\varphi)\,d\varphi d \rho + d\rho^2,
  \\
 &= \big(\rho^2 + {\mathcal{O}}(\rho)\big)\,d \varphi^2
  + {\mathcal{O}}(1)\,\extd\rho d\varphi + d\rho^2,
\end{align}
and
\begin{align}
  \tau &= \mu(t,\,\varphi)\,d t + \big(h(t,\,\varphi) - \rho\,
         g_1(t,\,\varphi)\big)\,d \varphi\,.
  \\
   &= \mu(t,\,\varphi)\, d t + {\mathcal{O}}(\rho)\, d\varphi\,.
\end{align}
The analysis of the asymptotic symmetries leads to conserved,
integrable and finite charges
and using a suitable Fourier decomposition these
lead to the asymptotic symmetry algebra
\begin{align}
  [\Jt,\,\Pt^a_n] &= \epsilon_{ab}\,\Pt^b_n\,, \\
  [\Jt,\,\Gt^a_n] &= \epsilon_{ab}\,\Gt^b_n\,, \\
  [\Pt^a_n,\,\Gt^b_m] &= -\big(\epsilon_{ab} + in \delta_{ab}\big)\,\Ht\,\delta_{n+m,\,0}\,.
\end{align}
It is interesting to note that the zero-mode generators
$\Jt$,$\Ht$, $\Pt^{a}_{0}$ and $\Gt^{a}_{0}$ span a
subalgebra that equals again the original Carroll algebra.
Thus this asymptotic symmetry algebra mirrors
constructions from three-dimensional
asymptotically flat and 
Anti-de Sitter spacetimes.
Two further sets of boundary conditions,
that can be seen as a limit from AdS,
were proposed
in \cite{Grumiller:2017sjh} one of which we will briefly describe in the following.\\
Assume again that Carroll gravity can be describe by a Chern-Simons action with gauge algebra \eqref{eq:CarrollAlgebraBulk} and invariant metric \eqref{eq:CarrollInvMetric}. Choosing a connection like
\begin{equation}\label{Carrollcon}
    a_{\varphi} =  \mathcal{K}(t,\varphi){\tt J} + \mathcal{J}(t,\varphi) {\tt H} + \mathcal{G}^a(t,\varphi) ({\tt G}_a  + {\tt P}_a)\,,\qquad a_t = \mu(t,\,\varphi)\,{\tt H}\,,
\end{equation}
it is straightforward to determine the asymptotic symmetry algebra that reads in terms of the Fourier modes of the state dependent functions $\mathcal{K}$, $\mathcal{J}$ and $\mathcal{G}^a$ 
\begin{subequations}\label{ASA_Carroll}
\begin{align}
[\Kt_n,\Jt_m] & = k n \delta_{m+n,0}, \\ 
[\Jt_n,\Gt^{\,a}_m] & = \epsilon^a{}_b \Gt_{n+m}^b, \\
[\Gt_n^a, \Gt_m^b] & = - 2 \epsilon^{ab} \Kt_{n+m} - 2nk\,\delta^{ab} \delta_{n+m,0}\,.
\end{align}
\end{subequations}
One puzzling aspect of these boundary conditions is that they appear to describe solutions that carry entropy despite (seemingly) having no horizon.\\
The spin-3 Carroll algebras (see Section \ref{sec:kinem-high-spin}),
like their spin-2 subalgebras,
admit an invariant metric
and thus can be written, without obstructions,
as a Chern-Simons theory.
While they have been
analyzed at a linear level no boundary conditions were proposed
so far.

\subsection{Invariant Metrics and Double Extensions}
\label{sec:invariant-metrics}

\begin{figure}[ht]
\begin{center}
\begin{tikzpicture}[
  txt/.style={draw,rectangle, minimum height=18},
  fund/.style={draw,rectangle, minimum height=18,rounded corners=3mm,
    top color=white, bottom color=black!20},
  indec/.style={draw,rectangle, minimum height=18, top color=white, bottom color=black!20},
  nonsuf/.style={->,densely dashed}
]

\node (simple) at (-3,0) [fund] {Simple};
\node (onedim) at (0,0) [fund] {One-dimensional};

\node (sdis) at (-3,-1) [label=center:$\oplus$] {};
\node (onedis) at (0,-1) [label=center:$\oplus$] {};

\draw[->] (simple) -- (sdis);
\draw[->] ($(simple.south) + (4mm,0)$) -- (sdis);
\draw[->] ($(simple.south) - (4mm,0)$) -- (sdis);
\draw[->] (onedim) -- (onedis);
\draw[->] ($(onedim.south) + (4mm,0) $)-- (onedis);
\draw[->] ($(onedim.south) - (4mm,0) $)-- (onedis);

\node (semisimple) at (-3,-2) [txt] {Semisimple};
\draw[->] (sdis) -- (semisimple);

\node (abelian) at (0,-2) [txt] {Abelian};
\draw[->] (onedis) -- (abelian);

  \node (de) at (3,0) [indec] {$D(\mathfrak{g},\mathfrak{h})$};

{\node (dedis) at (3,-1) [label=center:$\oplus$] {};}

{\node (doubleext) at (3,-2) [txt] {Double extensions};
\draw[->] (dedis) -- (doubleext);
}


{\node (redsssdis) at (0,-3) [label=center:$\oplus$] {};
\node (ssd) at (0,-4) [txt] {Symmetric self-dual};
\draw[->] (redsssdis) -- (ssd);

\draw[->] (abelian) -- (redsssdis);
\draw[->] (semisimple.south) -- (redsssdis.west);
\draw[->] (doubleext.south) -- (redsssdis.east) ;
}

{\draw[->] (de) -- (dedis);
\draw[->] ($(de.south) + (4mm,0) $) -- (dedis);
\draw[->] ($(de.south) - (4mm,0) $) -- (dedis);

\node (or) at (0,1) [circle,draw,inner sep=1pt] {or};
\draw[nonsuf] (simple.north) -- (-3,1) -- (or.west);
\draw[nonsuf] (onedim) -- (or);
\draw[nonsuf] (or.east) -- ($(3,1) + (3mm,0)$) -- ($(de.north) + (3mm,-1.5mm)$);

\node (abdesum) at (1.1,-2) [label=center:$\oplus$] {};
\draw[nonsuf] (abelian.east) -- (abdesum);
\draw[nonsuf] (doubleext.west) -- (abdesum);
\draw[nonsuf]  (abdesum) -- (1.1,-1) -- (1.8,-1) -- (1.8,0.75) -- ($(3,0.75)+(-0.3mm,0)$) -- ($(de.north) + (-0.3mm,-1.5mm)$);
}
\end{tikzpicture} 
\end{center}
\caption{This diagram shows how all Lie algebras with invariant metric,
  i.e., symmetric self-dual Lie algebras, are constructed.
  The fundamental indecomposable building blocks are the simple and
  the one-dimensional Lie algebras
  and they need to be accompanied by the operations of direct sums ($\oplus$)
  and double extensions ($D(\mathfrak{g},\mathfrak{h})$).
  Direct sums of simple and one-dimensional Lie algebras,
  lead to semisimple and abelian ones, respectively.
  To construct new indecomposable Lie algebras that admit an invariant metric
  one needs to double extend an abelian or an already double extended Lie algebra
  (as explained it should not have a simple factor).
}
\label{fig:double}
\end{figure}
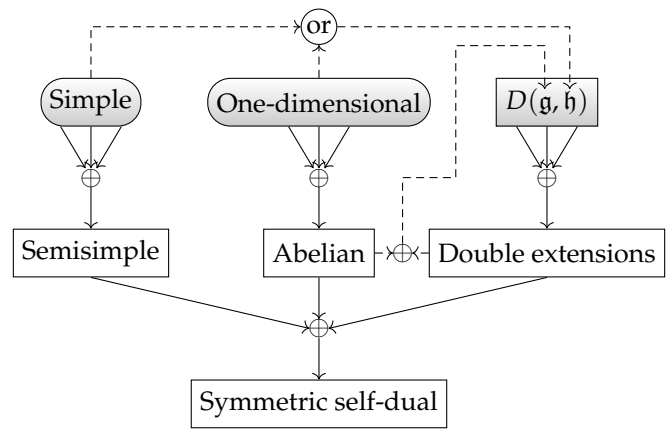

The connection between Lie algebras and their invariant metrics
has been greatly clarified by Mediny and Revoy~\cite{Medina1985}
(here we will follow \cite{FigueroaO'Farrill:1995cy}). They proved a structure theorem
that explains how all Lie algebras permitting such an invariant nondegenerate symmetric bilinear form\footnote{
These Lie algebras are sometimes called symmetric self-dual or quadratic.}
 can be constructed.
This provides a useful guiding principle for the construction of
Lie algebras with invariant metrics
and, as will be shown later, also explains why the Carroll algebras
inherit their invariant metric from the Poincar\'e algebra.\\
For that, one first restricts to indecomposable Lie algebras. We call a Lie algebra indecomposable if it cannot be decomposed as an orthogonal direct sum of two Lie algebras $\mathfrak{g}_{1}$ and $\mathfrak{g}_{2}$,
i.e., it cannot be written in such a way that the two algebras commute $[\mathfrak{g}_{1},\mathfrak{g}_{2}]=0$
and that they are orthogonal $\langle \mathfrak{g}_{1},\mathfrak{g}_{2}\rangle = 0$. Additionally one has to define double extensions~\cite{Medina1985}.
The Lie algebra $\mathfrak{d}=D(\mathfrak{g},\mathfrak{h})$ defined on
the vector space direct sum $\mathfrak{g} \oplus \mathfrak{h} \oplus
\mathfrak{h}^{*}$
(spanned by $\Gt_{i}$, $\Ht_{\alpha}$ and $\Ht^{\alpha}$, respectively) by 
\begin{align}
 [\Gt_{i},\Gt_{j}]                            & =f\indices{_{ij}^{k}} \Gt_{k}+f\indices{_{\alpha i}^{k}} \Omega_{kj}^{\mathfrak{g}} \Ht^{\alpha}, 
                    \label{eq:GG}           \\
 [\Ht_{\alpha},\Gt_{i}]                       & =f\indices{_{\alpha i}^{j}} \Gt_{j},
                         \label{eq:HG}      \\
 [\Ht_{\alpha},\Ht_{\beta}]                   & =f\indices{_{\alpha \beta}^{\gamma}} \Ht_{\gamma},
                             \label{eq:HH}  \\
 [\Ht_{\alpha},\Ht^{\beta}]                   & =-f\indices{_{\alpha \gamma}^{\beta}} \Ht^{\gamma},
                             \label{eq:HHd} \\
  [\Ht^{\alpha},\Gt_{j}]                      & =0,
                          \label{eq:HdG}    \\
  [\Ht^{\alpha},\Ht^{\beta}]                  & =0,
                              \label{eq:HdHd}
\end{align}
is a double extension of $\mathfrak{g}$ by $\mathfrak{h}$. It has the invariant metric
\begin{align}
  \Omega_{ab}^{\mathfrak{d}}= \bordermatrix{~ & \Gt_{j}                    & \Ht_{\beta}                       & \Ht^{\beta} \cr
                            \Gt_{i}           & \Omega_{ij}^{\mathfrak{g}} & 0                                 & 0  \cr
                             \Ht_{\alpha}     & 0                          & h_{\alpha\beta}                   & \delta\indices{_{\alpha}^{\beta}} \cr
                             \Ht^{\alpha}     & 0                          & \delta\indices{^{\alpha}_{\beta}} & 0 \cr},
\end{align}
where $\Omega_{ij}^{\mathfrak{g}}$ is an invariant metric on
$\mathfrak{g}$ and $h_{\alpha\beta}$ is some arbitrary (possibly degenerate)
symmetric invariant bilinear form on $\mathfrak{h}$.\\
An example is the Poincar\'e algebra,
see equation \eqref{eq:AllThreeAlgebras} with $\ell\rightarrow\infty$.
In this case $\mathfrak{g}$ is trivial and $\mathfrak{h}$ and $\mathfrak{h}^{*}$
is spanned by $\Jt_{a}$ and $\Pt_{a}$, respectively.
The invariant metric is then given by (\ref{eq:Isl2RBilinFormJanP}).
Similar considerations apply to the Carroll algebra of Section~\ref{sec:carr-grav}.
These two algebra are actually related by a 
natural generalization of the \IW contractions to
double extensions, see Section 5.3 in \cite{Prohazka:2017pkc}.
For that, one needs to apply the "dual contraction" on the dual
part of a subspace of $\mathfrak{h}$.
Explicitly this means that one takes the Poincar\'e algebra (see Table \ref{tab:KinSpin2AdS}) and rescales
$\Gt_a \mapsto \frac{1}{c} \Gt_a$.
Since $\Pt_a$ is in the dual part,
this can be read off of equation \eqref{eq:Isl2RBilinFormJanP}.
One then applies the dual contraction 
$\Pt_a \mapsto c \Pt_a$.
Using these rescalings leads to
\begin{align}\label{eq:invconr}
\left[\,\Jt  \comma \Gt_{a} \,\right]     &= \epsilon_{am}  \Gt_{m}, &
  \left[\, \Jt \comma \Pt_{a} \,\right]     &= \epsilon_{am}  \Pt_{m}, &
  \left[\, \Gt_{a} \comma \Pt_{b} \,\right] &= -\epsilon_{ab} \Ht,
  \\
  \left[\,\Gt_a  \comma \Gt_{b} \,\right]     &= -c^2 \epsilon_{ab}  \Jt, &
  \left[\, \Gt_a \comma \Ht \,\right]     &= -c^2 \epsilon_{am}  \Pt_{m}, &
\end{align}
and therefore to the Carroll algebra for $c \to 0$.
The part that makes this new interpretation
interesting is that it automatically
leaves the invariant metric untouched
since $\langle \Gt_a, \Pt_b \rangle \mapsto \langle \frac{1}{c}\Gt_a, c \Pt_b \rangle= \langle \Gt_a, \Pt_b \rangle$.
That this is not just a coincidence,
but that these non-simple Lie algebras actually have to be a double
extension is explained by the following
theorem.\\
Every indecomposable
Lie algebra which permits an invariant metric,
i.e., every indecomposable symmetric self-dual Lie algebra
is either~\cite{Medina1985,FigueroaO'Farrill:1995cy}:
\begin{enumerate}
\item A simple Lie algebra.
\item A one-dimensional Lie algebra.
\item A double extended Lie algebra $D(\mathfrak{g},\mathfrak{h})$ where:
  \begin{enumerate}
  \item $\mathfrak{g}$ has no factor $\mathfrak{p}$ for which
    the first and second cohomology group vanishes
    $H^{1}(\mathfrak{p},\mathbb{R}) = H^{2}(\mathfrak{p},\mathbb{R}) = 0$.
    This includes semisimple Lie algebra factors.
        \label{item:MRnosimp}
  \item $\mathfrak{h}$ is either simple or one-dimensional.
    \label{item:MRhsimp}
  \item $\mathfrak{h}$ acts on $\mathfrak{g}$ via outer derivations.
    \label{item:MRhout}
  \end{enumerate}
\end{enumerate}
Since every decomposable Lie algebra can be obtained from the
indecomposable ones this theorem describes how all of them can be generated,
see Figure \ref{fig:double} \cite{Prohazka:2017pkc}.

\subsection{Kinematical Higher Spin Algebras}
\label{sec:kinem-high-spin}

We now want to investigate in which sense the
kinematical Lie algebras can be generalized to higher spins,
specifically we will focus on spin-two and three fields.
For that it is again useful to start with the (semi)simple, (A)dS
algebras\footnote{
  Semisimple Lie algebras are a natural starting point for these
  kind of
  considerations since no (nontrivial) contraction can lead to
  a semisimple Lie algebra.
} explicitly given in Table \ref{tab:hsads}.
\begin{table}[H]
  \centering
$
\begin{array}{l r r r r r}
\toprule %
                                               &
                                                 \mathfrak{hs}_{3}\mathfrak{(A)dS}_{\parmp}
  \\ \midrule
   \text{Equations (\ref{eq:AdsSt}) to (\ref{eq:AdsEn})}\\ \midrule
  \left[\, \Jt \comma \Jt_{a} \,\right]        & \epsilon_{am} \Jt_{m}                               \\
  \left[\, \Jt \comma \Gt_{ab} \,\right]       & -\epsilon_{m(a}\Gt_{b)m}                            \\
  \left[\, \Jt \comma \Ht_{a} \,\right]        & \epsilon_{am} \Ht_{m}                               \\
  \left[\, \Jt \comma \Pt_{ab} \,\right]       & -\epsilon_{m(a}\Pt_{b)m}                            \\
  \left[\, \Gt_{a} \comma \Jt_{b}  \,\right]   & -(\epsilon_{am} \Gt_{bm}+ \epsilon_{ab} \Gt_{mm})   \\
  \left[\, \Gt_{a} \comma \Gt_{bc}  \,\right]  & -\epsilon_{a(b}\Jt_{c)}                             \\
  \left[\, \Gt_{a} \comma \Ht_{b}  \,\right]   & -(\epsilon_{am} \Pt_{bm}+ \epsilon_{ab} \Pt_{mm})   \\
  \left[\, \Gt_{a} \comma \Pt_{bc}  \,\right]  & -\epsilon_{a(b}\Ht_{c)}                             \\
  \left[\, \Ht \comma \Jt_{a} \,\right]        & \epsilon_{am} \Ht_{m}                               \\
  \left[\, \Ht \comma \Gt_{ab} \,\right]       & -\epsilon_{m(a}\Pt_{b)m}                            \\
  \left[\, \Ht \comma \Ht_{a} \,\right]        & \pm  \epsilon_{am} \Jt_{m}                          \\
  \left[\, \Ht \comma \Pt_{ab} \,\right]       & \mp \epsilon_{m(a}\Gt_{b)m}                         \\
  \left[\, \Pt_{a} \comma \Jt_{b}  \,\right]   & -(\epsilon_{am} \Pt_{bm}+ \epsilon_{ab} \Pt_{mm})   \\
  \left[\, \Pt_{a} \comma \Gt_{bc}  \,\right]  & -\epsilon_{a(b}\Ht_{c)}                             \\
  \left[\, \Pt_{a} \comma \Ht_{b}  \,\right]   & \mp(\epsilon_{am} \Gt_{bm}+ \epsilon_{ab} \Gt_{mm}) \\
  \left[\, \Pt_{a} \comma \Pt_{bc}  \,\right]  & \mp \epsilon_{a(b}\Jt_{c)}                          \\ \midrule
  \left[\, \Jt_{a} \comma \Jt_{b} \,\right]    & \epsilon_{ab} \Jt                                   \\
  \left[\, \Jt_{a} \comma \Gt_{bc} \,\right]   & \delta_{a(b}\epsilon_{c)m} \Gt_{m}                  \\
  \left[\, \Jt_{a} \comma \Ht_{b} \,\right]    & \epsilon_{ab} \Ht                                   \\
  \left[\, \Jt_{a} \comma \Pt_{bc} \,\right]   & \delta_{a(b}\epsilon_{c)m} \Pt_{m}                  \\
  \left[\, \Gt_{ab} \comma \Gt_{cd}  \,\right] & \delta_{(a(c}\epsilon_{d)b)}\Jt                     \\
  \left[\, \Gt_{ab} \comma \Ht_{c} \,\right]   & - \delta_{c(a}\epsilon_{b)m} \Pt_{m}                \\
  \left[\, \Gt_{ab} \comma \Pt_{cd}  \,\right] & \delta_{(a(c}\epsilon_{d)b)}\Ht                     \\
  \left[\, \Ht_{a} \comma \Ht_{b} \,\right]    & \pm \epsilon_{ab} \Jt                               \\
  \left[\, \Ht_{a} \comma \Pt_{bc} \,\right]   & \pm  \delta_{a(b}\epsilon_{c)m} \Gt_{m}             \\
  \left[\, \Pt_{ab} \comma \Pt_{cd}  \,\right] & \pm  \delta_{(a(c}\epsilon_{d)b)}\Jt                \\ \midrule
\end{array}
$
\caption{Higher spin versions of the (A)dS algebras. The upper sign is for AdS
  and the lower sign for dS.}
\label{tab:hsads}
\end{table}
The spin-2 part is a subalgebra and is extended by
the spin-3 generators $\Jt_{a}, \Ht_{a}, \Gt_{ab}, \Pt_{ab}$.
For the generalization of the contractions to the higher spin algebra
the following restrictions were imposed~\cite{Bergshoeff:2016soe}:
\begin{itemize}
\item The \IW contractions are restricted such that the contracted
  spin-2 Lie subalgebra of the contracted one coincides
  with the kinematical ones of Bacry and
  Levy-Leblond~\cite{Bacry:1968zf}
  (see Table \ref{tab:spin2contr} and Appendix \ref{sec:expl-kinem-algebra}).
\item The commutator of the spin-3 fields should be non-vanishing.
  This ensures that the spin-3 field also interacts with the spin-2 field.
\end{itemize}
Using this restrictions one can systematically examine the possible
contractions summarized in Table
\ref{tab:contr}~\cite{Bergshoeff:2016soe}\footnote{
  We ignore the traceless contractions in this review.
}.
\begin{table}[H]
  \centering
$
  \begin{array}{l r l l }
\toprule
    \text{Contraction }                                                                                                & \# & \multicolumn{1}{c}{\mathfrak{h}}    & \multicolumn{1}{c}{\mathfrak{i}}                     \\ \midrule
    \text{Space-time}                                                                                                  & 1  & 
                                           \{\Jt, \Gt_{a},\Jt_{a},\Gt_{ab}\}                                           & 
                                                                                \{\Ht, \Pt_{a},\Ht_{a}, \Pt_{ab} \}                                                                                                      \\
                                                                                                                       & 2  & 
                                                                                    \{\Jt, \Gt_{a},\Ht_{a}, \Pt_{ab}\} & 
                                                                                                                         \{\Ht, \Pt_{a},\Jt_{a},\Gt_{ab} \}                                                              \\ \midrule
    \text{Speed-space}                                                                                                 & 3  & \{\Jt,\Ht,\Jt_{a}, \Ht_{a} \}       & \{\Gt_{a},\Pt_{a},\Gt_{ab},\Pt_{ab} \}               \\
                                                                                                                       & 4  & \{\Jt,\Ht,\Gt_{ab},\Pt_{ab} \}      & \{\Gt_{a},\Pt_{a},\Jt_{a}, \Ht_{a} \}                \\ \midrule
    \text{Speed-time}                                                                                                  & 5  & \{\Jt, \Pt_{a},\Jt_{a},\Pt_{ab} \}  & \{\Gt_{a}, \Ht,\Ht_{a},\Gt_{ab} \}                   \\
                                                                                                                       & 6  & \{\Jt, \Pt_{a},\Ht_{a},\Gt_{ab}  \} & \{\Gt_{a}, \Ht,\Jt_{a},\Pt_{ab} \}                   \\ \midrule
                                                                                                                       & 7  & \{\Jt,\Jt_{a} \}                    & \{\Ht, \Pt_{a}, \Gt_{a},\Ht_{a},\Gt_{ab},\Pt_{ab} \} \\
     \text{General}                                                                                                    & 8  & \{\Jt,\Gt_{ab} \}                   & \{\Ht, \Pt_{a}, \Gt_{a},\Jt_{a},\Ht_{a} ,\Pt_{ab} \} \\
                                                                                                                       & 9  & \{\Jt,\Ht_{a} \}                    & \{\Ht, \Pt_{a}, \Gt_{a},\Jt_{a},\Gt_{ab},\Pt_{ab} \} \\
                                                                                                                       & 10 & \{\Jt,\Pt_{ab} \}                   & \{\Ht, \Pt_{a}, \Gt_{a},\Jt_{a},\Ht_{a} ,\Gt_{ab} \} \\ \bottomrule
  \end{array}
$
  \caption{The contractions to the kinematical higher spin algebras.
They can be summarized again as a (higher spin) cube,
see Figure~\ref{fig:hscube}.}
  \label{tab:contr}
\end{table}
Using this table it is straightforward to perform the contractions leading the the kinematical higher spin algebras. Consecutive contractions lead then to the (higher spin) cube
of Figure~\ref{fig:hscube}. 
\newcommand*{\shi}{1}
\newcommand*{\shih}{0.5}
\newcommand*{\shihh}{0.25}
\begin{figure}[ht]
  \centering
\tdplotsetmaincoords{60}{110}
\begin{tikzpicture}[
tdplot_main_coords,
dot/.style={circle,fill,scale=0.5},
linf/.style={thick,->,blue},
cinf/.style={thick,->,red},
tinf/.style={thick,->},
stinf/.style={ultra thick,->,gray},
scale=0.7
]


\node (ads) at (0,10,10) [dot, label=right:$\mathfrak{hs_3(A)}\mathfrak{dS}$] {};

\node (p) at (10,10,10) [label=below:$\mathfrak{hs}_{3}\mathfrak{poi}$] {};

\node (p1) at (10+\shi,10-\shi,10) [dot] {};
\node (p2) at (10-\shi,10+\shi,10) [dot] {};


\node (nh1) at (-\shi,10-\shi,0) [dot] {};
\node (nh2) at (\shi,10+\shi,0) [dot, label=right:$\mathfrak{hs}_{3}\mathfrak{nh}$] {};

\node (pp) at (0,0,10) [label=above:$\mathfrak{hs}_{3}\mathfrak{ppoi}$] {};

\node (pp1) at (-\shi,\shi,10) [dot] {};
\node (pp2) at (\shi,-\shi,10) [dot] {};

\node (g) at (10,10,0) [label=below:$\mathfrak{hs}_{3}\mathfrak{gal}$] {};

\node (g3) at (10-\shi-\shih,10+\shi+\shih,0) [dot] {};
\node (g4) at (10-\shi+\shih,10+\shi-\shih,0) [dot] {};

\node (g1) at (10+\shi+\shih,10-\shi-\shih,0) [dot] {};
\node (g2) at (10+\shi-\shih,10-\shi+\shih,0) [dot] {};

\node (pg) at (0,0,0) [label=above:$\mathfrak{hs}_{3}\mathfrak{pgal}$] {};


\node (pg1) at (-\shi+\shih,\shi-\shih,0) [dot] {};
\node (pg2) at (-\shi-\shih,\shi+\shih,0) [dot] {};


\node (pg3) at (\shi+\shih,-\shi-\shih,0) [dot] {};
\node (pg4) at (\shi-\shih,-\shi+\shih,0) [dot] {};


\node (car1) at (10+\shi-\shih,\shi-\shih,10) [dot] {};
\node (car2) at (10+\shi+\shih,\shi+\shih,10) [dot] {};

\node (car3) at (10-\shi-\shih,-\shi-\shih,10) [dot, label=left:$\mathfrak{hs}_{3}\mathfrak{car}$] {};
\node (car4) at (10-\shi+\shih,-\shi+\shih,10) [dot] {};

\node (st) at (10,0,0) [label=below left:$\mathfrak{hs}_{3}\mathfrak{st}$] {};

\node (st3) at (10-\shi-\shih-\shihh,-\shi-\shih-\shihh,0) [dot] {};
\node (st4) at (10-\shi-\shih+\shihh,-\shi-\shih+\shihh,0) [dot] {};
\node (st7) at (10+\shi+\shih+\shihh,\shi+\shih+\shihh,0) [dot] {};
\node (st8) at (10+\shi+\shih-\shihh,\shi+\shih-\shihh,0) [dot] {};

\node (st5) at (10+\shi-\shih+\shihh,\shi-\shih+\shihh,0) [dot] {};
\node (st6) at (10+\shi-\shih-\shihh,\shi-\shih-\shihh,0) [dot] {};

\node (st1) at (10-\shi+\shih+\shihh,-\shi+\shih+\shihh,0) [dot] {};
\node (st2) at (10-\shi+\shih-\shihh,-\shi+\shih-\shihh,0) [dot] {};

\draw[tinf] (p1) -- (car2);
\draw[tinf] (p1) -- (car1);
\draw[tinf] (p2) -- (car3);
\draw[tinf] (p2) -- (car4);

\draw[cinf] (ads) -- node [left] {\#3} (nh1);
\draw[cinf] (ads) -- node [right] {\#4}(nh2);

\draw[tinf] (ads) -- node [above] {\#5}(pp1);
\draw[tinf] (ads) -- node [below] {\#6}(pp2);

\draw[cinf] (p1) -- (g1);
\draw[cinf] (p1) -- (g2);
\draw[cinf] (p2) -- (g3);
\draw[cinf] (p2) -- (g4);

\draw[linf] (ads) -- node [left] {\#1} (p1);
\draw[linf] (ads) -- node [left] {\#2} (p2);

\draw[linf] (nh1) -- (g1);
\draw[linf] (nh1) -- (g2);
\draw[linf] (nh2) -- (g3);
\draw[linf] (nh2) -- (g4);

\draw[linf] (pp1) -- (car1);
\draw[linf] (pp1) -- (car2);
\draw[linf] (pp2) -- (car3);
\draw[linf] (pp2) -- (car4);

\draw[cinf] (car2) -- (st8);
\draw[cinf] (car2) -- (st7);

\draw[cinf] (car1) -- (st6);
\draw[cinf] (car1) -- (st5);

\draw[cinf] (car3) -- (st4);
\draw[cinf] (car3) -- (st3);
\draw[cinf] (car4) -- (st2);
\draw[cinf] (car4) -- (st1);

\draw[tinf,dashed] (nh1) -- (pg1);
\draw[tinf,dashed] (nh1) -- (pg2);
\draw[tinf,dashed] (nh2) -- (pg3);
\draw[tinf,dashed] (nh2) -- (pg4);

\draw[cinf,dashed] (pp1) -- (pg1);
\draw[cinf,dashed] (pp1) -- (pg2);
\draw[cinf,dashed] (pp2) -- (pg3);
\draw[cinf,dashed] (pp2) -- (pg4);

\draw[tinf] (g1) -- (st8);
\draw[tinf] (g1) -- (st7);
\draw[tinf] (g2) -- (st6);
\draw[tinf] (g2) -- (st5);
\draw[tinf] (g3) -- (st4);
\draw[tinf] (g3) -- (st3);
\draw[tinf] (g4) -- (st2);
\draw[tinf] (g4) -- (st1);

\draw[linf,dashed] (pg1) -- (st6);
\draw[linf,dashed] (pg1) -- (st5);
\draw[linf,dashed] (pg2) -- (st8);
\draw[linf,dashed] (pg2) -- (st7);

\draw[linf,dashed] (pg3) -- (st3);
\draw[linf,dashed] (pg3) -- (st4);
\draw[linf,dashed] (pg4) -- (st1);
\draw[linf,dashed] (pg4) -- (st2);

\end{tikzpicture}
\caption{
  This figure summarizes the contractions of Table \ref{tab:contr}.
  There are 2 space-time (blue; \#1,\#2), 2 speed-space (red; \#3,\#4)
  and 2 speed-time (black; \#5,\#6) contractions and combining them
  leads to the full cube.
  The explicit commutators of all algebras can be found in the
  Appendix of \cite{Bergshoeff:2016soe}.
  In comparison to Figure~\ref{fig:cube}, we have for clarity omitted the double lines.}
  \label{fig:hscube}
\end{figure}
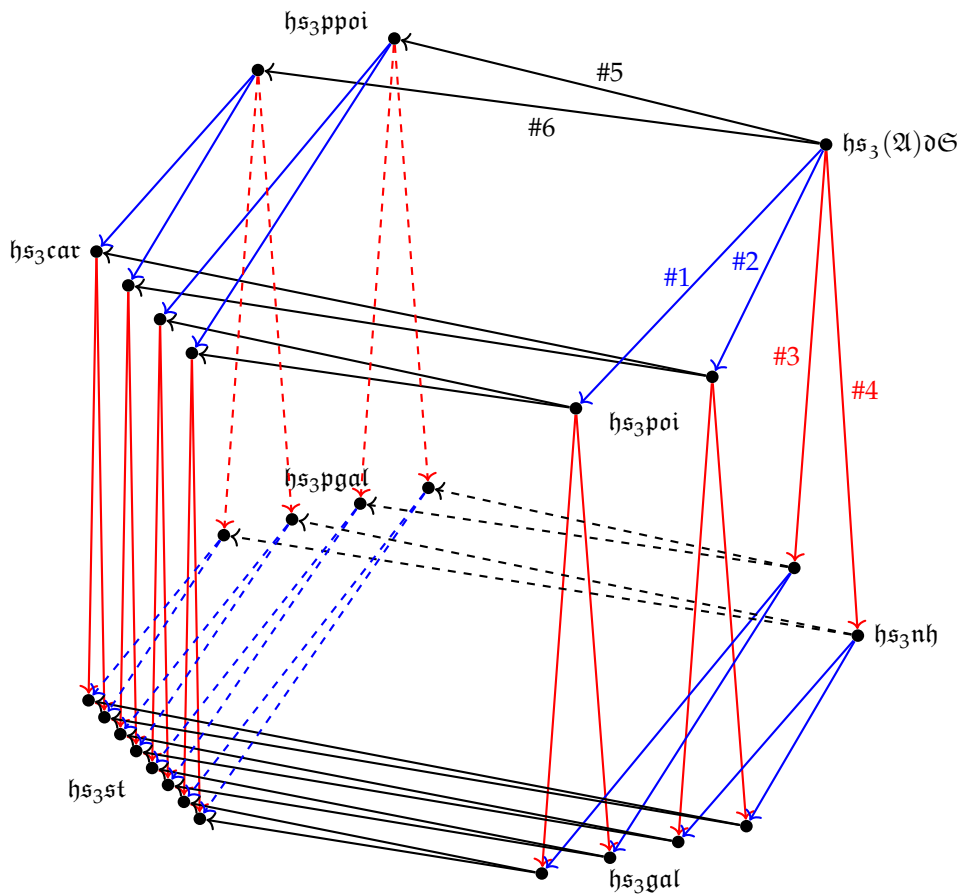
Given the zoo of higher spin algebras,
each corner of the higher spin cube represents one,
the question arises if they permit an invariant metric.
For the (semi)simple (A)dS algebras this is obvious,
but not so much for the other ones.
For the case of the higher spin Poincar\'e algebras
the considerations of Section~\ref{sec:non-ads-through}
generalize and for the higher spin Carroll algebras
there are again invariant metric preserving contractions~\cite{Prohazka:2017pkc}
analog to the ones discussed in Section~\ref{sec:kinem-high-spin}.\\
For the Galilei algebras the situation is different and
the knowledge of double extensions proves to be useful.
Already in the case of spin-2 the three-dimensional
Galilei algebra has no invariant metric~\cite{Papageorgiou:2009zc}.
However, it is possible to centrally extend the Galilei algebra by two nontrivial central
extensions -- out of three possible ones~\cite{levygalgr} --
to obtain a Lie algebra with an invariant metric.
One of these central extensions is possible in any dimension,
and corresponds to the mass in the so called Bargmann algebra.
Due to the second extension that is 
a peculiarity of three spacetime dimensions
this algebra is called extended Bargmann algebra
and  makes a Chern-Simons formulation
possible~\cite{Papageorgiou:2009zc}.\\
The higher spin Galilei generalizations also permit no invariant
metric~\cite{Bergshoeff:2016soe}.
In contrast to the spin-2 case central extensions are not sufficient to
provide an invariant metric, but double extensions provide guidance.
Interestingly the double extension of the spin-3 Galilei algebras
lead naturally to Lie algebras where the spin-2 part is exactly the
just mentioned extended Bargmann algebra.
Therefore, the higher spin generalization of the Galilei algebra can be considered the spin-3 extended Bargmann algebra.
Furthermore, the properties of the higher spin Carroll and extended Bargmann
theories have been studied in detail~\cite{Bergshoeff:2016soe}.

\section{Summary and Outlook}
\label{sec:outlook}

Three-dimensional higher spin theories beyond (Anti)-de Sitter
can be roughly separated by the gauge algebra that is used
in their Chern-Simons formulation.
As reviewed in Section \ref{sec:non-ads-through},
using the higher spin (A)dS gauge algebras
one is able to construct
backgrounds and boundary conditions for
Lobachevsky, Lifshitz, null warped and Schrödinger spacetimes.
Going beyond the (A)dS gauge algebra the best understood
cases so far are flat space higher spin theories that were discussed in Section
\ref{sec:flat-space-higher}.
The interesting question if there are higher spin algebras
beyond these cases has been answered by the construction
of the kinematical higher spin algebras reviewed in Section
\ref{sec:non-ads-through-1}.\\
While some progress has been made up until now there are certainly
interesting open problems that demand further investigation.

\subsection{Boundary Conditions and Boundary Theories}
\label{sec:boundary-conditions}

While it was shown that AdS higher spin gauge algebras permit backgrounds and boundary conditions beyond the standard AdS choices
their asymptotic symmetry algebras often turned out to be related
to already known ones.
It would be interesting to further investigate if these boundary conditions and their asymptotic symmetry algebras can be further specialized in order to yield, e.g.,  Lifshitz-like asymptotic symmetries.\\
While for the Carroll case boundary conditions have been proposed~\cite{Bergshoeff:2016soe,Grumiller:2017sjh}
for most of the other kinematical algebras,
and especially the higher spin generalizations,
no consistent boundary conditions have been established yet.
Further examination is also needed
for the mysterious result
that it seems that one can assign entropy
to
Carroll geometries~\cite{Grumiller:2017sjh}.
It would be interesting to see if this
result can be generalized to the
higher spin case.
A generalization and interpretation of higher spin entanglement entropy~\cite{deBoer:2013vca,Ammon:2013hba} in these setups
would be another intriguing option.\\
Another interesting generalization would
be the calculation of one loop partition functions.
Here, Newton--Hooke and Para-Poincar\'e seem to be intriguing options.
This is due 
to the still non-vanishing cosmological constant
and one might therefore hope that
they exhibit the ``box-like'' behavior of AdS.\\
For the higher spin cases it would be interesting to see
if the asymptotic symmetry algebras
lead to nonlinear generalizations similar to the
$\mathcal{W}_{N}$ algebras for AdS (some of them might be related to
the ones discussed in \cite{Rasmussen:2017eus}).\\
One interesting observation is
related to possible dual theories of the Chern-Simons theories treated in this review to
the Wess-Zumino-Novikov-Witten (WZNW) models~\cite{Witten:1988hf,Elitzur:1989nr}.
Here, again  WZNW models based on a Lie algebra
that admits an invariant metric play a distinguished role
since they admit a (generalized)
Sugawara construction~\cite{Mohammedi:1993rg}.
Double extensions and the Medina-Revoy theorem
are fundamental for the proof that the Sugawara construction
factorizes in a semisimple and a nonsemisimple one
\cite{FigueroaO'Farrill:1994hx}.

\subsection{Kinematical (higher spin) Algebras}
\label{sec:kinem-high-spin-1}

For the spin-2 extended Bargmann algebras
it was shown that they emerge as contractions of
(Anti)-de Sitter algebras that have been (trivially) centrally extended by
two one-dimensional algebras~\cite{Papageorgiou:2009zc}\footnote{
  See also Section 9.2 in \cite{Prohazka:2017pkc}.
}.
It is also not clear as of yet which (semisimple) algebra can be naturally contracted to yield the double extended higher spin versions of the Bargmann algebra. This is interesting, because the deformed theories are
often seen as more fundamental.
We are not aware of a systematic discussion of contractions and double extensions,
see however Section 5 in \cite{Prohazka:2017pkc} for a start.
Furthermore, it might be interesting to also look at the
Chern-Simons theories based on the Lie algebras that have not been
double extended.\\
For many considerations a generalization to the 
supersymmetric case seems possible.
Especially since the supersymmetric analog of
double extensions exists~\cite{99BB}
and
an analog of the structure theorem of Section \ref{sec:invariant-metrics}
has been proposed~\cite{07BBB}.
A Chern-Simons theory based on a supersymmetric
version of the extended Bargmann algebra has already been investigated in~\cite{Bergshoeff:2016lwr}.\\
Since the Chern-Simons theory based on the extended Bargmann algebra has been shown to be related to a specific version of Ho\v{r}ava-Lifshitz gravity~\cite{Hartong:2016yrf} if would be interesting to see if the higher spin extended Bargmann theories lead to a spin-3 Ho\v{r}ava-Lifshitz theory. \\
Last but not least,
two obvious generalizations to higher spin,
as well as to higher dimensions.

\vspace{6pt} 

\acknowledgments{
  The authors want to thank Daniel Grumiller for collaboration during the start of this work. In addition, we want to thank all our collaborators as well as all the other people that have been and are currently working on (non-)AdS higher spin holography for their valuable contributions to the field. Furthermore, the authors would like to thank Andrea Campoleoni and Nicolas Boulanger for inviting us to contribute to this special issue of Universe on Higher Spin Gauge Theories. The research of SP is supported by the ERC Advanced Grant ``High-Spin-Grav" and by FNRS-Belgium (convention FRFC PDR T.1025.14 and  convention IISN 4.4503.15). The research of MR is supported by the ERC Starting Grant 335146 "HoloBHC".}

\authorcontributions{Stefan Prohazka and Max Riegler contributed equally to this work.}

 \conflictsofinterest{The authors declare no conflict of interest.} 



\appendixtitles{yes} 
\appendixsections{multiple} 
\appendix

\section{Explicit Kinematical Algebra Relations}
\label{sec:expl-kinem-algebra}

\begin{table}[H]
  \centering
$
\begin{array}{l r r r r r}
\toprule %
                                            & \mathfrak{(A)dS}_{\parmp} & \mathfrak{poi}          & \mathfrak{nh}             & \mathfrak{ppoi}           \\ \midrule
  \left[\,\Jt  \comma \Jt \,\right]        & 0                         & 0                       & 0                         & 0                         \\
  \left[\,\Jt  \comma \Gt_{a} \,\right]     & \epsilon_{am}  \Gt_{m}    & \epsilon_{am}  \Gt_{m}  & \epsilon_{am}  \Gt_{m}    & \epsilon_{am}  \Gt_{m}    \\
  \left[\, \Jt\comma \Ht \,\right]          & 0                         & 0                       & 0                         & 0                         \\
  \left[\, \Jt \comma \Pt_{a} \,\right]     & \epsilon_{am}  \Pt_{m}    & \epsilon_{am}  \Pt_{m}  & \epsilon_{am}  \Pt_{m}    & \epsilon_{am}  \Pt_{m}    \\ 
  \left[\,\Gt_{a}  \comma \Gt_{b} \,\right] & - \epsilon_{ab}  \Jt      & - \epsilon_{ab}  \Jt    & 0                         & 0                         \\
  \left[\, \Gt_{a} \comma \Ht \,\right]     & -\epsilon_{am}  \Pt_{m}   & -\epsilon_{am}  \Pt_{m} & -\epsilon_{am}  \Pt_{m}   & 0                         \\
  \left[\, \Gt_{a} \comma \Pt_{b} \,\right] & -\epsilon_{ab} \Ht        & -\epsilon_{ab} \Ht      & 0                         & -\epsilon_{ab} \Ht        \\
  \left[\,\Ht  \comma \Pt_{a} \,\right]     & \pm \epsilon_{am} \Gt_{m} & 0                       & \pm \epsilon_{am} \Gt_{m} & \pm \epsilon_{am} \Gt_{m} \\
  \left[\,\Pt_{a}  \comma \Pt_{b} \,\right] & \mp  \epsilon_{ab}\Jt     & 0                       & 0                         & \mp  \epsilon_{ab}\Jt     \\ \midrule
\end{array}
$
\caption{(Anti-)de Sitter, Poincar\'e, Newton--Hooke and para-Poincar\'e algebras. The upper sign is for AdS (and contractions thereof) and the lower sign for dS (and contractions thereof).}
\label{tab:KinSpin2AdS}
\end{table}

\begin{table}[H]
  \centering
$
\begin{array}{l r r r r r}
\toprule %
                                            & \mathfrak{car}         & \mathfrak{gal}          & \mathfrak{pgal}           & \mathfrak{st}          \\ \midrule
 \left[\,\Jt  \comma \Jt \,\right]          & 0                      & 0                       & 0                         & 0                      \\
  \left[\,\Jt  \comma \Gt_{a} \,\right]     & \epsilon_{am}  \Gt_{m} & \epsilon_{am}  \Gt_{m}  & \epsilon_{am}  \Gt_{m}    & \epsilon_{am}  \Gt_{m} \\
  \left[\, \Jt\comma \Ht \,\right]          & 0                      & 0                       & 0                         & 0                      \\
  \left[\, \Jt \comma \Pt_{a} \,\right]     & \epsilon_{am}  \Pt_{m} & \epsilon_{am}  \Pt_{m}  & \epsilon_{am}  \Pt_{m}    & \epsilon_{am}  \Pt_{m} \\ 
  \left[\,\Gt_{a}  \comma \Gt_{b} \,\right] & 0                      & 0                       & 0                         & 0                      \\
  \left[\, \Gt_{a} \comma \Ht \,\right]     & 0                      & -\epsilon_{am}  \Pt_{m} & 0                         & 0                      \\
  \left[\, \Gt_{a} \comma \Pt_{b} \,\right] & -\epsilon_{ab} \Ht     & 0                       & 0                         & 0                      \\
  \left[\,\Ht  \comma \Pt_{a} \,\right]     & 0                      & 0                       & \pm \epsilon_{am} \Gt_{m} & 0                      \\
  \left[\,\Pt_{a}  \comma \Pt_{b} \,\right] & 0                      & 0                       & 0                         & 0                      \\ \midrule
\end{array}
$
\caption{Carroll, Galilei, para-Galilei and static algebra. The upper sign is for AdS (and contractions thereof) and the lower sign for dS (and contractions thereof).}
\label{tab:KinSpin2Car}
\end{table}




\externalbibliography{yes}
\bibliography{bibl}

\begin{thebibliography}{-------}
\providecommand{\natexlab}[1]{#1}

\bibitem[Klebanov and Polyakov(2002)]{Klebanov:2002ja}
Klebanov, I.; Polyakov, A.
\newblock {AdS dual of the critical O(N) vector model}.
\newblock {\em Phys.Lett.} {\bf 2002}, {\em B550},~213--219,
  \href{http://xxx.lanl.gov/abs/hep-th/0210114}{{\normalfont
  [arXiv:hep-th/hep-th/0210114]}}.

\bibitem[Mikhailov(2002)]{Mikhailov:2002bp}
Mikhailov, A.
\newblock {Notes on higher spin symmetries} {\bf 2002}.
\newblock  \href{http://xxx.lanl.gov/abs/hep-th/0201019}{{\normalfont
  [arXiv:hep-th/hep-th/0201019]}}.

\bibitem[Sezgin and Sundell(2002)]{Sezgin:2002rt}
Sezgin, E.; Sundell, P.
\newblock {Massless higher spins and holography}.
\newblock {\em Nucl.Phys.} {\bf 2002}, {\em B644},~303--370,
  \href{http://xxx.lanl.gov/abs/hep-th/0205131}{{\normalfont
  [arXiv:hep-th/hep-th/0205131]}}.

\bibitem[Fradkin and Vasiliev(1987{\natexlab{a}})]{Fradkin:1987ks}
Fradkin, E.; Vasiliev, M.A.
\newblock {On the Gravitational Interaction of Massless Higher Spin Fields}.
\newblock {\em Phys.Lett.} {\bf 1987}, {\em B189},~89--95.

\bibitem[Fradkin and Vasiliev(1987{\natexlab{b}})]{Fradkin:1986qy}
Fradkin, E.S.; Vasiliev, M.A.
\newblock {Cubic Interaction in Extended Theories of Massless Higher Spin
  Fields}.
\newblock {\em Nucl. Phys.} {\bf 1987}, {\em B291},~141--171.

\bibitem[Vasiliev(1990)]{Vasiliev:1990en}
Vasiliev, M.A.
\newblock {Consistent equation for interacting gauge fields of all spins in
  (3+1)-dimensions}.
\newblock {\em Phys. Lett.} {\bf 1990}, {\em B243},~378--382.

\bibitem[Sagnotti and Taronna(2011)]{Sagnotti:2010at}
Sagnotti, A.; Taronna, M.
\newblock {String Lessons for Higher-Spin Interactions}.
\newblock {\em Nucl. Phys.} {\bf 2011}, {\em B842},~299--361,
  \href{http://xxx.lanl.gov/abs/1006.5242}{{\normalfont
  [arXiv:hep-th/1006.5242]}}.

\bibitem[Vasiliev(2013)]{Vasiliev:2012vf}
Vasiliev, M.A.
\newblock {Holography, Unfolding and Higher-Spin Theory}.
\newblock {\em J.Phys.} {\bf 2013}, {\em A46},~214013,
  \href{http://xxx.lanl.gov/abs/1203.5554}{{\normalfont
  [arXiv:hep-th/1203.5554]}}.

\bibitem[Didenko and Skvortsov(2014)]{Didenko:2014dwa}
Didenko, V.; Skvortsov, E.
\newblock {Elements of Vasiliev theory} {\bf 2014}.
\newblock  \href{http://xxx.lanl.gov/abs/1401.2975}{{\normalfont
  [arXiv:hep-th/1401.2975]}}.

\bibitem[Giombi and Yin(2010)]{Giombi:2009wh}
Giombi, S.; Yin, X.
\newblock {Higher Spin Gauge Theory and Holography: The Three-Point Functions}.
\newblock {\em JHEP} {\bf 2010}, {\em 09},~115,
  \href{http://xxx.lanl.gov/abs/0912.3462}{{\normalfont
  [arXiv:hep-th/0912.3462]}}.

\bibitem[Giombi and Yin(2011)]{Giombi:2010vg}
Giombi, S.; Yin, X.
\newblock {Higher Spins in AdS and Twistorial Holography}.
\newblock {\em JHEP} {\bf 2011}, {\em 1104},~086,
  \href{http://xxx.lanl.gov/abs/1004.3736}{{\normalfont
  [arXiv:hep-th/1004.3736]}}.

\bibitem[de~Mello~Koch \em{et~al.}(2011)de~Mello~Koch, Jevicki, Jin, and
  Rodrigues]{Koch:2010cy}
de~Mello~Koch, R.; Jevicki, A.; Jin, K.; Rodrigues, J.P.
\newblock {$AdS_4/CFT_3$ Construction from Collective Fields}.
\newblock {\em Phys. Rev.} {\bf 2011}, {\em D83},~025006,
  \href{http://xxx.lanl.gov/abs/1008.0633}{{\normalfont
  [arXiv:hep-th/1008.0633]}}.

\bibitem[Giombi and Yin(2012)]{Giombi:2011ya}
Giombi, S.; Yin, X.
\newblock {On Higher Spin Gauge Theory and the Critical O(N) Model}.
\newblock {\em Phys. Rev.} {\bf 2012}, {\em D85},~086005,
  \href{http://xxx.lanl.gov/abs/1105.4011}{{\normalfont
  [arXiv:hep-th/1105.4011]}}.

\bibitem[Douglas \em{et~al.}(2011)Douglas, Mazzucato, and
  Razamat]{Douglas:2010rc}
Douglas, M.R.; Mazzucato, L.; Razamat, S.S.
\newblock {Holographic dual of free field theory}.
\newblock {\em Phys. Rev.} {\bf 2011}, {\em D83},~071701,
  \href{http://xxx.lanl.gov/abs/1011.4926}{{\normalfont
  [arXiv:hep-th/1011.4926]}}.

\bibitem[Giombi and Yin(2013)]{Giombi:2012ms}
Giombi, S.; Yin, X.
\newblock {The Higher Spin/Vector Model Duality}.
\newblock {\em J.Phys.} {\bf 2013}, {\em A46},~214003,
  \href{http://xxx.lanl.gov/abs/1208.4036}{{\normalfont
  [arXiv:hep-th/1208.4036]}}.

\bibitem[Sleight and Taronna(2016)]{Sleight:2016dba}
Sleight, C.; Taronna, M.
\newblock {Higher Spin Interactions from Conformal Field Theory: The Complete
  Cubic Couplings}.
\newblock {\em Phys. Rev. Lett.} {\bf 2016}, {\em 116},~181602,
  \href{http://xxx.lanl.gov/abs/1603.00022}{{\normalfont
  [arXiv:hep-th/1603.00022]}}.

\bibitem[Maldacena and Zhiboedov(2013{\natexlab{a}})]{Maldacena:2011jn}
Maldacena, J.; Zhiboedov, A.
\newblock {Constraining Conformal Field Theories with A Higher Spin Symmetry}.
\newblock {\em J. Phys.} {\bf 2013}, {\em A46},~214011,
  \href{http://xxx.lanl.gov/abs/1112.1016}{{\normalfont
  [arXiv:hep-th/1112.1016]}}.

\bibitem[Maldacena and Zhiboedov(2013{\natexlab{b}})]{Maldacena:2012sf}
Maldacena, J.; Zhiboedov, A.
\newblock {Constraining conformal field theories with a slightly broken higher
  spin symmetry}.
\newblock {\em Class. Quant. Grav.} {\bf 2013}, {\em 30},~104003,
  \href{http://xxx.lanl.gov/abs/1204.3882}{{\normalfont
  [arXiv:hep-th/1204.3882]}}.

\bibitem[Maldacena(1998)]{Maldacena:1997re}
Maldacena, J.M.
\newblock {The Large N limit of superconformal field theories and
  supergravity}.
\newblock {\em Adv.Theor.Math.Phys.} {\bf 1998}, {\em 2},~231--252,
  \href{http://xxx.lanl.gov/abs/hep-th/9711200}{{\normalfont
  [arXiv:hep-th/hep-th/9711200]}}.

\bibitem[Gubser \em{et~al.}(1998)Gubser, Klebanov, and Polyakov]{Gubser:1998bc}
Gubser, S.; Klebanov, I.R.; Polyakov, A.M.
\newblock {Gauge theory correlators from noncritical string theory}.
\newblock {\em Phys.Lett.} {\bf 1998}, {\em B428},~105--114,
  \href{http://xxx.lanl.gov/abs/hep-th/9802109}{{\normalfont
  [arXiv:hep-th/hep-th/9802109]}}.

\bibitem[Witten(1998)]{Witten:1998qj}
Witten, E.
\newblock {Anti-de Sitter space and holography}.
\newblock {\em Adv.Theor.Math.Phys.} {\bf 1998}, {\em 2},~253--291,
  \href{http://xxx.lanl.gov/abs/hep-th/9802150}{{\normalfont
  [arXiv:hep-th/hep-th/9802150]}}.

\bibitem[Aragone and Deser(1984)]{Aragone:1983sz}
Aragone, C.; Deser, S.
\newblock {Hypersymmetry in $D=3$ of Coupled Gravity Massless Spin 5/2 System}.
\newblock {\em Class. Quant. Grav.} {\bf 1984}, {\em 1},~L9.

\bibitem[Blencowe(1989)]{Blencowe:1988gj}
Blencowe, M.
\newblock {A Consistent Interacting Massless Higher Spin Field Theory in $D$ =
  (2+1)}.
\newblock {\em Class.Quant.Grav.} {\bf 1989}, {\em 6},~443.

\bibitem[Henneaux and Rey(2010)]{Henneaux:2010xg}
Henneaux, M.; Rey, S.J.
\newblock {Nonlinear $W_{\infty}$ as Asymptotic Symmetry of Three-Dimensional
  Higher Spin Anti-de Sitter Gravity}.
\newblock {\em JHEP} {\bf 2010}, {\em 1012},~007,
  \href{http://xxx.lanl.gov/abs/1008.4579}{{\normalfont
  [arXiv:hep-th/1008.4579]}}.

\bibitem[Campoleoni \em{et~al.}(2010)Campoleoni, Fredenhagen, Pfenninger, and
  Theisen]{Campoleoni:2010zq}
Campoleoni, A.; Fredenhagen, S.; Pfenninger, S.; Theisen, S.
\newblock {Asymptotic symmetries of three-dimensional gravity coupled to
  higher-spin fields}.
\newblock {\em JHEP} {\bf 2010}, {\em 1011},~007,
  \href{http://xxx.lanl.gov/abs/1008.4744}{{\normalfont
  [arXiv:hep-th/1008.4744]}}.

\bibitem[Gaberdiel and Hartman(2011)]{Gaberdiel:2011wb}
Gaberdiel, M.R.; Hartman, T.
\newblock {Symmetries of Holographic Minimal Models}.
\newblock {\em JHEP} {\bf 2011}, {\em 1105},~031,
  \href{http://xxx.lanl.gov/abs/1101.2910}{{\normalfont
  [arXiv:hep-th/1101.2910]}}.

\bibitem[Campoleoni \em{et~al.}(2011)Campoleoni, Fredenhagen, and
  Pfenninger]{Campoleoni:2011hg}
Campoleoni, A.; Fredenhagen, S.; Pfenninger, S.
\newblock {Asymptotic W-symmetries in three-dimensional higher-spin gauge
  theories}.
\newblock {\em JHEP} {\bf 2011}, {\em 1109},~113,
  \href{http://xxx.lanl.gov/abs/1107.0290}{{\normalfont
  [arXiv:hep-th/1107.0290]}}.

\bibitem[Henneaux \em{et~al.}(2013)Henneaux, Perez, Tempo, and
  Troncoso]{Henneaux:2013dra}
Henneaux, M.; Perez, A.; Tempo, D.; Troncoso, R.
\newblock {Chemical potentials in three-dimensional higher spin anti-de Sitter
  gravity}.
\newblock {\em JHEP} {\bf 2013}, {\em 1312},~048,
  \href{http://xxx.lanl.gov/abs/1309.4362}{{\normalfont
  [arXiv:hep-th/1309.4362]}}.

\bibitem[Castro \em{et~al.}(2011)Castro, Lepage-Jutier, and
  Maloney]{Castro:2010ce}
Castro, A.; Lepage-Jutier, A.; Maloney, A.
\newblock {Higher Spin Theories in AdS$_3$ and a Gravitational Exclusion
  Principle}.
\newblock {\em JHEP} {\bf 2011}, {\em 01},~142,
  \href{http://xxx.lanl.gov/abs/1012.0598}{{\normalfont
  [arXiv:hep-th/1012.0598]}}.

\bibitem[Ammon \em{et~al.}(2011)Ammon, Gutperle, Kraus, and
  Perlmutter]{Ammon:2011nk}
Ammon, M.; Gutperle, M.; Kraus, P.; Perlmutter, E.
\newblock {Spacetime Geometry in Higher Spin Gravity}.
\newblock {\em JHEP} {\bf 2011}, {\em 1110},~053,
  \href{http://xxx.lanl.gov/abs/1106.4788}{{\normalfont
  [arXiv:hep-th/1106.4788]}}.

\bibitem[Castro \em{et~al.}(2012)Castro, Hijano, Lepage-Jutier, and
  Maloney]{Castro:2011fm}
Castro, A.; Hijano, E.; Lepage-Jutier, A.; Maloney, A.
\newblock {Black Holes and Singularity Resolution in Higher Spin Gravity}.
\newblock {\em JHEP} {\bf 2012}, {\em 1201},~031,
  \href{http://xxx.lanl.gov/abs/1110.4117}{{\normalfont
  [arXiv:hep-th/1110.4117]}}.

\bibitem[Ammon \em{et~al.}(2012)Ammon, Kraus, and Perlmutter]{Ammon:2011ua}
Ammon, M.; Kraus, P.; Perlmutter, E.
\newblock {Scalar fields and three-point functions in D=3 higher spin gravity}.
\newblock {\em JHEP} {\bf 2012}, {\em 07},~113,
  \href{http://xxx.lanl.gov/abs/1111.3926}{{\normalfont
  [arXiv:hep-th/1111.3926]}}.

\bibitem[Henneaux \em{et~al.}(2012)Henneaux, Lucena~Gómez, Park, and
  Rey]{Henneaux:2012ny}
Henneaux, M.; Lucena~Gómez, G.; Park, J.; Rey, S.J.
\newblock {Super- W(infinity) Asymptotic Symmetry of Higher-Spin $AdS_3$
  Supergravity}.
\newblock {\em JHEP} {\bf 2012}, {\em 06},~037,
  \href{http://xxx.lanl.gov/abs/1203.5152}{{\normalfont
  [arXiv:hep-th/1203.5152]}}.

\bibitem[Campoleoni \em{et~al.}(2013)Campoleoni, Fredenhagen, Pfenninger, and
  Theisen]{Campoleoni:2012hp}
Campoleoni, A.; Fredenhagen, S.; Pfenninger, S.; Theisen, S.
\newblock {Towards metric-like higher-spin gauge theories in three dimensions}.
\newblock {\em J.Phys.} {\bf 2013}, {\em A46},~214017,
  \href{http://xxx.lanl.gov/abs/1208.1851}{{\normalfont
  [arXiv:hep-th/1208.1851]}}.

\bibitem[de~Boer and Jottar(2014)]{deBoer:2013gz}
de~Boer, J.; Jottar, J.I.
\newblock {Thermodynamics of higher spin black holes in $AdS_3$}.
\newblock {\em JHEP} {\bf 2014}, {\em 1401},~023,
  \href{http://xxx.lanl.gov/abs/1302.0816}{{\normalfont
  [arXiv:hep-th/1302.0816]}}.

\bibitem[Compère \em{et~al.}(2013)Compère, Jottar, and Song]{Compere:2013nba}
Compère, G.; Jottar, J.I.; Song, W.
\newblock {Observables and Microscopic Entropy of Higher Spin Black Holes}.
\newblock {\em JHEP} {\bf 2013}, {\em 1311},~054,
  \href{http://xxx.lanl.gov/abs/1308.2175}{{\normalfont
  [arXiv:hep-th/1308.2175]}}.

\bibitem[de~Boer and Jottar(2016)]{deBoer:2014fra}
de~Boer, J.; Jottar, J.I.
\newblock {Boundary conditions and partition functions in higher spin
  AdS$_{3}$/CFT$_{2}$}.
\newblock {\em JHEP} {\bf 2016}, {\em 04},~107,
  \href{http://xxx.lanl.gov/abs/1407.3844}{{\normalfont
  [arXiv:hep-th/1407.3844]}}.

\bibitem[Campoleoni and Fredenhagen(2013)]{Campoleoni:2013iha}
Campoleoni, A.; Fredenhagen, S.
\newblock {On the higher-spin charges of conical defects}.
\newblock {\em Phys. Lett.} {\bf 2013}, {\em B726},~387--389,
  \href{http://xxx.lanl.gov/abs/1307.3745}{{\normalfont
  [arXiv:hep-th/1307.3745]}}.

\bibitem[Campoleoni and Henneaux(2015)]{Campoleoni:2014tfa}
Campoleoni, A.; Henneaux, M.
\newblock {Asymptotic symmetries of three-dimensional higher-spin gravity: the
  metric approach}.
\newblock {\em JHEP} {\bf 2015}, {\em 03},~143,
  \href{http://xxx.lanl.gov/abs/1412.6774}{{\normalfont
  [arXiv:hep-th/1412.6774]}}.

\bibitem[Castro \em{et~al.}(2012)Castro, Gopakumar, Gutperle, and
  Raeymaekers]{Castro:2011iw}
Castro, A.; Gopakumar, R.; Gutperle, M.; Raeymaekers, J.
\newblock {Conical Defects in Higher Spin Theories}.
\newblock {\em JHEP} {\bf 2012}, {\em 1202},~096,
  \href{http://xxx.lanl.gov/abs/1111.3381}{{\normalfont
  [arXiv:hep-th/1111.3381]}}.

\bibitem[Castro and Llabrés(2015)]{Castro:2014mza}
Castro, A.; Llabrés, E.
\newblock {Unravelling Holographic Entanglement Entropy in Higher Spin
  Theories}.
\newblock {\em JHEP} {\bf 2015}, {\em 03},~124,
  \href{http://xxx.lanl.gov/abs/1410.2870}{{\normalfont
  [arXiv:hep-th/1410.2870]}}.

\bibitem[de~Boer \em{et~al.}(2015)de~Boer, Castro, Hijano, Jottar, and
  Kraus]{deBoer:2014sna}
de~Boer, J.; Castro, A.; Hijano, E.; Jottar, J.I.; Kraus, P.
\newblock {Higher spin entanglement and $ {\mathcal{W}}_{\mathrm{N}} $
  conformal blocks}.
\newblock {\em JHEP} {\bf 2015}, {\em 07},~168,
  \href{http://xxx.lanl.gov/abs/1412.7520}{{\normalfont
  [arXiv:hep-th/1412.7520]}}.

\bibitem[Bañados \em{et~al.}(2016)Bañados, Castro, Faraggi, and
  Jottar]{Banados:2015tft}
Bañados, M.; Castro, A.; Faraggi, A.; Jottar, J.I.
\newblock {Extremal Higher Spin Black Holes}.
\newblock {\em JHEP} {\bf 2016}, {\em 04},~077,
  \href{http://xxx.lanl.gov/abs/1512.00073}{{\normalfont
  [arXiv:hep-th/1512.00073]}}.

\bibitem[Gaberdiel and Gopakumar(2011)]{Gaberdiel:2010pz}
Gaberdiel, M.R.; Gopakumar, R.
\newblock {An $\text{AdS}_3$ Dual for Minimal Model CFTs}.
\newblock {\em Phys.Rev.} {\bf 2011}, {\em D83},~066007,
  \href{http://xxx.lanl.gov/abs/1011.2986}{{\normalfont
  [arXiv:hep-th/1011.2986]}}.

\bibitem[Gaberdiel and Gopakumar(2013)]{Gaberdiel:2012uj}
Gaberdiel, M.R.; Gopakumar, R.
\newblock {Minimal Model Holography}.
\newblock {\em J.Phys.} {\bf 2013}, {\em A46},~214002,
  \href{http://xxx.lanl.gov/abs/1207.6697}{{\normalfont
  [arXiv:hep-th/1207.6697]}}.

\bibitem[Candu \em{et~al.}(2013)Candu, Gaberdiel, Kelm, and
  Vollenweider]{Candu:2012ne}
Candu, C.; Gaberdiel, M.R.; Kelm, M.; Vollenweider, C.
\newblock {Even spin minimal model holography}.
\newblock {\em JHEP} {\bf 2013}, {\em 1301},~185,
  \href{http://xxx.lanl.gov/abs/1211.3113}{{\normalfont
  [arXiv:hep-th/1211.3113]}}.

\bibitem[Gutperle and Kraus(2011)]{Gutperle:2011kf}
Gutperle, M.; Kraus, P.
\newblock {Higher Spin Black Holes}.
\newblock {\em JHEP} {\bf 2011}, {\em 1105},~022,
  \href{http://xxx.lanl.gov/abs/1103.4304}{{\normalfont
  [arXiv:hep-th/1103.4304]}}.

\bibitem[Ammon \em{et~al.}(2013)Ammon, Gutperle, Kraus, and
  Perlmutter]{Ammon:2012wc}
Ammon, M.; Gutperle, M.; Kraus, P.; Perlmutter, E.
\newblock {Black holes in three dimensional higher spin gravity: A review}.
\newblock {\em J.Phys.} {\bf 2013}, {\em A46},~214001,
  \href{http://xxx.lanl.gov/abs/1208.5182}{{\normalfont
  [arXiv:hep-th/1208.5182]}}.

\bibitem[Bunster \em{et~al.}(2014)Bunster, Henneaux, Perez, Tempo, and
  Troncoso]{Bunster:2014mua}
Bunster, C.; Henneaux, M.; Perez, A.; Tempo, D.; Troncoso, R.
\newblock {Generalized Black Holes in Three-dimensional Spacetime}.
\newblock {\em JHEP} {\bf 2014}, {\em 1405},~031,
  \href{http://xxx.lanl.gov/abs/1404.3305}{{\normalfont
  [arXiv:hep-th/1404.3305]}}.

\bibitem[Ammon \em{et~al.}(2013)Ammon, Castro, and Iqbal]{Ammon:2013hba}
Ammon, M.; Castro, A.; Iqbal, N.
\newblock {Wilson Lines and Entanglement Entropy in Higher Spin Gravity}.
\newblock {\em JHEP} {\bf 2013}, {\em 10},~110,
  \href{http://xxx.lanl.gov/abs/1306.4338}{{\normalfont
  [arXiv:hep-th/1306.4338]}}.

\bibitem[de~Boer and Jottar(2014)]{deBoer:2013vca}
de~Boer, J.; Jottar, J.I.
\newblock {Entanglement Entropy and Higher Spin Holography in AdS$_3$}.
\newblock {\em JHEP} {\bf 2014}, {\em 04},~089,
  \href{http://xxx.lanl.gov/abs/1306.4347}{{\normalfont
  [arXiv:hep-th/1306.4347]}}.

\bibitem[Brown and Henneaux(1986)]{Brown:1986nw}
Brown, J.D.; Henneaux, M.
\newblock {Central Charges in the Canonical Realization of Asymptotic
  Symmetries: An Example from Three-Dimensional Gravity}.
\newblock {\em Commun.Math.Phys.} {\bf 1986}, {\em 104},~207--226.

\bibitem[Riegler(2012)]{Riegler:2012fa}
Riegler, M.
\newblock {Asymptotic Symmetry Algebras in Non-Anti-de-Sitter Higher-Spin Gauge
  Theories} {\bf 2012}.
\newblock  \href{http://xxx.lanl.gov/abs/1210.6500}{{\normalfont
  [arXiv:hep-th/1210.6500]}}.

\bibitem[Afshar \em{et~al.}(2012)Afshar, Gary, Grumiller, Rashkov, and
  Riegler]{Afshar:2012nk}
Afshar, H.; Gary, M.; Grumiller, D.; Rashkov, R.; Riegler, M.
\newblock {Non-AdS holography in 3-dimensional higher spin gravity - General
  recipe and example}.
\newblock {\em JHEP} {\bf 2012}, {\em 1211},~099,
  \href{http://xxx.lanl.gov/abs/1209.2860}{{\normalfont
  [arXiv:hep-th/1209.2860]}}.

\bibitem[Afshar \em{et~al.}(2013)Afshar, Gary, Grumiller, Rashkov, and
  Riegler]{Afshar:2012hc}
Afshar, H.; Gary, M.; Grumiller, D.; Rashkov, R.; Riegler, M.
\newblock {Semi-classical unitarity in 3-dimensional higher-spin gravity for
  non-principal embeddings}.
\newblock {\em Class. Quant. Grav.} {\bf 2013}, {\em 30},~104004,
  \href{http://xxx.lanl.gov/abs/1211.4454}{{\normalfont
  [arXiv:hep-th/1211.4454]}}.

\bibitem[Son(2008)]{Son:2008ye}
Son, D.
\newblock {Toward an AdS/cold atoms correspondence: A Geometric realization of
  the Schrodinger symmetry}.
\newblock {\em Phys.Rev.} {\bf 2008}, {\em D78},~046003,
  \href{http://xxx.lanl.gov/abs/0804.3972}{{\normalfont
  [arXiv:hep-th/0804.3972]}}.

\bibitem[Balasubramanian and McGreevy(2008)]{Balasubramanian:2008dm}
Balasubramanian, K.; McGreevy, J.
\newblock {Gravity duals for non-relativistic CFTs}.
\newblock {\em Phys.Rev.Lett.} {\bf 2008}, {\em 101},~061601,
  \href{http://xxx.lanl.gov/abs/0804.4053}{{\normalfont
  [arXiv:hep-th/0804.4053]}}.

\bibitem[Adams \em{et~al.}(2008)Adams, Balasubramanian, and
  McGreevy]{Adams:2008wt}
Adams, A.; Balasubramanian, K.; McGreevy, J.
\newblock {Hot Spacetimes for Cold Atoms}.
\newblock {\em JHEP} {\bf 2008}, {\em 0811},~059,
  \href{http://xxx.lanl.gov/abs/0807.1111}{{\normalfont
  [arXiv:hep-th/0807.1111]}}.

\bibitem[Breunhölder \em{et~al.}(2015)Breunhölder, Gary, Grumiller, and
  Prohazka]{Breunhoelder:2015waa}
Breunhölder, V.; Gary, M.; Grumiller, D.; Prohazka, S.
\newblock {Null warped AdS in higher spin gravity}.
\newblock {\em JHEP} {\bf 2015}, {\em 12},~021,
  \href{http://xxx.lanl.gov/abs/1509.08487}{{\normalfont
  [arXiv:hep-th/1509.08487]}}.

\bibitem[Lei and Ross(2015)]{Lei:2015ika}
Lei, Y.; Ross, S.F.
\newblock {Connection versus metric description for non-AdS solutions in
  higher-spin theories}.
\newblock {\em Class. Quant. Grav.} {\bf 2015}, {\em 32},~185005,
  \href{http://xxx.lanl.gov/abs/1504.07252}{{\normalfont
  [arXiv:hep-th/1504.07252]}}.

\bibitem[Kachru \em{et~al.}(2008)Kachru, Liu, and Mulligan]{Kachru:2008yh}
Kachru, S.; Liu, X.; Mulligan, M.
\newblock {Gravity Duals of Lifshitz-like Fixed Points}.
\newblock {\em Phys.Rev.} {\bf 2008}, {\em D78},~106005,
  \href{http://xxx.lanl.gov/abs/0808.1725}{{\normalfont
  [arXiv:hep-th/0808.1725]}}.

\bibitem[Gutperle \em{et~al.}(2014)Gutperle, Hijano, and
  Samani]{Gutperle:2013oxa}
Gutperle, M.; Hijano, E.; Samani, J.
\newblock {Lifshitz black holes in higher spin gravity}.
\newblock {\em JHEP} {\bf 2014}, {\em 1404},~020,
  \href{http://xxx.lanl.gov/abs/1310.0837}{{\normalfont
  [arXiv:hep-th/1310.0837]}}.

\bibitem[Gary \em{et~al.}(2014)Gary, Grumiller, Prohazka, and
  Rey]{Gary:2014mca}
Gary, M.; Grumiller, D.; Prohazka, S.; Rey, S.J.
\newblock {Lifshitz Holography with Isotropic Scale Invariance}.
\newblock {\em JHEP} {\bf 2014}, {\em 1408},~001,
  \href{http://xxx.lanl.gov/abs/1406.1468}{{\normalfont
  [arXiv:hep-th/1406.1468]}}.

\bibitem[Afshar \em{et~al.}(2013)Afshar, Bagchi, Fareghbal, Grumiller, and
  Rosseel]{Afshar:2013vka}
Afshar, H.; Bagchi, A.; Fareghbal, R.; Grumiller, D.; Rosseel, J.
\newblock {Spin-3 Gravity in Three-Dimensional Flat Space}.
\newblock {\em Phys.Rev.Lett.} {\bf 2013}, {\em 111},~121603,
  \href{http://xxx.lanl.gov/abs/1307.4768}{{\normalfont
  [arXiv:hep-th/1307.4768]}}.

\bibitem[Gonzalez \em{et~al.}(2013)Gonzalez, Matulich, Pino, and
  Troncoso]{Gonzalez:2013oaa}
Gonzalez, H.A.; Matulich, J.; Pino, M.; Troncoso, R.
\newblock {Asymptotically flat spacetimes in three-dimensional higher spin
  gravity}.
\newblock {\em JHEP} {\bf 2013}, {\em 1309},~016,
  \href{http://xxx.lanl.gov/abs/1307.5651}{{\normalfont
  [arXiv:hep-th/1307.5651]}}.

\bibitem[Grumiller \em{et~al.}(2014)Grumiller, Riegler, and
  Rosseel]{Grumiller:2014lna}
Grumiller, D.; Riegler, M.; Rosseel, J.
\newblock {Unitarity in three-dimensional flat space higher spin theories}.
\newblock {\em JHEP} {\bf 2014}, {\em 07},~015,
  \href{http://xxx.lanl.gov/abs/1403.5297}{{\normalfont
  [arXiv:hep-th/1403.5297]}}.

\bibitem[Gary \em{et~al.}(2015)Gary, Grumiller, Riegler, and
  Rosseel]{Gary:2014ppa}
Gary, M.; Grumiller, D.; Riegler, M.; Rosseel, J.
\newblock {Flat space (higher spin) gravity with chemical potentials}.
\newblock {\em JHEP} {\bf 2015}, {\em 01},~152,
  \href{http://xxx.lanl.gov/abs/1411.3728}{{\normalfont
  [arXiv:hep-th/1411.3728]}}.

\bibitem[Krishnan \em{et~al.}(2014)Krishnan, Raju, Roy, and
  Thakur]{Krishnan:2013zya}
Krishnan, C.; Raju, A.; Roy, S.; Thakur, S.
\newblock {Higher Spin Cosmology}.
\newblock {\em Phys. Rev.} {\bf 2014}, {\em D89},~045007,
  \href{http://xxx.lanl.gov/abs/1308.6741}{{\normalfont
  [arXiv:hep-th/1308.6741]}}.

\bibitem[Krishnan and Roy(2013)]{Krishnan:2013cra}
Krishnan, C.; Roy, S.
\newblock {Higher Spin Resolution of a Toy Big Bang}.
\newblock {\em Phys. Rev.} {\bf 2013}, {\em D88},~044049,
  \href{http://xxx.lanl.gov/abs/1305.1277}{{\normalfont
  [arXiv:hep-th/1305.1277]}}.

\bibitem[Basu(2015)]{Basu:2015exa}
Basu, R.
\newblock {Higher Spin de Sitter Quantum Gravity}.
\newblock {\em JHEP} {\bf 2015}, {\em 10},~151,
  \href{http://xxx.lanl.gov/abs/1507.04757}{{\normalfont
  [arXiv:hep-th/1507.04757]}}.

\bibitem[Achucarro and Townsend(1986)]{Achucarro:1987vz}
Achucarro, A.; Townsend, P.
\newblock {A Chern-Simons Action for Three-Dimensional anti-De Sitter
  Supergravity Theories}.
\newblock {\em Phys.Lett.} {\bf 1986}, {\em B180},~89.

\bibitem[Witten(1988)]{Witten:1988hc}
Witten, E.
\newblock {(2+1)-Dimensional Gravity as an Exactly Soluble System}.
\newblock {\em Nucl.Phys.} {\bf 1988}, {\em B311},~46.

\bibitem[Riegler(2016)]{Riegler:2016hah}
Riegler, M.
\newblock {How General Is Holography?}
\newblock PhD thesis, Vienna, Tech. U.,  2016,
  \href{http://xxx.lanl.gov/abs/1609.02733}{{\normalfont
  [arXiv:hep-th/1609.02733]}}.

\bibitem[Riegler and Zwikel(2017)]{Riegler:2017fqv}
Riegler, M.; Zwikel, C.
\newblock {Canonical Charges in Flatland} {\bf 2017}.
\newblock  \href{http://xxx.lanl.gov/abs/1709.09871}{{\normalfont
  [arXiv:hep-th/1709.09871]}}.

\bibitem[Prohazka(2017)]{Prohazka:2017pkc}
Prohazka, S.
\newblock {Chern-Simons Holography: Boundary Conditions, Contractions and
  Double Extensions for a Journey Beyond Anti-de Sitter}.
\newblock PhD thesis, Vienna, Tech. U.,  2017,
  \href{http://xxx.lanl.gov/abs/1710.11110}{{\normalfont
  [arXiv:hep-th/1710.11110]}}.

\bibitem[Afshar \em{et~al.}(2015)Afshar, Bagchi, Detournay, Grumiller,
  Prohazka, and Riegler]{Afshar:2014rwa}
Afshar, H.; Bagchi, A.; Detournay, S.; Grumiller, D.; Prohazka, S.; Riegler, M.
\newblock {Holographic Chern-Simons Theories}.
\newblock {\em Lect. Notes Phys.} {\bf 2015}, {\em 892},~311--329,
  \href{http://xxx.lanl.gov/abs/1404.1919}{{\normalfont
  [arXiv:hep-th/1404.1919]}}.

\bibitem[Gary \em{et~al.}(2012)Gary, Grumiller, and Rashkov]{Gary:2012ms}
Gary, M.; Grumiller, D.; Rashkov, R.
\newblock {Towards non-AdS holography in 3-dimensional higher spin gravity}.
\newblock {\em JHEP} {\bf 2012}, {\em 1203},~022,
  \href{http://xxx.lanl.gov/abs/1201.0013}{{\normalfont
  [arXiv:hep-th/1201.0013]}}.

\bibitem[Henneaux and Teitelboim(1992)]{Henneaux:1992ig}
Henneaux, M.; Teitelboim, C.
\newblock {\em {Quantization of gauge systems}}; Princeton University Press,
  Princeton, NJ,  1992.

\bibitem[Blagojevic(2010)]{Blagojevic:2002du}
Blagojevic, M.
\newblock {\em Gravitation and gauge symmetries}; CRC Press,  2010.

\bibitem[Regge and Teitelboim(1974)]{Regge:1974zd}
Regge, T.; Teitelboim, C.
\newblock {Role of Surface Integrals in the Hamiltonian Formulation of General
  Relativity}.
\newblock {\em Annals Phys.} {\bf 1974}, {\em 88},~286.

\bibitem[Barnich and Troessaert(2011)]{Barnich:2011mi}
Barnich, G.; Troessaert, C.
\newblock {BMS charge algebra}.
\newblock {\em JHEP} {\bf 2011}, {\em 12},~105,
  \href{http://xxx.lanl.gov/abs/1106.0213}{{\normalfont
  [arXiv:hep-th/1106.0213]}}.

\bibitem[Gaberdiel and Gopakumar(2012)]{Gaberdiel:2012ku}
Gaberdiel, M.R.; Gopakumar, R.
\newblock {Triality in Minimal Model Holography}.
\newblock {\em JHEP} {\bf 2012}, {\em 1207},~127,
  \href{http://xxx.lanl.gov/abs/1205.2472}{{\normalfont
  [arXiv:hep-th/1205.2472]}}.

\bibitem[Polyakov(1990)]{Polyakov:1989dm}
Polyakov, A.M.
\newblock {Gauge Transformations and Diffeomorphisms}.
\newblock {\em Int. J. Mod. Phys.} {\bf 1990}, {\em A5},~833.

\bibitem[Bershadsky(1991)]{Bershadsky:1990bg}
Bershadsky, M.
\newblock {Conformal field theories via Hamiltonian reduction}.
\newblock {\em Commun. Math. Phys.} {\bf 1991}, {\em 139},~71--82.

\bibitem[Afshar \em{et~al.}(2014)Afshar, Creutzig, Grumiller, Hikida, and
  Ronne]{Afshar:2014cma}
Afshar, H.; Creutzig, T.; Grumiller, D.; Hikida, Y.; Ronne, P.B.
\newblock {Unitary W-algebras and three-dimensional higher spin gravities with
  spin one symmetry}.
\newblock {\em JHEP} {\bf 2014}, {\em 06},~063,
  \href{http://xxx.lanl.gov/abs/1404.0010}{{\normalfont
  [arXiv:hep-th/1404.0010]}}.

\bibitem[Feigin and Semikhatov(2004)]{Feigin:2004wb}
Feigin, B.L.; Semikhatov, A.M.
\newblock {W(2)(n) algebras}.
\newblock {\em Nucl. Phys.} {\bf 2004}, {\em B698},~409--449,
  \href{http://xxx.lanl.gov/abs/math/0401164}{{\normalfont
  [arXiv:math-qa/math/0401164]}}.

\bibitem[Castro \em{et~al.}(2012)Castro, Hijano, and
  Lepage-Jutier]{Castro:2012bc}
Castro, A.; Hijano, E.; Lepage-Jutier, A.
\newblock {Unitarity Bounds in AdS$_3$ Higher Spin Gravity}.
\newblock {\em JHEP} {\bf 2012}, {\em 06},~001,
  \href{http://xxx.lanl.gov/abs/1202.4467}{{\normalfont
  [arXiv:hep-th/1202.4467]}}.

\bibitem[Gutperle and Li(2015)]{Gutperle:2014aja}
Gutperle, M.; Li, Y.
\newblock {Higher Spin Lifshitz Theory and Integrable Systems}.
\newblock {\em Phys. Rev.} {\bf 2015}, {\em D91},~046012,
  \href{http://xxx.lanl.gov/abs/1412.7085}{{\normalfont
  [arXiv:hep-th/1412.7085]}}.

\bibitem[Beccaria \em{et~al.}(2015)Beccaria, Gutperle, Li, and
  Macorini]{Beccaria:2015iwa}
Beccaria, M.; Gutperle, M.; Li, Y.; Macorini, G.
\newblock {Higher spin Lifshitz theories and the Korteweg-de Vries hierarchy}.
\newblock {\em Phys. Rev.} {\bf 2015}, {\em D92},~085005,
  \href{http://xxx.lanl.gov/abs/1504.06555}{{\normalfont
  [arXiv:hep-th/1504.06555]}}.

\bibitem[Gutperle and Li(2017)]{Gutperle:2017ewo}
Gutperle, M.; Li, Y.
\newblock {Higher Spin Chern-Simons Theory and the Super Boussinesq hierarchy}
  {\bf 2017}.
\newblock  \href{http://xxx.lanl.gov/abs/1709.02345}{{\normalfont
  [arXiv:hep-th/1709.02345]}}.

\bibitem[Lei and Peng(2016)]{Lei:2015gza}
Lei, Y.; Peng, C.
\newblock {Higher spin holography with Galilean symmetry in general
  dimensions}.
\newblock {\em Class. Quant. Grav.} {\bf 2016}, {\em 33},~135008,
  \href{http://xxx.lanl.gov/abs/1507.08293}{{\normalfont
  [arXiv:hep-th/1507.08293]}}.

\bibitem[Lei(2016)]{Lei:2016pfu}
Lei, Y.
\newblock {Singularities in holographic non-relativistic spacetimes}.
\newblock PhD thesis, Durham U.,  2016.

\bibitem[Grumiller and Riegler(2016)]{Grumiller:2016pqb}
Grumiller, D.; Riegler, M.
\newblock {Most general AdS$_{3}$ boundary conditions}.
\newblock {\em JHEP} {\bf 2016}, {\em 10},~023,
  \href{http://xxx.lanl.gov/abs/1608.01308}{{\normalfont
  [arXiv:hep-th/1608.01308]}}.

\bibitem[Krishnan and Raju(2017)]{Krishnan:2017xct}
Krishnan, C.; Raju, A.
\newblock {Chiral Higher Spin Gravity}.
\newblock {\em Phys. Rev.} {\bf 2017}, {\em D95},~126004,
  \href{http://xxx.lanl.gov/abs/1703.01769}{{\normalfont
  [arXiv:hep-th/1703.01769]}}.

\bibitem[Polchinski(1999)]{Polchinski:1999ry}
Polchinski, J.
\newblock {S matrices from AdS space-time} {\bf 1999}.
\newblock  \href{http://xxx.lanl.gov/abs/hep-th/9901076}{{\normalfont
  [arXiv:hep-th/hep-th/9901076]}}.

\bibitem[Susskind(1998)]{Susskind:1998vk}
Susskind, L.
\newblock {Holography in the flat space limit} {\bf 1998}.
\newblock pp. 98--112,
  \href{http://xxx.lanl.gov/abs/hep-th/9901079}{{\normalfont
  [arXiv:hep-th/hep-th/9901079]}}.
\newblock [AIP Conf. Proc.493,98(1999)].

\bibitem[Giddings(2000)]{Giddings:1999jq}
Giddings, S.B.
\newblock {Flat space scattering and bulk locality in the AdS / CFT
  correspondence}.
\newblock {\em Phys. Rev.} {\bf 2000}, {\em D61},~106008,
  \href{http://xxx.lanl.gov/abs/hep-th/9907129}{{\normalfont
  [arXiv:hep-th/hep-th/9907129]}}.

\bibitem[Barnich and Compere(2007)]{Barnich:2006av}
Barnich, G.; Compere, G.
\newblock {Classical central extension for asymptotic symmetries at null
  infinity in three spacetime dimensions}.
\newblock {\em Class. Quant. Grav.} {\bf 2007}, {\em 24},~F15--F23,
  \href{http://xxx.lanl.gov/abs/gr-qc/0610130}{{\normalfont
  [arXiv:gr-qc/gr-qc/0610130]}}.

\bibitem[Strominger(2014)]{Strominger:2013jfa}
Strominger, A.
\newblock {On BMS Invariance of Gravitational Scattering}.
\newblock {\em JHEP} {\bf 2014}, {\em 07},~152,
  \href{http://xxx.lanl.gov/abs/1312.2229}{{\normalfont
  [arXiv:hep-th/1312.2229]}}.

\bibitem[Kapec \em{et~al.}(2015)Kapec, Lysov, Pasterski, and
  Strominger]{Kapec:2015vwa}
Kapec, D.; Lysov, V.; Pasterski, S.; Strominger, A.
\newblock {Higher-Dimensional Supertranslations and Weinberg's Soft Graviton
  Theorem} {\bf 2015}.
\newblock  \href{http://xxx.lanl.gov/abs/1502.07644}{{\normalfont
  [arXiv:gr-qc/1502.07644]}}.

\bibitem[Prohazka \em{et~al.}(2017)Prohazka, Salzer, and
  Schöller]{Prohazka:2017equ}
Prohazka, S.; Salzer, J.; Schöller, F.
\newblock {Linking Past and Future Null Infinity in Three Dimensions}.
\newblock {\em Phys. Rev.} {\bf 2017}, {\em D95},~086011,
  \href{http://xxx.lanl.gov/abs/1701.06573}{{\normalfont
  [arXiv:hep-th/1701.06573]}}.

\bibitem[Ashtekar \em{et~al.}(1997)Ashtekar, Bicak, and
  Schmidt]{Ashtekar:1996cd}
Ashtekar, A.; Bicak, J.; Schmidt, B.G.
\newblock {Asymptotic structure of symmetry reduced general relativity}.
\newblock {\em Phys. Rev.} {\bf 1997}, {\em D55},~669--686,
  \href{http://xxx.lanl.gov/abs/gr-qc/9608042}{{\normalfont
  [arXiv:gr-qc/gr-qc/9608042]}}.

\bibitem[Bondi \em{et~al.}(1962)Bondi, van~der Burg, and Metzner]{Bondi:1962px}
Bondi, H.; van~der Burg, M.G.J.; Metzner, A.W.K.
\newblock {Gravitational waves in general relativity. 7. Waves from
  axisymmetric isolated systems}.
\newblock {\em Proc. Roy. Soc. Lond.} {\bf 1962}, {\em A269},~21--52.

\bibitem[Sachs(1962)]{Sachs:1962zza}
Sachs, R.
\newblock {Asymptotic symmetries in gravitational theory}.
\newblock {\em Phys. Rev.} {\bf 1962}, {\em 128},~2851--2864.

\bibitem[Barnich \em{et~al.}(2014)Barnich, Donnay, Matulich, and
  Troncoso]{Barnich:2014cwa}
Barnich, G.; Donnay, L.; Matulich, J.; Troncoso, R.
\newblock {Asymptotic symmetries and dynamics of three-dimensional flat
  supergravity}.
\newblock {\em JHEP} {\bf 2014}, {\em 08},~071,
  \href{http://xxx.lanl.gov/abs/1407.4275}{{\normalfont
  [arXiv:hep-th/1407.4275]}}.

\bibitem[Barnich \em{et~al.}(2015)Barnich, Lambert, and Mao]{Barnich:2015jua}
Barnich, G.; Lambert, P.H.; Mao, P.J.
\newblock {Three-dimensional asymptotically flat Einstein–Maxwell theory}.
\newblock {\em Class. Quant. Grav.} {\bf 2015}, {\em 32},~245001,
  \href{http://xxx.lanl.gov/abs/1503.00856}{{\normalfont
  [arXiv:gr-qc/1503.00856]}}.

\bibitem[Detournay and Riegler(2017)]{Detournay:2016sfv}
Detournay, S.; Riegler, M.
\newblock {Enhanced Asymptotic Symmetry Algebra of 2+1 Dimensional Flat Space}.
\newblock {\em Phys. Rev.} {\bf 2017}, {\em D95},~046008,
  \href{http://xxx.lanl.gov/abs/1612.00278}{{\normalfont
  [arXiv:hep-th/1612.00278]}}.

\bibitem[Setare and Adami(2017)]{Setare:2017mry}
Setare, M.R.; Adami, H.
\newblock {Enhanced asymptotic $BMS_3$ algebra of the flat spacetime solutions
  of generalized minimal massive gravity} {\bf 2017}.
\newblock  \href{http://xxx.lanl.gov/abs/1703.00936}{{\normalfont
  [arXiv:hep-th/1703.00936]}}.

\bibitem[Basu \em{et~al.}(2017)Basu, Detournay, and Riegler]{Basu:2017aqn}
Basu, R.; Detournay, S.; Riegler, M.
\newblock {Spectral Flow in 3D Flat Spacetimes} {\bf 2017}.
\newblock  \href{http://xxx.lanl.gov/abs/1706.07438}{{\normalfont
  [arXiv:hep-th/1706.07438]}}.

\bibitem[Fuentealba \em{et~al.}(2017)Fuentealba, Matulich, and
  Troncoso]{Fuentealba:2017fck}
Fuentealba, O.; Matulich, J.; Troncoso, R.
\newblock {Asymptotic structure of $\mathcal{N}=2$ supergravity in 3D: extended
  super-BMS$_3$ and nonlinear energy bounds}.
\newblock {\em JHEP} {\bf 2017}, {\em 09},~030,
  \href{http://xxx.lanl.gov/abs/1706.07542}{{\normalfont
  [arXiv:hep-th/1706.07542]}}.

\bibitem[Afshar \em{et~al.}(2016{\natexlab{a}})Afshar, Detournay, Grumiller,
  and Oblak]{Afshar:2015wjm}
Afshar, H.; Detournay, S.; Grumiller, D.; Oblak, B.
\newblock {Near-Horizon Geometry and Warped Conformal Symmetry}.
\newblock {\em JHEP} {\bf 2016}, {\em 03},~187,
  \href{http://xxx.lanl.gov/abs/1512.08233}{{\normalfont
  [arXiv:hep-th/1512.08233]}}.

\bibitem[Afshar \em{et~al.}(2016{\natexlab{b}})Afshar, Grumiller, Merbis,
  Perez, Tempo, and Troncoso]{Afshar:2016kjj}
Afshar, H.; Grumiller, D.; Merbis, W.; Perez, A.; Tempo, D.; Troncoso, R.
\newblock {Soft hairy horizons in three spacetime dimensions} {\bf 2016}.
\newblock  \href{http://xxx.lanl.gov/abs/1611.09783}{{\normalfont
  [arXiv:hep-th/1611.09783]}}.

\bibitem[Grumiller \em{et~al.}(2017)Grumiller, Merbis, and
  Riegler]{Grumiller:2017sjh}
Grumiller, D.; Merbis, W.; Riegler, M.
\newblock {Most general flat space boundary conditions in three-dimensional
  Einstein gravity}.
\newblock {\em Class. Quant. Grav.} {\bf 2017}, {\em 34},~184001,
  \href{http://xxx.lanl.gov/abs/1704.07419}{{\normalfont
  [arXiv:hep-th/1704.07419]}}.

\bibitem[Bagchi \em{et~al.}(2013)Bagchi, Detournay, Fareghbal, and
  Simón]{Bagchi:2012xr}
Bagchi, A.; Detournay, S.; Fareghbal, R.; Simón, J.
\newblock {Holography of 3D Flat Cosmological Horizons}.
\newblock {\em Phys. Rev. Lett.} {\bf 2013}, {\em 110},~141302,
  \href{http://xxx.lanl.gov/abs/1208.4372}{{\normalfont
  [arXiv:hep-th/1208.4372]}}.

\bibitem[Barnich(2012)]{Barnich:2012xq}
Barnich, G.
\newblock {Entropy of three-dimensional asymptotically flat cosmological
  solutions}.
\newblock {\em JHEP} {\bf 2012}, {\em 10},~095,
  \href{http://xxx.lanl.gov/abs/1208.4371}{{\normalfont
  [arXiv:hep-th/1208.4371]}}.

\bibitem[Barnich \em{et~al.}(2012)Barnich, Gomberoff, and
  Gonzalez]{Barnich:2012aw}
Barnich, G.; Gomberoff, A.; Gonzalez, H.A.
\newblock {The Flat limit of three dimensional asymptotically anti-de Sitter
  spacetimes}.
\newblock {\em Phys. Rev.} {\bf 2012}, {\em D86},~024020,
  \href{http://xxx.lanl.gov/abs/1204.3288}{{\normalfont
  [arXiv:gr-qc/1204.3288]}}.

\bibitem[Bagchi and Fareghbal(2012)]{Bagchi:2012cy}
Bagchi, A.; Fareghbal, R.
\newblock {BMS/GCA Redux: Towards Flatspace Holography from Non-Relativistic
  Symmetries}.
\newblock {\em JHEP} {\bf 2012}, {\em 10},~092,
  \href{http://xxx.lanl.gov/abs/1203.5795}{{\normalfont
  [arXiv:hep-th/1203.5795]}}.

\bibitem[Bagchi \em{et~al.}(2013)Bagchi, Detournay, Grumiller, and
  Simon]{Bagchi:2013lma}
Bagchi, A.; Detournay, S.; Grumiller, D.; Simon, J.
\newblock {Cosmic evolution from phase transition of 3-dimensional flat space}.
\newblock {\em Phys.Rev.Lett.} {\bf 2013}, {\em 111},~181301,
  \href{http://xxx.lanl.gov/abs/1305.2919}{{\normalfont
  [arXiv:hep-th/1305.2919]}}.

\bibitem[Fareghbal and Naseh(2014)]{Fareghbal:2013ifa}
Fareghbal, R.; Naseh, A.
\newblock {Flat-Space Energy-Momentum Tensor from BMS/GCA Correspondence}.
\newblock {\em JHEP} {\bf 2014}, {\em 03},~005,
  \href{http://xxx.lanl.gov/abs/1312.2109}{{\normalfont
  [arXiv:hep-th/1312.2109]}}.

\bibitem[Krishnan \em{et~al.}(2014)Krishnan, Raju, and Roy]{Krishnan:2013wta}
Krishnan, C.; Raju, A.; Roy, S.
\newblock {A Grassmann path from $AdS_3$ to flat space}.
\newblock {\em JHEP} {\bf 2014}, {\em 03},~036,
  \href{http://xxx.lanl.gov/abs/1312.2941}{{\normalfont
  [arXiv:hep-th/1312.2941]}}.

\bibitem[Bagchi and Basu(2014)]{Bagchi:2013qva}
Bagchi, A.; Basu, R.
\newblock {3D Flat Holography: Entropy and Logarithmic Corrections}.
\newblock {\em JHEP} {\bf 2014}, {\em 03},~020,
  \href{http://xxx.lanl.gov/abs/1312.5748}{{\normalfont
  [arXiv:hep-th/1312.5748]}}.

\bibitem[Detournay \em{et~al.}(2014)Detournay, Grumiller, Schöller, and
  Simón]{Detournay:2014fva}
Detournay, S.; Grumiller, D.; Schöller, F.; Simón, J.
\newblock {Variational principle and one-point functions in three-dimensional
  flat space Einstein gravity}.
\newblock {\em Phys. Rev.} {\bf 2014}, {\em D89},~084061,
  \href{http://xxx.lanl.gov/abs/1402.3687}{{\normalfont
  [arXiv:hep-th/1402.3687]}}.

\bibitem[Barnich and Oblak(2014)]{Barnich:2014kra}
Barnich, G.; Oblak, B.
\newblock {Notes on the BMS group in three dimensions: I. Induced
  representations}.
\newblock {\em JHEP} {\bf 2014}, {\em 06},~129,
  \href{http://xxx.lanl.gov/abs/1403.5803}{{\normalfont
  [arXiv:hep-th/1403.5803]}}.

\bibitem[Bagchi \em{et~al.}(2015)Bagchi, Basu, Grumiller, and
  Riegler]{Bagchi:2014iea}
Bagchi, A.; Basu, R.; Grumiller, D.; Riegler, M.
\newblock {Entanglement entropy in Galilean conformal field theories and flat
  holography}.
\newblock {\em Phys. Rev. Lett.} {\bf 2015}, {\em 114},~111602,
  \href{http://xxx.lanl.gov/abs/1410.4089}{{\normalfont
  [arXiv:hep-th/1410.4089]}}.

\bibitem[Fareghbal and Naseh(2015)]{Fareghbal:2014qga}
Fareghbal, R.; Naseh, A.
\newblock {Aspects of Flat/CCFT Correspondence}.
\newblock {\em Class. Quant. Grav.} {\bf 2015}, {\em 32},~135013,
  \href{http://xxx.lanl.gov/abs/1408.6932}{{\normalfont
  [arXiv:hep-th/1408.6932]}}.

\bibitem[Fareghbal and Hosseini(2015)]{Fareghbal:2014kfa}
Fareghbal, R.; Hosseini, S.M.
\newblock {Holography of 3D Asymptotically Flat Black Holes}.
\newblock {\em Phys. Rev.} {\bf 2015}, {\em D91},~084025,
  \href{http://xxx.lanl.gov/abs/1412.2569}{{\normalfont
  [arXiv:hep-th/1412.2569]}}.

\bibitem[Barnich \em{et~al.}(2015)Barnich, Gonzalez, Maloney, and
  Oblak]{Barnich:2015mui}
Barnich, G.; Gonzalez, H.A.; Maloney, A.; Oblak, B.
\newblock {One-loop partition function of three-dimensional flat gravity}.
\newblock {\em JHEP} {\bf 2015}, {\em 04},~178,
  \href{http://xxx.lanl.gov/abs/1502.06185}{{\normalfont
  [arXiv:hep-th/1502.06185]}}.

\bibitem[Barnich and Oblak(2015)]{Barnich:2015uva}
Barnich, G.; Oblak, B.
\newblock {Notes on the BMS group in three dimensions: II. Coadjoint
  representation}.
\newblock {\em JHEP} {\bf 2015}, {\em 03},~033,
  \href{http://xxx.lanl.gov/abs/1502.00010}{{\normalfont
  [arXiv:hep-th/1502.00010]}}.

\bibitem[Bagchi \em{et~al.}(2016)Bagchi, Grumiller, and Merbis]{Bagchi:2015wna}
Bagchi, A.; Grumiller, D.; Merbis, W.
\newblock {Stress tensor correlators in three-dimensional gravity}.
\newblock {\em Phys. Rev.} {\bf 2016}, {\em D93},~061502,
  \href{http://xxx.lanl.gov/abs/1507.05620}{{\normalfont
  [arXiv:hep-th/1507.05620]}}.

\bibitem[Asadi \em{et~al.}(2016)Asadi, Baghchesaraei, and
  Fareghbal]{Asadi:2016plj}
Asadi, M.; Baghchesaraei, O.; Fareghbal, R.
\newblock {Stress Tensor Correlators of CCFT$_2$ using Flat-Space Holography}
  {\bf 2016}.
\newblock  \href{http://xxx.lanl.gov/abs/1701.00063}{{\normalfont
  [arXiv:hep-th/1701.00063]}}.

\bibitem[Barnich \em{et~al.}(2017)Barnich, Gonzalez, and
  Salgado-Rebolledo]{Barnich:2017jgw}
Barnich, G.; Gonzalez, H.A.; Salgado-Rebolledo, P.
\newblock {Geometric actions for three-dimensional gravity} {\bf 2017}.
\newblock  \href{http://xxx.lanl.gov/abs/1707.08887}{{\normalfont
  [arXiv:hep-th/1707.08887]}}.

\bibitem[Fareghbal and Karimi(2017)]{Fareghbal:2017ujy}
Fareghbal, R.; Karimi, P.
\newblock {Logarithmic Correction to BMSFT Entanglement Entropy} {\bf 2017}.
\newblock  \href{http://xxx.lanl.gov/abs/1709.01804}{{\normalfont
  [arXiv:hep-th/1709.01804]}}.

\bibitem[Coleman and Mandula(1967)]{Coleman:1967ad}
Coleman, S.R.; Mandula, J.
\newblock {All Possible Symmetries of the S Matrix}.
\newblock {\em Phys. Rev.} {\bf 1967}, {\em 159},~1251--1256.

\bibitem[Pelc and Horwitz(1997)]{Pelc:1996vg}
Pelc, O.; Horwitz, L.P.
\newblock {Generalization of the Coleman-Mandula theorem to higher dimension}.
\newblock {\em J. Math. Phys.} {\bf 1997}, {\em 38},~139--172,
  \href{http://xxx.lanl.gov/abs/hep-th/9605147}{{\normalfont
  [arXiv:hep-th/hep-th/9605147]}}.

\bibitem[Aragone and Deser(1979)]{Aragone:1979hx}
Aragone, C.; Deser, S.
\newblock {Consistency Problems of Hypergravity}.
\newblock {\em Phys.Lett.} {\bf 1979}, {\em B86},~161.

\bibitem[Weinberg and Witten(1980)]{Weinberg:1980kq}
Weinberg, S.; Witten, E.
\newblock {Limits on Massless Particles}.
\newblock {\em Phys. Lett.} {\bf 1980}, {\em 96B},~59--62.

\bibitem[Bekaert \em{et~al.}(2012)Bekaert, Boulanger, and
  Sundell]{Bekaert:2010hw}
Bekaert, X.; Boulanger, N.; Sundell, P.
\newblock {How higher-spin gravity surpasses the spin two barrier: no-go
  theorems versus yes-go examples}.
\newblock {\em Rev.Mod.Phys.} {\bf 2012}, {\em 84},~987--1009,
  \href{http://xxx.lanl.gov/abs/1007.0435}{{\normalfont
  [arXiv:hep-th/1007.0435]}}.

\bibitem[Sleight and Taronna(2017)]{Sleight:2016xqq}
Sleight, C.; Taronna, M.
\newblock {Higher-Spin Algebras, Holography and Flat Space}.
\newblock {\em JHEP} {\bf 2017}, {\em 02},~095,
  \href{http://xxx.lanl.gov/abs/1609.00991}{{\normalfont
  [arXiv:hep-th/1609.00991]}}.

\bibitem[Ponomarev and Tseytlin(2016)]{Ponomarev:2016jqk}
Ponomarev, D.; Tseytlin, A.A.
\newblock {On quantum corrections in higher-spin theory in flat space}.
\newblock {\em JHEP} {\bf 2016}, {\em 05},~184,
  \href{http://xxx.lanl.gov/abs/1603.06273}{{\normalfont
  [arXiv:hep-th/1603.06273]}}.

\bibitem[Ponomarev and Skvortsov(2017)]{Ponomarev:2016lrm}
Ponomarev, D.; Skvortsov, E.D.
\newblock {Light-Front Higher-Spin Theories in Flat Space}.
\newblock {\em J. Phys.} {\bf 2017}, {\em A50},~095401,
  \href{http://xxx.lanl.gov/abs/1609.04655}{{\normalfont
  [arXiv:hep-th/1609.04655]}}.

\bibitem[Campoleoni \em{et~al.}(2017)Campoleoni, Francia, and
  Heissenberg]{Campoleoni:2017mbt}
Campoleoni, A.; Francia, D.; Heissenberg, C.
\newblock {On higher-spin supertranslations and superrotations}.
\newblock {\em JHEP} {\bf 2017}, {\em 05},~120,
  \href{http://xxx.lanl.gov/abs/1703.01351}{{\normalfont
  [arXiv:hep-th/1703.01351]}}.

\bibitem[Inonu and Wigner(1953)]{Inonu:1953sp}
Inonu, E.; Wigner, E.P.
\newblock {On the Contraction of groups and their represenations}.
\newblock {\em Proc. Nat. Acad. Sci.} {\bf 1953}, {\em 39},~510--524.

\bibitem[Bagchi and Gopakumar(2009)]{Bagchi:2009my}
Bagchi, A.; Gopakumar, R.
\newblock {Galilean Conformal Algebras and AdS/CFT}.
\newblock {\em JHEP} {\bf 2009}, {\em 07},~037,
  \href{http://xxx.lanl.gov/abs/0902.1385}{{\normalfont
  [arXiv:hep-th/0902.1385]}}.

\bibitem[Bagchi and Mandal(2009)]{Bagchi:2009ca}
Bagchi, A.; Mandal, I.
\newblock {On Representations and Correlation Functions of Galilean Conformal
  Algebras}.
\newblock {\em Phys. Lett.} {\bf 2009}, {\em B675},~393--397,
  \href{http://xxx.lanl.gov/abs/0903.4524}{{\normalfont
  [arXiv:hep-th/0903.4524]}}.

\bibitem[Bagchi \em{et~al.}(2010)Bagchi, Gopakumar, Mandal, and
  Miwa]{Bagchi:2009pe}
Bagchi, A.; Gopakumar, R.; Mandal, I.; Miwa, A.
\newblock {GCA in 2d}.
\newblock {\em JHEP} {\bf 2010}, {\em 08},~004,
  \href{http://xxx.lanl.gov/abs/0912.1090}{{\normalfont
  [arXiv:hep-th/0912.1090]}}.

\bibitem[Bagchi(2010)]{Bagchi:2010eg}
Bagchi, A.
\newblock {Correspondence between Asymptotically Flat Spacetimes and
  Nonrelativistic Conformal Field Theories}.
\newblock {\em Phys. Rev. Lett.} {\bf 2010}, {\em 105},~171601,
  \href{http://xxx.lanl.gov/abs/1006.3354}{{\normalfont
  [arXiv:hep-th/1006.3354]}}.

\bibitem[Campoleoni \em{et~al.}(2016)Campoleoni, Gonzalez, Oblak, and
  Riegler]{Campoleoni:2016vsh}
Campoleoni, A.; Gonzalez, H.A.; Oblak, B.; Riegler, M.
\newblock {BMS Modules in Three Dimensions}.
\newblock {\em Int. J. Mod. Phys.} {\bf 2016}, {\em A31},~1650068,
  \href{http://xxx.lanl.gov/abs/1603.03812}{{\normalfont
  [arXiv:hep-th/1603.03812]}}.

\bibitem[Riegler(2015)]{Riegler:2014bia}
Riegler, M.
\newblock {Flat space limit of higher-spin Cardy formula}.
\newblock {\em Phys. Rev.} {\bf 2015}, {\em D91},~024044,
  \href{http://xxx.lanl.gov/abs/1408.6931}{{\normalfont
  [arXiv:hep-th/1408.6931]}}.

\bibitem[Zamolodchikov(1985)]{Zamolodchikov:1985wn}
Zamolodchikov, A.
\newblock {Infinite Additional Symmetries in Two-Dimensional Conformal Quantum
  Field Theory}.
\newblock {\em Theor.Math.Phys.} {\bf 1985}, {\em 65},~1205--1213.

\bibitem[Cornalba and Costa(2002)]{Cornalba:2002fi}
Cornalba, L.; Costa, M.S.
\newblock {A New cosmological scenario in string theory}.
\newblock {\em Phys. Rev.} {\bf 2002}, {\em D66},~066001,
  \href{http://xxx.lanl.gov/abs/hep-th/0203031}{{\normalfont
  [arXiv:hep-th/hep-th/0203031]}}.

\bibitem[Cornalba and Costa(2004)]{Cornalba:2003kd}
Cornalba, L.; Costa, M.S.
\newblock {Time dependent orbifolds and string cosmology}.
\newblock {\em Fortsch. Phys.} {\bf 2004}, {\em 52},~145--199,
  \href{http://xxx.lanl.gov/abs/hep-th/0310099}{{\normalfont
  [arXiv:hep-th/hep-th/0310099]}}.

\bibitem[Matulich \em{et~al.}(2015)Matulich, Perez, Tempo, and
  Troncoso]{Matulich:2014hea}
Matulich, J.; Perez, A.; Tempo, D.; Troncoso, R.
\newblock {Higher spin extension of cosmological spacetimes in 3D:
  asymptotically flat behaviour with chemical potentials and thermodynamics}.
\newblock {\em JHEP} {\bf 2015}, {\em 05},~025,
  \href{http://xxx.lanl.gov/abs/1412.1464}{{\normalfont
  [arXiv:hep-th/1412.1464]}}.

\bibitem[Basu and Riegler(2016)]{Basu:2015evh}
Basu, R.; Riegler, M.
\newblock {Wilson Lines and Holographic Entanglement Entropy in Galilean
  Conformal Field Theories}.
\newblock {\em Phys. Rev.} {\bf 2016}, {\em D93},~045003,
  \href{http://xxx.lanl.gov/abs/1511.08662}{{\normalfont
  [arXiv:hep-th/1511.08662]}}.

\bibitem[David \em{et~al.}(2012)David, Ferlaino, and Kumar]{David:2012iu}
David, J.R.; Ferlaino, M.; Kumar, S.P.
\newblock {Thermodynamics of higher spin black holes in 3D}.
\newblock {\em JHEP} {\bf 2012}, {\em 1211},~135,
  \href{http://xxx.lanl.gov/abs/1210.0284}{{\normalfont
  [arXiv:hep-th/1210.0284]}}.

\bibitem[Bousso and Porrati(2017{\natexlab{a}})]{Bousso:2017dny}
Bousso, R.; Porrati, M.
\newblock {Soft Hair as a Soft Wig}.
\newblock {\em Class. Quant. Grav.} {\bf 2017}, {\em 34},~204001,
  \href{http://xxx.lanl.gov/abs/1706.00436}{{\normalfont
  [arXiv:hep-th/1706.00436]}}.

\bibitem[Bousso and Porrati(2017{\natexlab{b}})]{Bousso:2017rsx}
Bousso, R.; Porrati, M.
\newblock {Observable Supertranslations} {\bf 2017}.
\newblock  \href{http://xxx.lanl.gov/abs/1706.09280}{{\normalfont
  [arXiv:hep-th/1706.09280]}}.

\bibitem[Hawking \em{et~al.}(2016{\natexlab{a}})Hawking, Perry, and
  Strominger]{Hawking:2016msc}
Hawking, S.W.; Perry, M.J.; Strominger, A.
\newblock {Soft Hair on Black Holes}.
\newblock {\em Phys. Rev. Lett.} {\bf 2016}, {\em 116},~231301,
  \href{http://xxx.lanl.gov/abs/1601.00921}{{\normalfont
  [arXiv:hep-th/1601.00921]}}.

\bibitem[Hawking \em{et~al.}(2016{\natexlab{b}})Hawking, Perry, and
  Strominger]{Hawking:2016sgy}
Hawking, S.W.; Perry, M.J.; Strominger, A.
\newblock {Superrotation Charge and Supertranslation Hair on Black Holes} {\bf
  2016}.
\newblock  \href{http://xxx.lanl.gov/abs/1611.09175}{{\normalfont
  [arXiv:hep-th/1611.09175]}}.

\bibitem[Afshar \em{et~al.}(2016{\natexlab{a}})Afshar, Detournay, Grumiller,
  Merbis, Perez, Tempo, and Troncoso]{Afshar:2016wfy}
Afshar, H.; Detournay, S.; Grumiller, D.; Merbis, W.; Perez, A.; Tempo, D.;
  Troncoso, R.
\newblock {Soft Heisenberg hair on black holes in three dimensions}.
\newblock {\em Phys. Rev.} {\bf 2016}, {\em D93},~101503,
  \href{http://xxx.lanl.gov/abs/1603.04824}{{\normalfont
  [arXiv:hep-th/1603.04824]}}.

\bibitem[Afshar \em{et~al.}(2016{\natexlab{b}})Afshar, Grumiller, and
  Sheikh-Jabbari]{Afshar:2016uax}
Afshar, H.; Grumiller, D.; Sheikh-Jabbari, M.M.
\newblock {Black Hole Horizon Fluffs: Near Horizon Soft Hairs as Microstates of
  Three Dimensional Black Holes} {\bf 2016}.
\newblock  \href{http://xxx.lanl.gov/abs/1607.00009}{{\normalfont
  [arXiv:hep-th/1607.00009]}}.

\bibitem[Grumiller \em{et~al.}(2016)Grumiller, Perez, Prohazka, Tempo, and
  Troncoso]{Grumiller:2016kcp}
Grumiller, D.; Perez, A.; Prohazka, S.; Tempo, D.; Troncoso, R.
\newblock {Higher Spin Black Holes with Soft Hair}.
\newblock {\em JHEP} {\bf 2016}, {\em 10},~119,
  \href{http://xxx.lanl.gov/abs/1607.05360}{{\normalfont
  [arXiv:hep-th/1607.05360]}}.

\bibitem[Grumiller \em{et~al.}(2017)Grumiller, Perez, Tempo, and
  Troncoso]{Grumiller:2017jft}
Grumiller, D.; Perez, A.; Tempo, D.; Troncoso, R.
\newblock {Log corrections to entropy of three dimensional black holes with
  soft hair}.
\newblock {\em JHEP} {\bf 2017}, {\em 08},~107,
  \href{http://xxx.lanl.gov/abs/1705.10605}{{\normalfont
  [arXiv:hep-th/1705.10605]}}.

\bibitem[Setare and Adami(2016)]{Setare:2016vhy}
Setare, M.R.; Adami, H.
\newblock {The Heisenberg algebra as near horizon symmetry of the black flower
  solutions of Chern-Simons-like theories of gravity} {\bf 2016}.
\newblock  \href{http://xxx.lanl.gov/abs/1606.05260}{{\normalfont
  [arXiv:hep-th/1606.05260]}}.

\bibitem[Ammon \em{et~al.}(2017)Ammon, Grumiller, Prohazka, Riegler, and
  Wutte]{Ammon:2017vwt}
Ammon, M.; Grumiller, D.; Prohazka, S.; Riegler, M.; Wutte, R.
\newblock {Higher-Spin Flat Space Cosmologies with Soft Hair}.
\newblock {\em JHEP} {\bf 2017}, {\em 05},~031,
  \href{http://xxx.lanl.gov/abs/1703.02594}{{\normalfont
  [arXiv:hep-th/1703.02594]}}.

\bibitem[Gaberdiel \em{et~al.}(2011{\natexlab{a}})Gaberdiel, Gopakumar, and
  Saha]{Gaberdiel:2010ar}
Gaberdiel, M.R.; Gopakumar, R.; Saha, A.
\newblock {Quantum $W$-symmetry in $AdS_3$}.
\newblock {\em JHEP} {\bf 2011}, {\em 02},~004,
  \href{http://xxx.lanl.gov/abs/1009.6087}{{\normalfont
  [arXiv:hep-th/1009.6087]}}.

\bibitem[Gaberdiel \em{et~al.}(2011{\natexlab{b}})Gaberdiel, Gopakumar,
  Hartman, and Raju]{Gaberdiel:2011zw}
Gaberdiel, M.R.; Gopakumar, R.; Hartman, T.; Raju, S.
\newblock {Partition Functions of Holographic Minimal Models}.
\newblock {\em JHEP} {\bf 2011}, {\em 08},~077,
  \href{http://xxx.lanl.gov/abs/1106.1897}{{\normalfont
  [arXiv:hep-th/1106.1897]}}.

\bibitem[Creutzig \em{et~al.}(2012)Creutzig, Hikida, and
  Ronne]{Creutzig:2011fe}
Creutzig, T.; Hikida, Y.; Ronne, P.B.
\newblock {Higher spin AdS$_3$ supergravity and its dual CFT}.
\newblock {\em JHEP} {\bf 2012}, {\em 02},~109,
  \href{http://xxx.lanl.gov/abs/1111.2139}{{\normalfont
  [arXiv:hep-th/1111.2139]}}.

\bibitem[Giombi and Klebanov(2013)]{Giombi:2013fka}
Giombi, S.; Klebanov, I.R.
\newblock {One Loop Tests of Higher Spin AdS/CFT}.
\newblock {\em JHEP} {\bf 2013}, {\em 12},~068,
  \href{http://xxx.lanl.gov/abs/1308.2337}{{\normalfont
  [arXiv:hep-th/1308.2337]}}.

\bibitem[Giombi \em{et~al.}(2014{\natexlab{a}})Giombi, Klebanov, and
  Safdi]{Giombi:2014iua}
Giombi, S.; Klebanov, I.R.; Safdi, B.R.
\newblock {Higher Spin AdS$_{d+1}$/CFT$_d$ at One Loop}.
\newblock {\em Phys. Rev.} {\bf 2014}, {\em D89},~084004,
  \href{http://xxx.lanl.gov/abs/1401.0825}{{\normalfont
  [arXiv:hep-th/1401.0825]}}.

\bibitem[Giombi \em{et~al.}(2014{\natexlab{b}})Giombi, Klebanov, and
  Tseytlin]{Giombi:2014yra}
Giombi, S.; Klebanov, I.R.; Tseytlin, A.A.
\newblock {Partition Functions and Casimir Energies in Higher Spin
  AdS$_{d+1}$/CFT$_d$}.
\newblock {\em Phys. Rev.} {\bf 2014}, {\em D90},~024048,
  \href{http://xxx.lanl.gov/abs/1402.5396}{{\normalfont
  [arXiv:hep-th/1402.5396]}}.

\bibitem[Beccaria and Tseytlin(2014)]{Beccaria:2014xda}
Beccaria, M.; Tseytlin, A.A.
\newblock {Higher spins in AdS$_{5}$ at one loop: vacuum energy, boundary
  conformal anomalies and AdS/CFT}.
\newblock {\em JHEP} {\bf 2014}, {\em 11},~114,
  \href{http://xxx.lanl.gov/abs/1410.3273}{{\normalfont
  [arXiv:hep-th/1410.3273]}}.

\bibitem[Beccaria and Tseytlin(2015)]{Beccaria:2015vaa}
Beccaria, M.; Tseytlin, A.A.
\newblock {On higher spin partition functions}.
\newblock {\em J. Phys.} {\bf 2015}, {\em A48},~275401,
  \href{http://xxx.lanl.gov/abs/1503.08143}{{\normalfont
  [arXiv:hep-th/1503.08143]}}.

\bibitem[Campoleoni \em{et~al.}(2016)Campoleoni, Gonzalez, Oblak, and
  Riegler]{Campoleoni:2015qrh}
Campoleoni, A.; Gonzalez, H.A.; Oblak, B.; Riegler, M.
\newblock {Rotating Higher Spin Partition Functions and Extended BMS
  Symmetries}.
\newblock {\em JHEP} {\bf 2016}, {\em 04},~034,
  \href{http://xxx.lanl.gov/abs/1512.03353}{{\normalfont
  [arXiv:hep-th/1512.03353]}}.

\bibitem[Krishnan and Roy(2014)]{Krishnan:2013tza}
Krishnan, C.; Roy, S.
\newblock {Desingularization of the Milne Universe}.
\newblock {\em Phys. Lett.} {\bf 2014}, {\em B734},~92--95,
  \href{http://xxx.lanl.gov/abs/1311.7315}{{\normalfont
  [arXiv:hep-th/1311.7315]}}.

\bibitem[Gonzalez and Pino(2014)]{Gonzalez:2014tba}
Gonzalez, H.A.; Pino, M.
\newblock {Boundary dynamics of asymptotically flat 3D gravity coupled to
  higher spin fields}.
\newblock {\em JHEP} {\bf 2014}, {\em 05},~127,
  \href{http://xxx.lanl.gov/abs/1403.4898}{{\normalfont
  [arXiv:hep-th/1403.4898]}}.

\bibitem[Barnich \em{et~al.}(2013)Barnich, Gomberoff, and
  González]{Barnich:2012rz}
Barnich, G.; Gomberoff, A.; González, H.A.
\newblock {Three-dimensional Bondi-Metzner-Sachs invariant two-dimensional
  field theories as the flat limit of Liouville theory}.
\newblock {\em Phys. Rev.} {\bf 2013}, {\em D87},~124032,
  \href{http://xxx.lanl.gov/abs/1210.0731}{{\normalfont
  [arXiv:hep-th/1210.0731]}}.

\bibitem[Taylor(2016)]{Taylor:2015glc}
Taylor, M.
\newblock {Lifshitz holography}.
\newblock {\em Class. Quant. Grav.} {\bf 2016}, {\em 33},~033001,
  \href{http://xxx.lanl.gov/abs/1512.03554}{{\normalfont
  [arXiv:hep-th/1512.03554]}}.

\bibitem[Bacry and Levy-Leblond(1968)]{Bacry:1968zf}
Bacry, H.; Levy-Leblond, J.
\newblock {Possible kinematics}.
\newblock {\em J. Math. Phys.} {\bf 1968}, {\em 9},~1605--1614.

\bibitem[Saletan(1961)]{Saletan:61}
Saletan, E.J.
\newblock Contraction of Lie Groups.
\newblock {\em Journal of Mathematical Physics} {\bf 1961}, {\em 2},~1--21.

\bibitem[Bergshoeff \em{et~al.}(2017)Bergshoeff, Grumiller, Prohazka, and
  Rosseel]{Bergshoeff:2016soe}
Bergshoeff, E.; Grumiller, D.; Prohazka, S.; Rosseel, J.
\newblock {Three-dimensional Spin-3 Theories Based on General Kinematical
  Algebras}.
\newblock {\em JHEP} {\bf 2017}, {\em 01},~114,
  \href{http://xxx.lanl.gov/abs/1612.02277}{{\normalfont
  [arXiv:hep-th/1612.02277]}}.

\bibitem[Duval \em{et~al.}(2014)Duval, Gibbons, and Horvathy]{Duval:2014uva}
Duval, C.; Gibbons, G.W.; Horvathy, P.A.
\newblock {Conformal Carroll groups and BMS symmetry}.
\newblock {\em Class. Quant. Grav.} {\bf 2014}, {\em 31},~092001,
  \href{http://xxx.lanl.gov/abs/1402.5894}{{\normalfont
  [arXiv:gr-qc/1402.5894]}}.

\bibitem[Hartong(2015)]{Hartong:2015xda}
Hartong, J.
\newblock {Gauging the Carroll Algebra and Ultra-Relativistic Gravity}.
\newblock {\em JHEP} {\bf 2015}, {\em 08},~069,
  \href{http://xxx.lanl.gov/abs/1505.05011}{{\normalfont
  [arXiv:hep-th/1505.05011]}}.

\bibitem[Medina and Revoy(1985)]{Medina1985}
Medina, A.; Revoy, P.
\newblock Alg\`{e}bres de Lie et produit scalaire invariant.
\newblock {\em Annales scientifiques de l'\'{E}cole Normale Sup\'{e}rieure}
  {\bf 1985}, {\em 18},~553--561.

\bibitem[Figueroa-O'Farrill and Stanciu(1996)]{FigueroaO'Farrill:1995cy}
Figueroa-O'Farrill, J.M.; Stanciu, S.
\newblock {On the structure of symmetric selfdual Lie algebras}.
\newblock {\em J. Math. Phys.} {\bf 1996}, {\em 37},~4121--4134,
  \href{http://xxx.lanl.gov/abs/hep-th/9506152}{{\normalfont
  [arXiv:hep-th/hep-th/9506152]}}.

\bibitem[Papageorgiou and Schroers(2009)]{Papageorgiou:2009zc}
Papageorgiou, G.; Schroers, B.J.
\newblock {A Chern-Simons approach to Galilean quantum gravity in 2+1
  dimensions}.
\newblock {\em JHEP} {\bf 2009}, {\em 11},~009,
  \href{http://xxx.lanl.gov/abs/0907.2880}{{\normalfont
  [arXiv:hep-th/0907.2880]}}.

\bibitem[Levy-L\'{e}blond(1971)]{levygalgr}
Levy-L\'{e}blond, J.M.
\newblock Galilei Group and Galilean Invariance. In {\em Group Theory and its
  Applications}; Loebl, E.M., Ed.; Academic Press,  1971; pp. 221 -- 299.

\bibitem[Rasmussen and Raymond(2017)]{Rasmussen:2017eus}
Rasmussen, J.; Raymond, C.
\newblock {Galilean contractions of $W$-algebras}.
\newblock {\em Nucl. Phys.} {\bf 2017}, {\em B922},~435--479,
  \href{http://xxx.lanl.gov/abs/1701.04437}{{\normalfont
  [arXiv:hep-th/1701.04437]}}.

\bibitem[Witten(1989)]{Witten:1988hf}
Witten, E.
\newblock {Quantum Field Theory and the Jones Polynomial}.
\newblock {\em Commun.Math.Phys.} {\bf 1989}, {\em 121},~351--399.

\bibitem[Elitzur \em{et~al.}(1989)Elitzur, Moore, Schwimmer, and
  Seiberg]{Elitzur:1989nr}
Elitzur, S.; Moore, G.W.; Schwimmer, A.; Seiberg, N.
\newblock {Remarks on the Canonical Quantization of the Chern-Simons-Witten
  Theory}.
\newblock {\em Nucl. Phys.} {\bf 1989}, {\em B326},~108.

\bibitem[Mohammedi(1994)]{Mohammedi:1993rg}
Mohammedi, N.
\newblock {On bosonic and supersymmetric current algebras for nonsemisimple
  groups}.
\newblock {\em Phys. Lett.} {\bf 1994}, {\em B325},~371--376,
  \href{http://xxx.lanl.gov/abs/hep-th/9312182}{{\normalfont
  [arXiv:hep-th/hep-th/9312182]}}.

\bibitem[Figueroa-O'Farrill and Stanciu(1994)]{FigueroaO'Farrill:1994hx}
Figueroa-O'Farrill, J.M.; Stanciu, S.
\newblock {Nonsemisimple Sugawara constructions}.
\newblock {\em Phys. Lett.} {\bf 1994}, {\em B327},~40--46,
  \href{http://xxx.lanl.gov/abs/hep-th/9402035}{{\normalfont
  [arXiv:hep-th/hep-th/9402035]}}.

\bibitem[Benamor and Benayadi(1999)]{99BB}
Benamor, H.; Benayadi, S.
\newblock Double extension of quadratic lie superalgebras.
\newblock {\em Communications in Algebra} {\bf 1999}, {\em 27},~67--88,
  \href{http://xxx.lanl.gov/abs/http://dx.doi.org/10.1080/00927879908826421}{{\normalfont
  [http://dx.doi.org/10.1080/00927879908826421]}}.

\bibitem[{Bajo} \em{et~al.}(2007){Bajo}, {Benayadi}, and {Bordemann}]{07BBB}
{Bajo}, I.; {Benayadi}, S.; {Bordemann}, M.
\newblock {Generalized double extension and descriptions of qadratic Lie
  superalgebras}.
\newblock {\em ArXiv e-prints} {\bf 2007},
  \href{http://xxx.lanl.gov/abs/0712.0228}{{\normalfont
  [arXiv:math-ph/0712.0228]}}.

\bibitem[Bergshoeff and Rosseel(2016)]{Bergshoeff:2016lwr}
Bergshoeff, E.A.; Rosseel, J.
\newblock {Three-Dimensional Extended Bargmann Supergravity}.
\newblock {\em Phys. Rev. Lett.} {\bf 2016}, {\em 116},~251601,
  \href{http://xxx.lanl.gov/abs/1604.08042}{{\normalfont
  [arXiv:hep-th/1604.08042]}}.

\bibitem[Hartong \em{et~al.}(2016)Hartong, Lei, and Obers]{Hartong:2016yrf}
Hartong, J.; Lei, Y.; Obers, N.A.
\newblock {Nonrelativistic Chern-Simons theories and three-dimensional
  Hořava-Lifshitz gravity}.
\newblock {\em Phys. Rev.} {\bf 2016}, {\em D94},~065027,
  \href{http://xxx.lanl.gov/abs/1604.08054}{{\normalfont
  [arXiv:hep-th/1604.08054]}}.

\end{thebibliography}


\end{document}